\documentclass[prd,nofootinbib,preprint,superscriptaddress]{revtex4}

\usepackage{amsmath, amssymb, amsthm, graphicx, epsfig, fancyhdr,epsfig, slashed, mathrsfs}

\usepackage{tikzsymbols}
\usepackage{tikz,xcolor
}
\usepackage{natbib}
\usepackage{float}

\usepackage{mathtools}

\definecolor{cadmiumred}{rgb}{0.89, 0.0, 0.13}
\definecolor{candyapplered}{rgb}{1.0, 0.03, 0.0}
\definecolor{copper}{rgb}{0.72, 0.45, 0.2}
\definecolor{darkmagenta}{rgb}{0.55, 0.0, 0.55}

\usepackage{hyperref}
\hypersetup{colorlinks,linkcolor={green!50!black},citecolor={cyan},urlcolor={lavender(floral)}}
\definecolor{lavender(floral)}{rgb}{0.71, 0.49, 0.86}
\definecolor{mediumblue}{rgb}{0.0, 0.0, 0.8}

\definecolor{lime}{HTML}{A6CE39}
\DeclareRobustCommand{\orcidicon}{
	\begin{tikzpicture}
	\draw[lime, fill=lime] (0,0) 
	circle [radius=0.2] 
	node[white] {{\fontfamily{qag}\selectfont \tiny ID}};
	\draw[white, fill=white] (-0.0625,0.095) 
	circle [radius=0.007];
	\end{tikzpicture}
	\hspace{-2mm}
}

\foreach \x in {Anish,Maxim, Shila,Lalak}{\expandafter\xdef\csname orcid\x\endcsname{\noexpand\href{https://orcid.org/\csname orcidauthor\x\endcsname}
			{\noexpand\orcidicon}}
}

\newcommand{\be}{\begin{equation}}
\newcommand{\ee}{\end{equation}}
\newcommand{\bea}{\begin{eqnarray}}
\newcommand{\eea}{\end{eqnarray}}

\definecolor{dukeblue}{rgb}{0.0, 0.0, 0.61}

\def\hds#1{\href{https://doi.org/#1}}

\newcommand{\spp}[1]{\textcolor{red}{[Shila: #1]}}

 \usepackage{
 ulem,cancel,
 multirow,
 slashed,
 upgreek, 
 physics, cases}
 \usepackage{easyReview}
\usepackage{bbold}

\newcommand{\eq}[1]{\begin{align}#1\end{align}}

\definecolor{amber(sae/ece)}{rgb}{1.0, 0.49, 0.0}

\definecolor{blue(ncs)}{rgb}{0.0, 0.53, 0.74}

\definecolor{darkviolet}{rgb}{0.58, 0.0, 0.83}

\def\mg#1{\color{magenta}{\bf #1}}

\definecolor{americanrose}{rgb}{1.0, 0.01, 0.24}

\definecolor{jazzberryjam}{rgb}{0.65, 0.04, 0.37}
\definecolor{jonquil}{rgb}{0.98, 0.85, 0.37}
\definecolor{mikadoyellow}{rgb}{1.0, 0.77, 0.05}
\definecolor{schoolbusyellow}{rgb}{1.0, 0.85, 0.0}

\newcommand{\ba}{\begin{eqnarray}}
\newcommand{\ea}{\end{eqnarray}}

\newcommand{\unit}[1]{\, \text{#1}}

\newcommand{\lt}{\left }
\newcommand{\rt}{\right }
\newcommand{\seq}{\simeq}
\newcommand{\gsim}{\gtrsim}
\newcommand{\lsim}{\lesssim}

\def\gs{g_{\star}}
\def\gss{g_{\star, s}}
\def\mpl{M_P}
\def\GeV{\unit{GeV}}
\def\MeV{\unit{MeV}}

\def\cmb{\text{CMB}}

\def\ncmb{N_{\cmb}}

\newcommand{\Planck}{\textit{Planck}}
\newcommand{\Keck}{\textit{Keck}}
\newcommand{\KeckArray}{\textit{Keck Array}}
\newcommand{\BICEP}{\textsc{Bicep}}

\newcommand{\cmbsfour}{{CMB-S4}}
\newcommand{\LB}{\textit{LiteBIRD}}

\newcommand{\WMAP}{\textsl{WMAP}}
\newcommand{\wmap}{{\WMAP}}

\def\Trh{T_{\rm rh}}
\def\rh{{\rm rh}}

\def\Br{\text{Br}}
\def\cc{\bar{\chi}\chi}
\def\ccp{\chi\chi}

\def\sm{\text{SM}}
\def\bsm{\text{BSM}}
\def\dm{\text{DM}}
\def\cdm{CDM}
\def\bbn{BBN}

\def\ncmb{{\cal N}_{\cmb}}

\def\td{\text{d}}
\def\pd{\partial}

\def\g{\gamma}
\def\n{\nu}
\def\r{\rho}
\def\a{\alpha}
\def\b{\beta}
\def\m{\mu}
\def\G{\Gamma}
\def\L{\Lambda}
\def\c{\chi}

\def\cw{\text{CW}}

\def\dz{\mathbb{x}}
\def\lop{{\rm 1-loop}}

\def\cs{\boldsymbol{a}}
\def\hubble{{\cal H}}

\def\cc{\bar{\chi}\chi}
\def\lH{\lambda_H}

\def\mH{m_H}
\def\mc{m_{\chi}}
\def\yc{y_\c}
\def\Yc{Y_\c}
\def\Yco{Y_{\c,0}}

\def\lO{\lambda_{12}}
\def\ld{\lambda_{12}}
\def\lT{\lambda_{22}}
\def\ls{\lambda_{22}}
\def\HH{H^\dagger H}

\def\Trh{T_{\rm rh}}
\def\Tmax{T_{max}}

\def\Br{\text{Br}}

\def\one{2-to-2 scattering of non-relativistic inflaton with graviton as the mediator}
\def\two{2-to-2 scattering of \sm~particles with graviton as mediator with $\mc\ll\Trh$}
\def\three{2-to-2 scattering of \sm~particles with graviton as mediator with $\Tmax\gg \mc\gg\Trh$}
\def\four{2-to-2 scattering of \sm~particles with inflaton as mediator with $\Trh \ll m_\phi$}

\def\vgh{V_{\rm GH}}
\def\vcw{V_{\cw}}

\def\cq{{\cal Q}}

\usepackage[capitalise]{cleveref}
\Crefname{figure}{Fig.}{Figs.}
\Crefname{section}{Sec.}{Secs.}

\begin{document}


\title{Post-inflationary production of particle Dark Matter: \\ \it{Hilltop and Coleman-Weinberg inflation}}

\author{Anish Ghoshal\orcidAnish{}}
\email{anish.ghoshal@fuw.edu.pl}
\affiliation{Institute of Theoretical Physics, Faculty of Physics, University of Warsaw, ul. Pasteura 5, 02-093 Warsaw, Poland}

\author{Maxim~Yu.~Khlopov\orcidMaxim{}}
\email{khlopov@apc.in2p3.fr}
\affiliation{Virtual Institute of Astroparticle physics, 75018 Paris, France}
\affiliation{Institute of Physics, Southern Federal University, 344090 Rostov on Don, Russia}
\affiliation{National Research Nuclear University “MEPHI”, 115409 Moscow, Russia}

\author{Zygmunt Lalak\orcidLalak{}}
\email{zygmunt.lalak@fuw.edu.pl}
\affiliation{Institute of Theoretical Physics, Faculty of Physics, University of Warsaw, ul. Pasteura 5, 02-093 Warsaw, Poland}

\author{Shiladitya Porey\orcidShila{}}
\email{shiladityamailbox@gmail.com}
\affiliation{Department of Physics, Novosibirsk State University, Pirogova 2, 630090 Novosibirsk, Russia}

\begin{abstract}

\textit{
We investigate the production of non-thermal dark matter (DM), $\chi$, during post-inflationary reheating era. 
For inflation, we consider two slow roll single field inflationary scenarios – generalized version of Hilltop (GH) inflation, and Coleman-Weinberg (CW) inflation. 
Using a set of benchmark values that comply with the current constraints from Cosmic Microwave Background Radiation (CMBR) data for each inflationary model, we explored the parameter space involving mass of dark matter particles, $m_\chi$, and coupling between inflaton and $\chi$, $y_\chi$. For these benchmarks, we find that tensor-to-scalar ratio $r$ can be as small as $2.69\times 10^{-6}$ for GH and $1.91\times 10^{-3}$ for CW inflation, both well inside $1-\sigma$ contour on scalar spectral index versus $r$ plane from \Planck2018+\BICEP3+\KeckArray2018 dataset, and testable by future cosmic microwave background (CMB) observations e.g. Simons Observatory.  
For the production of $\chi$ from the inflaton decay satisfying CMB and other cosmological bounds and successfully explaining total cold dark matter density of the present universe, we find that $y_\chi$ should be within this range 
 ${\cal O}\left(10^{-4}\right) \gtrsim y_\chi\gtrsim {\cal O}\left(10^{-20}\right)$ for both inflationary scenarios. We also show that, even for the same inflationary scenario, the allowed parameter space on reheating temperature versus $m_\chi$ plane alters with inflationary parameters including scalar spectral index, $r$,  and energy scale of inflation.  }
\end{abstract}

\maketitle

\section{Introduction}
The $\Lambda$CDM cosmological model of the universe is undoubtedly successful in explaining not only the origin of the universe and its transformation from a very hot to a cold stage but also accurately depicts each cosmological epoch from the era of BBN up to the present-day universe. After the observation and subsequent analysis of CMB data obtained from COBE, \WMAP
, and later from the \Planck~mission, this cosmological model was bolstered. However, the $\Lambda$CDM model, along with the standard model of particle physics, fails to shed light on the nature of \dm. There is a possibility that DM can potentially exist in the form of primordial black holes (PBHs) or massive compact halo objects (MACHOs), but it cannot account for the total density of cold dark matter (CDM)~\cite{Carr:2020xqk,Dolgov:2019vlq}. Therefore, DM is believed to be in the form of particles, specifically beyond the standard model (BSM) particles, such as the popular candidate known as weakly interacting massive particles (WIMPs). These particles were expected to be in thermal equilibrium with the standard model particles, also known as radiation, in the early hot universe and subsequently decoupled at a later stage, depending on the temperature of radiation, the mass of the WIMP, and the cross-section of its interaction with standard model particles. However, the failure of particle detectors to detect the presence of WIMPs has led to the consideration of alternative scenarios, such as feebly interacting massive particles (FIMPs), which were never in thermal equilibrium with radiation~\cite{McDonald:2001vt, Choi:2005vq, Kusenko:2006rh, Petraki:2007gq, Hall:2009bx, Bernal:2017kxu,Haque:2021mab,Haque:2022kez,Haque:2023yra}. Consequently, the number density of FIMP is independent of initial number density and can be produced either from decay of massive particles, such as moduli field or curvaton~\cite{Baer:2014eja} or inflaton, or from the scattering of standard model (\sm) particle or inflaton via gravitational interaction\cite{Garny:2015sjg, Tang:2016vch, Tang:2017hvq, Garny:2017kha, Bernal:2018qlk}. 
If these feebly interacting particles are not massive, they 
can also contribute to other observables like the measurements of dark radiation as $\Delta N_{\rm eff}$ in CMB etc. \cite{Paul:2018njm,Ghoshal:2023phi}.

In addition to supporting $\Lambda$CDM, precise measurements of the cosmic microwave background (CMB) also reveal that the proper explanation for some features of the universe $\Lambda$CDM model fails to incorporate. These problems include large-scale homogeneity, nearly small values of inhomogeneity, and  description of the formation of large-scale structures. In this context, a short period of exponential expansion known as cosmic inflation~\cite{Starobinsky:1980te, Guth:1980zm, Linde:1981mu, Albrecht:1982wi} during the infant universe has the ability to address these issues. In general, dark matter and inflation are unrelated scenarios. Nonetheless, inflation and
reheating both form the frameworks for the production of particles beyond the SM of particles. Since we know the SM cannot accommodate DM particle, and it must be from beyond the SM sector, it is naturally very interesting to look for the possibility
of the production of dark matter particles during inflation. Following the inflationary era, a period of reheating is also required to make the universe hot and dominated by relativistic standard model particles, commonly referred to as radiation. The simplest possibility is that the universe is driven by a single scalar field, the inflaton, which is a Standard Model gauge singlet studied in several works~\cite{Lerner:2009xg, Kahlhoefer:2015jma}. After the BICEP+PLANCK observations, inflationary potentials with concave shapes or flat potentials for the inflaton have gained favorability. Models incorporating non-minimal couplings between inflaton and Ricci scalar have been extensively explored (e.g. Refs.~\cite{Clark:2009dc, Khoze:2013uia, Almeida:2018oid, Bernal:2018hjm, Aravind:2015xst, Ballesteros:2016xej, Borah:2018rca,  Hamada:2014xka, Choubey:2017hsq, Cline:2020mdt, Tenkanen:2016twd, Abe:2020ldj}). Additionally, there are several models that consider flat potentials without coupling, such as inflection point inflation. These models can also incorporate dark matter, such as  SMART $U(1)_X$ \cite{Okada:2020cvq} (see also~\cite{Das:2022oyx,Das:2016zue,Das:2015nwk}),  model with a single axion-like particle~\cite{Daido:2017wwb}, the $\nu$MSM~\cite{Shaposhnikov:2006xi}, the NMSM~\cite{Davoudiasl:2004be}, the  WIMPflation~\cite{Hooper:2018buz}, and extension with a complex flavon field \cite{Ema:2016ops}. Very recently testability of FIMP DM involving long-lived particle searches at laboratories involving various portals (categorized via spin of the mediators) \cite{Barman:2023ktz,Barman:2022scg,Barman:2022njh,Barman:2021yaz,Barman:2021lot,Ghosh:2023tyz,Chakraborty:2023ocr} and involving primordial Gravitational Waves of inflationary origin have been proposed \cite{Ghoshal:2022ruy,Berbig:2023yyy,Paul:2018njm,Ghoshal:2022jdt}.

In this work, alongside our previous work~\cite{Ghoshal:2023jvf}, we address both the inflation and dark matter including two different BSM fields, one is boson and another one is fermion.
For the inflationary part we have considered Hilltop inflation. Any inflationary potential can be approximated as Hilltop near its maximum. Then we have considered Coleman-Weinberg inflation. We derive the conditions on DM parameter space from the analysis of inflationary constraint and constraints on inflaton-DM couplings, inflaton and DM masses based on DM phenomenology.

	\textit{The paper is organized as follows}: we begin in~\cref{Sec:Inflationary models} where we introduce the Lagrangian density of our model. In~\cref{Sec:Generalized Hilltop}, we study the slow-roll inflationary scenario, reheating, and production of \dm~in Generalized Hilltop scenario with canonical kinetic energy of the inflaton. Following this,
 in~\cref{Sec:CW}, we explore  the same aspects for Coleman-Weinberg inflation.
 Finally, we summarize our results in~\cref{Sec:Conclusion}.

In this work, we 
assume that the spacetime metric is diagonal 
with signature $(+,-,-,-)$. We also use $\hbar=c=k_B=1$ unit in which reduced Planck mass is $\mpl=2.4\times 10^{18}\GeV$.


\section{Lagrangian Density}
\label{Sec:Inflationary models}
%
%
In addition to the Standard Model (\sm)~Higgs field $H$ our model includes two other beyond the standard model (\bsm)~fields: a real scalar inflaton $\phi$ and a vector-like singlet (under \sm~gauge groups) fermionic field $\chi$ that plays the role of a non-thermal \dm. As a result, we write the action as~\cite{Bernal:2021qrl,Ghoshal:2022jeo,Ghoshal:2023noe,Ghoshal:2022aoh,Ghoshal:2024ycp}
\be\label{Eq:Action}
{\cal S}= \int \td x^4 \sqrt{-g}\, \lt( {\cal L}_{\rm INF} ({\cal R},\phi) + {\cal L}_\chi + {\cal L}_H + {\cal L}_{\rh} \rt)\,.
\ee 
Here, $g$ and ${\cal R}$ are the determinant of spacetime metric and curvature scalar, respectively, and ${\cal L}_{\rm INF} ({\cal R},\phi)$ is the Lagrangian density of the slow-roll inflationary scenario driven by $\phi$, and the Lagrangian density of $\chi$ and \sm~Higgs doublet $H$ are as follows: 
\eq{
{\cal L}_\chi &= i \bar{\c}\slashed{\partial} \chi - m_\c \cc \,, \label{Eq:Lagrangian-for-chi}\\
{\cal L}_H &= \lt(\partial H\rt)^2 +\mH^2 \HH  - \lH \lt( \HH \rt)^2 \,,\label{Eq:Lagrangian-for-H}
}
where $\slashed{\partial}$ is the Dirac operator in Feynman's slash notation, $\mc$ and $\mH$ both have the mass dimension whereas $\lH$ is dimensionless. Since we consider production of \dm~particle, $\chi$, and \sm~Higgs particle, $h$, during reheating era, we
can write the interaction Lagrangian as 
\be\label{Eq:reheating lagrangian}
\mathcal{L}_{\rh} = - \yc \phi \cc - \lO \phi \HH - \lT \phi^2 \HH +{\cal L}_{\rm scatter} + \text{h.c.} \,.
\ee 
where ${\cal L}_{\rm scatter}$ includes higher order terms that account for the scattering of $\c$ by inflaton, or \sm~particles (including $H$). The couplings $\yc$ and $\ls$ are also dimensionless but $\ld$ has mass dimension.
Now, in the following sections, we analyze two inflationary models while considering benchmark values that satisfy current constraints from \cmb. Additionally, with the assumption that the \dm~is produced during reheating era, we explore the parameter space involving $\yc$ and $\mc$ such that $\c$ becomes accountable for the total \cdm~density of the present universe.


\section{Generalized Hilltop}
\label{Sec:Generalized Hilltop}
One of the papers~\cite{Planck:2018jri} from \Planck~2018 collaboration features a set of single field slow-roll inflationary scenarios, with Hilltop inflation being one of them. This specific model with potential of the form $\L^4 \lt(1-\phi^4/\m_4^4+ \cdots \rt)$ and with $-2<\log_{10}(\m_4/\mpl)<2$ satisfies predictions within $1-\sigma$ C.L. on $(n_s,r)$ plane. 
The three dots represent the missing higher order terms, which are expected to stabilize the potential from below just after the end of inflation and prevent the universe from collapsing.
Although, for $\phi_0\ll \mpl$, terms of higher order of $\qty(\phi/\phi_0)$ can be added to the potential to stabilize without altering the $(n_s,r)$ predictions, the predicted value of $n_s$ of this inflationary model is small compared to the current best fit value obtained from \Planck~2018~\cite{Kallosh:2019jnl,Kallosh:2019hzo,Kallosh:2021mnu}. 
However, the most recent combined data of $B$-mode of polarization of \cmb~from \Planck, \wmap, and \BICEP/\Keck~strongly favors a concave rather than convex shape of the inflationary potential. Any concave potential can be approximated as a Hilltop potential around the local maximum~\cite{Lillepalu:2022knx}. Thus, in this work, we consider a generalized form of the Hilltop potential and thus this inflationary model is referred as Generalized Hilltop (abbreviated as GH) inflationary scenario. 
The Lagrangian density and potential density of this inflationary model are given by~\cite{Hoffmann:2021vty,Lillepalu:2022knx} (see also~\cite{Kallosh:2019jnl,Kallosh:2019hzo,Kallosh:2021mnu,Boubekeur:2005zm,German:2020rpn, Dimopoulos:2020kol})
\eq{
&{\cal L}_{\rm INF} ({\cal R},\phi)= \frac{\mpl^2}{2} {\cal R} + \frac{1}{2}\pd_\m \phi \pd^\n \phi - \vgh(\phi) \,,   \label{Eq:GH-Lagrangian}\\
&\vgh(\phi)=V_0 \left(1- \left( \frac{\phi}{\phi_0}\right)^m \right)^n\,,  \label{Eq:GH-potential}
}
where $V_0$ and $\phi_0$ have the mass dimension of $4$ and $1$, respectively. We expect both $V_0$ and $\phi_0$ are positive. In~\cref{Eq:GH-potential}, if either of $m$, $n$, or both are fractions, the potential has discontinuity either at $\phi=0$ or at $\phi=\phi_0$. However, if both $m$, and $n$ are positive integers, the potential is continuous at both $\phi=0$ and $\phi=\phi_0$. Henceforth, we consider only positive integer values of $m$ and $n$.
Moreover, when $n$ is positive integer (and for $\phi_0>0$), the inflaton descends along the slope of the $\vgh(\phi)$ from small to large values of $\phi$. Additionally, for positive integer value of $n$, the potential is symmetric (asymmetric) about the origin if $m$ is an even (odd) number. 
If $n$ is an even number (with a positive integer value of $m$), then the potential is bounded from below. The potential in~\cref{Eq:GH-potential} can have two stationary points: $\vgh^\prime(\phi)=0$ at $\phi=0$ if $m>1$,  $\vgh^{\prime}(\phi)=0$ at $\phi=\phi_0$ if $n>1$. Here (and also throughout this article), prime denotes derivative with respect to $\phi$. Additionally, $\vgh^{\prime\prime}(\phi)=0$ at $\phi=0$ for $m>2$ and $\vgh^{\prime\prime}(\phi)=0$ at $\phi=\phi_0$ for $n>2$.
%
%
%
%
%
%
Therefore, we choose $n=2$ 
such that $\vgh(\phi)$ remains continuous, finite, real, and bounded from below for $\phi\geq \phi_0$ and also there exists a minimum at $\phi=\phi_0$. 
When $n=m=2$, $(n_s,r)$ prediction for that potential does not match well with the \cmb~data
~\cite{Lillepalu:2022knx}.
Thus, we choose $m>n$ which is also needed to satisfy $n_s$ from the current \cmb~bound, as shown in Ref.~\cite{Lillepalu:2022knx}. The chosen benchmark values for GH inflation are shown in~\cref{Table:benchmark_values_GH}. Benchmark `GH-BM1' is for large field inflation, while `GH-BM2' is for small field inflation. Benchmark `GH-BM3' is for the inflationary scenario described in Ref.~\cite{Hoffmann:2021vty}

\begin{table}[H]
    \centering
        \caption{ \it Benchmark values for GH 
        inflation ($\ncmb\approx 60$).}
    \label{Table:benchmark_values_GH}
\vspace{-10pt}
\begin{center}
\vspace{0.5pt}
\begin{tabular}{ |c||c|c| c |c|c| c |c|c|c|}
\hline
{\it Benchmark} & &
$\phi_{0}/M_P$ &
$\phi_{*}/M_P$ & $\phi_{\rm end}/M_P$ &
$V_0$ &
$n_s$ &
$r$ \\
\hline 
\hline 
GH-BM1 & $m=10, n=2$& $16$ & $10.0544$ & $14.9705 $  
&$
9.4255\times 10^{-11} $ & $0.96473$ & $2.9754\times 10^{-3}$\\
\hline
GH-BM2 & $ m=15, n=2$ & $1$ & $ 0.4606$ & $ 0.6284$ & $
8.3643\times 10^{-14}$ & $0.964736$ & $2.6918\times 10^{-6} $ \\
\hline
GH-BM3 & $m=4, n=2$ & $20$ & $9.7845$ & $18.7226$ & $
6.9045\times10^{-10}$ & $0.964597$ & $1.9747\times 10^{-2}$ \\
\hline
\end{tabular}
\end{center}
\end{table}

The slow-rolling condition of the inflation is generally expressed in terms of potential-slow-roll parameters. The first two potential-slow-roll parameters for a single inflaton whose kinetic energy is minimally connected to gravity, as in~\cref{Eq:GH-Lagrangian}, are defined as 
\eq{
 & \epsilon_V (\phi) =\frac{\mpl^2}{2} \left(\frac{\vgh'(\phi )}{\vgh(\phi )}\right)^2 =\mpl^2\frac{m^2 \, n^2  \, \phi ^{2 m-2}}{2 \left(\phi^m-\phi_0^m\right)^2}
  \,, \label{Eq:GH-EpsilonV}\\
 & \eta_V (\phi) = \mpl^2\frac{\vgh''(\phi )}{\vgh(\phi )} 
 =\mpl^2\frac{m \, n \, \phi ^{m-2} \left((m n-1) \phi^m-(m-1) \phi_0^m\right)}{\left(\phi
   ^m-\phi_0^m\right)^2}
 \,. \label{Eq:GH-EtaV}
}
In order to maintain slow roll inflation, it is required that both $\epsilon_V$, and $\lt|\eta_V\rt|$ are $<1$. If either condition is violated, it signals the end of the slow roll inflationary epoch. The duration of inflation is parameterized in terms of the number of e-folds, $\ncmb$, which 
indicates the amount of exponential expansion of the cosmological scale factor $\cs$
or the amount of reduction of comoving Hubble radius ($(\cs\, \hubble)^{-1}$, $\hubble$ being the Hubble parameter) that occurs during inflation, is defined as 
\be
\ncmb = {1\over \mpl^2}\int_{\phi_{\rm end}}^{\phi_*} {\vgh(\phi)\over \vgh'(\phi)} \,\td \phi\, .
\ee
 Here, $\phi_{\rm end}$ is the value of the inflaton when inflation ends, and $\phi_*$ is the value of inflaton at which the cosmic length-scale of \cmb~observation corresponding to e-fold $\ncmb$ leaves the comoving Hubble radius during inflation. $\ncmb\geq 60$ is required to solve Horizon problem~\cite{Baumann:2022mni}. Therefore, in this work, we choose benchmark values corresponding to $\ncmb\approx 60$ 

\begin{figure}[H]
    \centering    
    \includegraphics[width=0.48\linewidth]{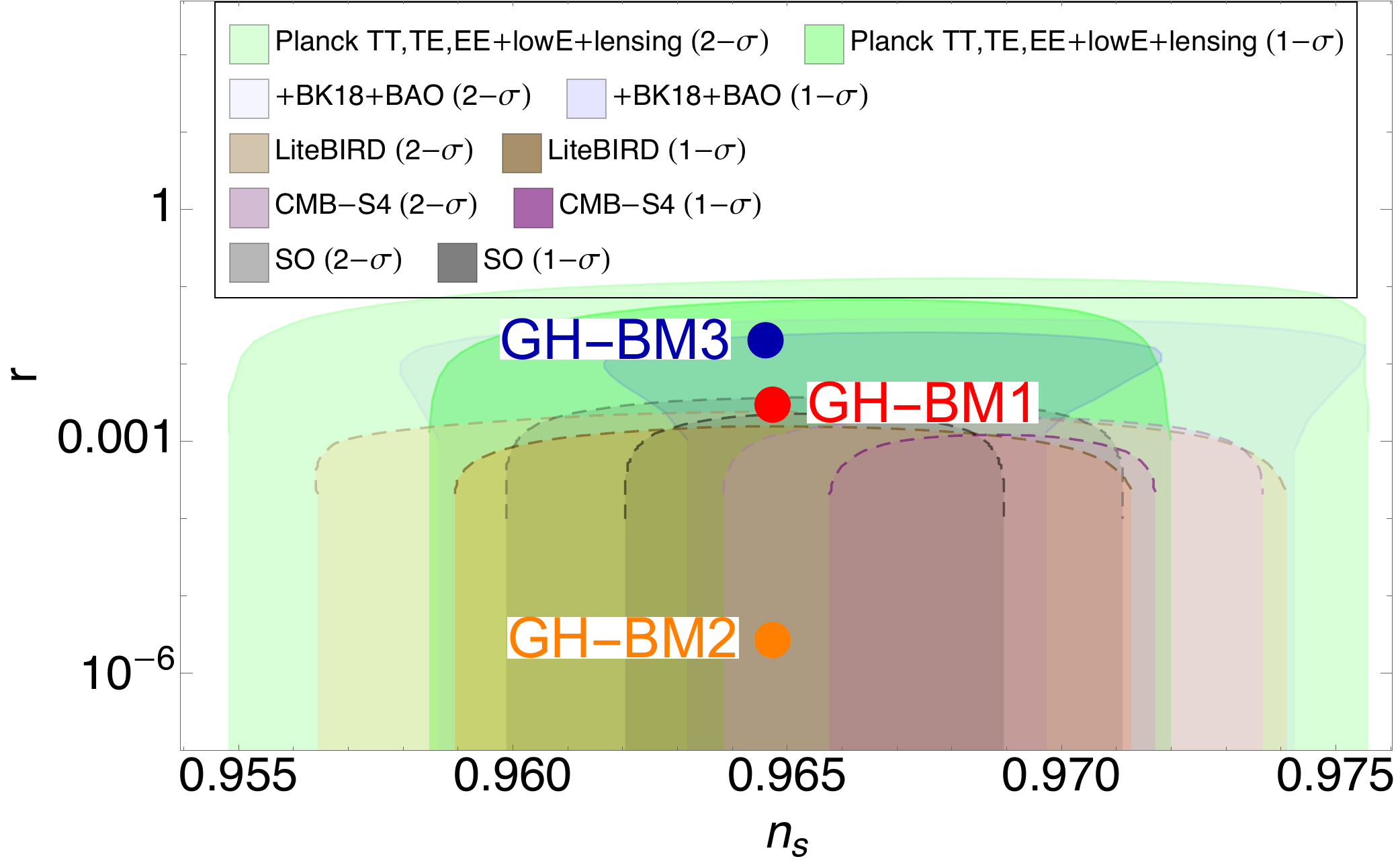} 
    \caption{\it \raggedright \label{Fig:GH-ns-r-bound} 
    The predicted values of $n_s, r$ for the three benchmark values from~\cref{Table:benchmark_values_GH} are displayed as colored circular dots. The $n_s-r$ contours at $1-\sigma$ and $2-\sigma$ C.L. from present and future \cmb~observations are also displayed in the background. We use deep colored region (located inside) for $95\%$ ($1-\sigma$ C.L.) and light colored region (located outside) for $68\%$ ($2 -\sigma$ C.L.) best-fit contours. 
    The bound on $(n_s, r)$ plane from the \Planck~2018 is indicated by green colored region, while bound from combined \Planck~2018+\BICEP3~(2022)+\KeckArray2018~is indicated by blue-shaded region. Other colored regions, with dashed lines as their boundary, indicate contours from future \cmb~experiments with higher sensitivity - such as \LB, \cmbsfour, and Simons Observatory (SO)~\cite{LiteBIRD:2020khw,CMB-S4:2016ple,SimonsObservatory:2018koc}.
    }
\end{figure}

Cosmic inflation, on the other hand, conjures scalar and tensor perturbations. The $'k'$-th Fourier mode of the quantum fluctuations departs the Hubble horizon when it becomes $k<\hubble$. This happens because the radius of comoving Hubble horizon shrinks during inflation. Following that, $'k'$-th mode becomes super horizon and frozen. When the inflation ends and the comoving Horizon begins to expand again during radiation or matter domination, this $'k'$-th mode may reenter the causal area.
After reentering, the statistical nature of the $k$-th Fourier mode of 
scalar (or density perturbation) and tensor perturbation can be expressed in terms of the power spectrum (parameterized in power law form) as 
\eq{
&\mathcal{P}_s \left( k \right) = A_s \left(  \frac{k}{k_*} \right)^{n_s -1 + (1/2) \alpha_s \ln(k/k_*) + (1/6)\beta_s (\ln(k/k_*))^2 }  \label{eq:define scalar power spectrum}\,,\\
& \mathcal{P}_h \left( k \right) = A_t \left(  \frac{k}{k_*} \right)^{n_t + (1/2) d n_t/d \ln k \ln(k/k_*) + \cdots } \,,\label{eq:define tensor power spectrum}
}
where $A_s$, $A_t$, $n_s$, $n_t$, $\a_s$, and $\b_s$ are respectively amplitude of scalar and tensor primordial power spectrum,  scalar 
 and tensor spectral index, running of scalar spectrum index, and running of running of scalar spectral index. Moreover, $k_*$ is the pivot scale at which $A_s$ is independent of $n_s$. Additionally, at this scale, constraints on the inflationary observables are drawn from \cmb~measurements. $k_*$ also corresponds to $\phi_*$ around which the inflationary potential is constructed. 
Now, $n_s$ can be estimated using potential-slow-roll parameters at leading order as
\be
 n_s = 1-6\, \epsilon_V(\phi_*) + 2 \, \eta_V(\phi_*)\,.
\ee
On the other hand, the definition of tensor-to-scalar ratio under the assumption of slow roll approximation, is
\be\label{Eq:def-r}
 r =\frac{A_t}{A_s}\approx 16 \, \epsilon_V(\phi_*)\,.
\ee
Using\cref{Eq:def-r}, we can define $A_s$ for $\vgh(\phi)$ as
\eq{
A_s \approx \frac{\vgh (\phi_*)}{24 \pi^2  \mpl^4\, \epsilon_V} \approx \frac{2 \vgh(\phi_*)}{3 \pi^2 \mpl^4 \, r} \,.
}
 The latest bound on $A_s$, $n_s$, and $r$ are listed in~\cref{Table:PlanckData}, where T and E stand for the temperature and E-mode polarization of \cmb. We choose the benchmark values of the inflationary scenario of~\cref{Eq:GH-Lagrangian,Eq:GH-potential} in~\cref{Table:benchmark_values_GH} such that the conditions from~\cref{Table:PlanckData} are satisfied, and~\cref{Fig:GH-ns-r-bound} displays the $(n_s, r)$ predictions for those benchmark values along with the $1-\sigma$ ($98\%$ C.L.) and $2-\sigma$ ($65\%$ C.L.) contour (current bound and prospective future reach) on $n_s-r$ plane from current and future \cmb~observations. The predicted value of $(n_s.r)$ of both `GH-BM1' and `GH-BM3' are within $1-\sigma$ contour of \Planck2018+\BICEP3 (2022) +\KeckArray2018. Furthermore, the predicted value of $(n_s.r)$ of `GH-BM1' falls even within the region of prospective future reach of SO at $2-\sigma$ best-fit contour. 
 \cref{Fig:GH-ns-r-bound} also shows that the estimated value of $r$ for benchmark `GH-BM2' is very small, which is expected as it is derived for small field inflationary scenario. Due to the small value of $r$, this benchmark can only be tested by future \cmb~experiments, e.g. \cmbsfour.

\begin{table}[H]
\begin{center}
\caption{ \centering\it Constraints on inflationary parameters from \cmb~experiments.} \label{Table:PlanckData}
\begin{tabular}{ |c|| c| c||c| }
\hline
$\ln(10^{10} A_s)$ & $
3.044\pm 0.014$ & $68\%$, TT,TE,EE+lowE+lensing+BAO & 
\cite{Aghanim:2018eyx,ParticleDataGroup:2022pth}  \\
 \hline
 $n_s$ & $0.9647\pm 0.0043$ & $68\%$, TT,TE,EE+lowE+lensing+BAO & 
 \cite{Aghanim:2018eyx} \\ 
 \hline 
 $r$ & $0.014^{+0.010}_{-0.011}\, \text{and}$ &  $ 95 \%  \,, \text{BK18, \textsc{Bicep}3, \textit{Keck Array}~2020,}$& \cite{Aghanim:2018eyx, BICEPKeck:2022mhb,BICEP:2021xfz}\\
   & $ <0.036 $ & and \textsl{WMAP} and \textit{Planck}~CMB polarization & (see also \cite{Campeti:2022vom}) \\
 \hline 
\end{tabular}
\end{center}
\end{table}%
%
%
%
%

%
%


\subsection{Stability analysis}
\label{Sec:GH-Stability analysis}
%
%
In this subsection, we estimate the maximum permissible value of $\yc$ and $\lambda_{12}$%
\footnote{
{
While the quadratic coupling $\lambda_{22}$ does play a role during reheating, the stability analysis presented in this section indicates a very small permissible upper value for $\lambda_{22}$~\cite{Drees:2021wgd}. Therefore, our primary focus in this article has been on the coupling $\lambda_{12}$ since we are interested in perturbative approach of reheating.
}
}
(defined in~\cref{Eq:reheating lagrangian}) such that the flatness of the inflationary potential does not get destabilized from the radiative correction emerging from interaction of inflaton with other fields e.g. $\c$ and $H$. %
%
For this, we need to consider 
Coleman–Weinberg radiative correction at 1-loop order to the inflaton-potential which is
~\cite{Coleman:1973jx}
\be
V_{\lop}
=\sum_{j
} \frac{n_j}{64\pi^2} (-1)^{2s_j}\widetilde{m}_j^4
\left[ \ln\left( \frac{\widetilde{m}_j^2 
}{\mu^2} \right) - c_j \right]  \,.
\ee

Here, $j\equiv H, \chi, \phi$, and $c_j=\frac{3}{2}$; $n_{H,\chi}=4$, $n_{\phi}=1$; $s_H =0$, $s_\chi=1/2$, and $s_{\phi}=0$. 
In this work, we assume two values of the renormalization scale: $\mu=\phi_{*}$ and $\mu=\phi_{\rm end}$. Meanwhile, 
the inflaton dependent mass of $H$ and $\chi$ are
\begin{align}
\widetilde{m}_{\chi}^2 (\phi) = \left( m_\chi + y_\chi \phi \right)^2  \,, 
&& \widetilde{m}_{H}^2 (\phi) = m_H^2 + \lambda_{12} \phi \,.  \label{Eq:Inflaton-dependent-mass-GH}
\end{align}
%
Now, the first and 
second derivatives of the Coleman–Weinberg term for $\chi$ and $H$ with respect to $\phi$ are
\begin{align}
	 & V_{\lop}^\prime = \sum\limits_{j 
	 }\frac{n_j}{32 \pi^2}  (-1)^{2 s_j} 
	\widetilde{m}_j^2 \, \lt(\widetilde{m}_j^{2}\right)^{\prime}
	\left[ \ln \left(\frac{\widetilde{m}_j^2}{\mu^2} \right)
	- 1 \right]\,,  \label{Eq:first derivative test}\\
	& V_{\lop}^{\prime\prime} =  \sum\limits_{j
	} \frac{n_j}{32 \pi^2} (-1)^{2 s_j} 
	\left\{ \left[ \left(\left(\widetilde{m}_j^{2}\right)^{\prime} \right)^2
	+ \widetilde{m}_j^2 \left(\widetilde{m}_j^{2}\right)^{\prime\prime} \right]
	\ln \left(\frac{\widetilde{m}_j^2}{\mu^2} \right)
	- \widetilde{m}_j^2 \left(\widetilde{m}_j^{2}\right)^{\prime\prime}  \right\}\,.   \label{Eq:second derivative test}
\end{align}
%
Using~\cref{Eq:Inflaton-dependent-mass-GH} in~\cref{Eq:first derivative test,Eq:second derivative test}, we get Coleman–Weinberg correction term for $H$ and $\chi$ as (with $\mc=m_H=0$)
\eq{
& \lt| V_{{\lop},H}^\prime  
\rt|=\frac{\lO^2 \phi }{8 \pi ^2}  \left(\ln
\left(\frac{\lO \phi }{\mu^2}\right)-1\right)  \, , \label{Eq:stability-H-first-derivative}
%
&&\lt| V_{{\lop},\c}^\prime  
\rt|  = \frac{\Phi ^3 \yc^4 }{4 \pi ^2} \left( 1 - \ln
\left(\frac{\phi ^2 \yc^2}{\mu^2}\right)\right) \, ,\\
%
%
%
%
%
%
& \lt|V_{{\lop},H}^{\prime \prime}
\rt|= \frac{\lO^2 }{8 \pi ^2} \ln
\left(\frac{\lO \phi }{\mu^2}\right) \, .\label{Eq:stability-H-second-derivative}
&& \lt|V_{{\lop},\c}^{\prime \prime} 
\rt|= \frac{1}{8 \pi ^2} \lt(6 \phi ^2 \yc^4  \ln
\left(\frac{\phi ^2 \yc^2}{\mu^2}\right)-2 \phi ^2 \yc^4 \rt)\, .
}
Let us define tree level potential $V_{\rm tree}(\phi)$ as $V_{\rm tree}(\phi)\equiv \vgh(\phi)$. Then 
\begin{align}
 V_{\rm tree}^{\prime} (\phi) &\equiv \vgh'(\phi) =  -m \, n \, \frac{V_0}{\phi } \left(\frac{\phi }{\phi_0}\right)^m
   \left(1-\left(\frac{\phi }{{\phi _0}}\right)^m\right)^{n-1}\,, \\
V_{\rm tree}^{\prime\prime} (\phi) &\equiv \vgh''(\phi) = 
m\, n \, \frac{V_0}{\phi ^2} \left(\frac{\phi }{\phi_0}\right)^m \left(1-\left(\frac{\phi
   }{\phi_0}\right)^m\right)^{n-2} \left((m n-1) \left(\frac{\phi }{\phi_0}\right)^m-m+1\right)
   \,,
\end{align}
Maintaining the stability of the inflation-potential, the maximum allowed value of $\lO$ and $\yc$ can be obtained when all the following conditions are satisfied at $\phi=\mu$
\eq{
 \lt| V_{{\lop},H}^\prime (\phi=\mu) \rt|<V_{\rm tree}^{\prime} (\phi=\mu) \,,
 \quad
 &&\lt| V_{{\lop},\c}^\prime (\phi=\mu) \rt|< V_{\rm tree}^{\prime} (\phi=\mu) \,,\\
 \lt|V_{{\lop},H}^{\prime \prime} (\phi=\mu)\rt| <V_{\rm tree}^{\prime\prime} (\phi=\mu)\,,
 \quad
 &&\lt|V_{{\lop},\c}^{\prime \prime} (\phi=\mu)\rt|<V_{\rm tree}^{\prime\prime} (\phi=\mu)\,.
}
The permitted upper limit for the couplings $\yc$ and $\lO$ for GH inflationary scenario are listed in~\cref{Table:GH_stability} for $\mu=\phi_*$ and $\mu=\phi_{\rm end}$. From this table, we conclude that the permissible upper limit of the couplings are: $\yc<2.5051\times 10^{-4}, <2.7183\times 10^{-4}, <3.7039\times 10^{-4}$ and $\lO/\mpl<1.2757\times 10^{-6}, <1.2413\times 10^{-7}, <1.1649\times 10^{-6}$ for `GH-BM1', `GH-BM2', and `GH-BM3', respectively. 
\begin{table}[H]
\begin{center}
\caption{Allowed upper limit of $\yc$ and $\lO$ for the benchmark values from~\cref{Table:benchmark_values_GH}.}
\label{Table:GH_stability}
\begin{tabular}{|c|| c| c| c|c|} 
 \hline
{\it Benchmark}  &\multicolumn{2}{|c|}{stability for $\yc$} & \multicolumn{2}{|c|}{stability for $\lO$}\\ [0.5ex] 
  \cline{2-5} 
   & about $\mu=\phi_*$ & about $\mu=\phi_{\rm end}$ & about $\mu=\phi_*$ & about $\mu=\phi_{\rm end}$ \\ [0.5ex] 
 \hline\hline
 GH-BM1 & $\yc<2.5051\times 10^{-4}$ & $\yc<
 3.8572\times 10^{-4}$ & $\lO/\mpl<1.2757\times 10^{-6}$ & $\lO/\mpl<2.1619\times 10^{-6}$ \\ 
 \hline
 GH-BM2 & $\yc<2.7183\times 10^{-4}$ & $\yc<6.5637\times 10^{-4}$ & $\lO/\mpl<1.2413\times 10^{-7}$ & $\lO/\mpl<9.9151\times 10^{-7}$ \\
 \hline
 GH-BM3 & $\yc<5.3167\times 10^{-4}$ & $\yc<3.7039\times 10^{-4}$ & $\lO/\mpl<5.3244\times 10^{-6}$ & $\lO/\mpl<1.1649\times 10^{-6}$\\
 \hline
\end{tabular}
\end{center}
\end{table}


\subsection{Reheating era and production of \dm}
\label{Sec:GH-Reheating}
%
%
As soon as the slow-roll inflationary phase terminates, inflaton quickly descends to the minimum of the potential and starts coherent oscillations about the minimum of the potential. The minimum of the potential of~\cref{Eq:GH-potential} is $\phi_0$ (for the chosen benchmark values of~\cref{Table:benchmark_values_GH}) and thus, the physical mass of the inflaton is~\cite{Enqvist:2012qc}:
\begin{eqnarray}\label{Eq:mass_of_inflaton_GH}
\frac{m_{\phi }}{\mpl} = 
\left(\mpl^{-2}\, \vgh'' (\phi) |_{\phi=\phi_{0}} \right)^{1/2} \,.
\end{eqnarray} 
If we define two dimensionless variables $\dz=\phi/\mpl$ and $\dz_0/\mpl$, then, we are assuming that
expanding the potential $\vgh(\dz)$ about the minimum, $\dz_0$, we get (for $(\phi-\phi_0)\ll 1 \implies (\dz-\dz_0)\ll 1$)
\be  
\vgh(\dz-\dz_0)=- m \,V_0\, \frac{\lt( \dz-\dz_0\rt)^n}{\dz_0} + \quad \text{(smaller terms)}\,.
\ee 

Since we choose $n=2$ for three benchmark values%
\footnote{
For example, we assume $\frac{17}{4} \frac{(\dz-\dz_0)^2}{dz_0^2} \ll 3 \frac{(\dz-\dz_0)}{\dz_0} \ll 1$ for `GH-BM3'.
}%
,  the energy density of inflaton and pressure, averaging over an oscillating cycle during reheating, behaves as~\cite{Enqvist:2012qc} (see also~\cite{Garcia:2020eof}) 
\begin{align}
\label{Eq:arXiv:1201.6164v1}
\rho_{\phi} \propto \cs^{-3}\,,  \qquad  \lt<p\rt>=0\,.
\end{align}
This oscillating inflaton begins to produce $\c$ and relativistic Higgs particles $h$ following~\cref{Eq:reheating lagrangian} and initiates the reheating era. This is a highly adiabatic epoch during which the universe changes from being cold and dominated by the energy density of oscillating inflaton to hot visible universe. The relativistic \sm~particles produced during reheating era eventually thermalize among themselves and develop the local-thermal fluid of the universe. As a result of that the energy density of radiation, $\rho_{\rm rad}$, and temperature of the universe, $T$, both increase, while $\rho_\phi$ decreases. However, $\chi$ being feebly interacting with the \sm~particles, it may not share the same temperature as that of \sm~plasma. Sooner, the energy density of oscillating inflaton becomes equal to that of relativistic \sm~particles. The temperature of the universe at that particular moment is referred to as the reheating temperature, denoted by $\Trh$, which can be estimated as~\cite{Bernal:2021qrl}:
\begin{align}
\Trh 
&= \sqrt{\frac{2}{\pi}} \left(\frac{10}{\gs}\right)^{1/4} \sqrt{\mpl} \sqrt{\Gamma_\phi}\,,   \label{Eq:definition of reheating temperature}
\end{align}
where $\gs=106.75$ is effective number of degrees of freedom of relativistic fluid of the universe and  $\Gamma_\phi$ is the total decay width of inflaton. 
At the beginning of reheating era, on the other hand, $\hubble$ is greater than $\Gamma_\phi$. Then, $\hubble$ continues to decrease and it becomes $\hubble(\Trh)\sim \Gamma_\phi$, where $\hubble(\Trh)$ is the value of Hubble parameter when the temperature of the universe is $\Trh$. After this, $\rho_\phi$ is transferred completely and almost immediately to $\rho_{\rm rad}$ and the universe becomes radiation dominated. 

It is expected that at the beginning of the reheating epoch, $\rho_\phi$ decreases relatively at a faster rate, leading a quick rise of the temperature of the universe. However, after some initial increase of temperature, the Hubble expansion comes into play, causing the temperature to decrease. The maximum attainable temperature during reheating process can be estimated as~\cite{Giudice:2000ex,Chung:1998rq}:
\eq{ \label{Eq:TMAX}
\Tmax = 
\G_\phi^{1/4} \lt( \frac{60}{\gs\, \pi^2} \rt)^{1/4} \lt( \frac{3}{8}\rt)^{2/5} {\cal H}_I^{1/4} \mpl^{1/2} \,,
}
where ${\cal H}_I$ is the value of Hubble parameter at the beginning of reheating era when $\rho_{\rm rad}=0$~\cite{Giudice:2000ex} i.e. 
\be\label{Eq:HI-GH}
{\cal H}_I= \sqrt{\frac{\vgh(\phi_{\rm end})}{3\mpl^2}}\,,
\ee
i.e.
we assume in this work that ${\cal H}_I$ is the value Hubble parameter when slow roll inflation ends. $\Tmax>\Trh$ indicates that a particle of mass$>\Trh$ can still be produced during reheating. Furthermore, in many cases, the amount of \dm~produced during reheating depends on the ratio $\Tmax/\Trh$ (for example, see Refs.~\cite{Mambrini:2021zpp,Barman:2022tzk}). Then, from~\cref{Eq:definition of reheating temperature} and~\cref{Eq:TMAX} we get:
\be\label{Eq:TmaxTrh-ratio}
\frac{\Tmax}{\Trh} = \lt(\frac{3}{8}\rt)^{2/5}  \lt( \frac{{\cal H}_I}{{\cal H}(\Trh)} \rt)^{1/4}\,,
\ee
where we have used $\hubble(\Trh)= (2/3)\G_\phi$~\cite{Bernal:2021qrl}. At $T=\Trh$, the universe is radiation dominated and thus 
\be 
\hubble^2 (\Trh)=\frac{1}{3\mpl^2}\lt(\frac{\pi^2}{30} \gs \, \Trh^4\rt)\,.
\ee 
Since, we are not considering any variation of $\gs$ during reheating era, $\lt(\Tmax/\Trh\rt)\propto {\cal H}_I^{1/4} \Trh^{-1/2}$. As a result, for a specific benchmark, ${\cal H}_I$ is fixed and, therefore, $\lt(\Tmax/\Trh\rt)$ decreases as $\Trh$ increases. In this work, we confine our discussion to perturbative approach of reheating~\cite{Lozanov:2019jxc}. 
The decay width of inflaton to $\chi$ and $h$ and total decay width of inflaton are~\cite{Bernal:2021qrl} (here we neglect the effect of thermal mass~\cite{Kolb:2003ke})
\begin{align}\label{Eq:decay-width-of-inflaton}
	\Gamma_{\phi\to {h h}}  \simeq \frac{\lambda_{12}^2}{8\pi\, m_{\phi}}\,, 
 \qquad 
 \Gamma_{\phi \to \chi\chi}  \simeq \frac{y_\chi^2\, m_{\phi}}{8\pi}\,, %
 \qquad \G_\phi=\Gamma_{\phi\to {h h}}+ \Gamma_{\phi \to \chi\chi} \approx \Gamma_{\phi\to {h h}}\,.
\end{align}	
The assumption $\G_\phi\sim \Gamma_{\phi\to {h h}}$ is necessary to avoid of \dm~domination from occurring just after the reheating era. The branching fraction for the production of $\c$ from the decay-channel then can be defined as
\be\label{Eq:Br}
\Br = \frac{\G_{\phi \to \ccp} }{\G_{\phi \to \ccp} + \G_{\phi \to hh} } \simeq  \frac{\G_{\phi \to \ccp} }{ \G_{\phi \to hh} }   = m_{\phi}^2  \lt( \frac{\yc}{\lO}\rt)^2\,.
\ee
Now,~\cref{Table:reheating_values_GH}
presents the values of $m_\phi, \G_\phi, \Trh$, and $\hubble_I$ for the benchmark values mentioned in~\cref{Table:benchmark_values_GH}. Out of three benchmark values, ${\cal H}_I$, and $m_\phi$ are the least for `GH-BM2'. As a result of this, $\Trh/\lO$ is the max for this particular benchmark. 

\begin{figure}[H]
    \centering     
   \includegraphics[width=0.48\linewidth]{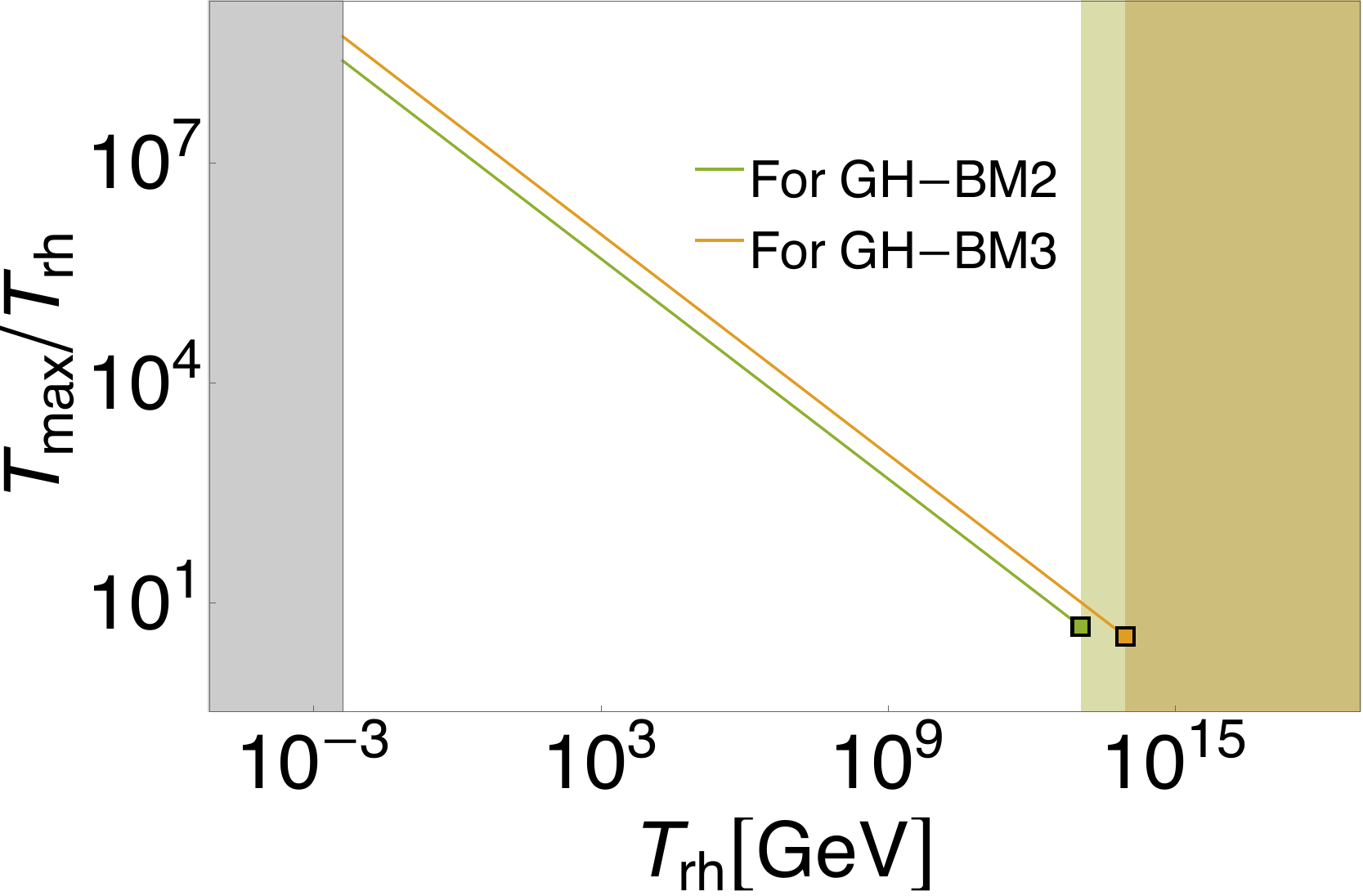}
    \caption{\it \raggedright \label{Fig:Tmax/Trh-GH}
    This figure showcases the variation of $T_{max}/\Trh$ against $\Trh$ for two benchmark values `GH-BM2' and `GH-BM3' from~\cref{Table:benchmark_values_GH}. The square colored boxes on the lines represent the maximum permissible values of $\Trh$ which correspond to the maximum allowable values of $\lO$ from~\cref{Table:GH_stability}. The vertical stripe on the left of the plot, highlighted in gray, displays lower bound on $\Trh$ i.e. $\Trh\gsim 4 \unit{MeV}$. The colorful stripes on the right side of the plot show that there are no permissible values of $\Trh$ based on stability analysis of the corresponding benchmark values.  
    }
\end{figure}

Now, using $\G_\phi$ from~\cref{Eq:decay-width-of-inflaton} in~\cref{Eq:TmaxTrh-ratio}, we get
\eq{
\frac{\Tmax}{\Trh}\approx1.67 \qty(\frac{{\cal H}_I\, m_\phi}{\lO^2})^{1/4}\,.
}
Thus, for a specific benchmark value, $\Tmax/\Trh$ is highest for the lowest value of $\lO$. The lower bound on $\lO$ can be obtained from the condition $\Trh\gsim 4 \unit{Mev}$ (or, ${\cal O}(1\unit{MeV})$, depending on the theory) which is required for successful Big Bang nucleosynthesis (\bbn) and this leads to $\lO/\mpl\gsim 5.5449\times 10^{-23}$ (for `GH-BM1'), $\gsim 4.6885\times 10^{-23}$ (for `GH-BM2'), and $\gsim 5.1603\times 10^{-23}$ (for `GH-BM3'). For these lower limits of $\lO$, the highest possible value of $\Tmax/\Trh \sim 4.9418 \times 10^8$ (for `GH-BM1'), $\sim 2.4586 \times 10^8$ (for `GH-BM2'), and $\sim 5.2695 \times 10^8$ (for `GH-BM3'). 
For higher values of $\lO$ i.e. for larger values of $\Trh$, $\Tmax/\Trh$ decreases as mentioned earlier, and it is displayed in~\cref{Fig:Tmax/Trh-GH}. 
These values can also be approximated using~\cref{Fig:Tmax/Trh-GH} which shows how $\Tmax/\Trh$ declines as $\Trh$ increases for two benchmark values – `GH-BM2', and `GH-BM3'. Since $\Tmax/\Trh$ is almost the same for `GH-BM1' and `GH-BM3', `GH-BM1' is not included in that figure. 
The gray-colored vertical stripe on the left of the figure indicates bound on the $\Trh$ coming from the fact that $\Trh\geq 4 \unit{MeV}$ (for successfully occurring of Big Bang nucleosynthesis (\bbn))~\cite{Giudice:2000ex}. The colored boxes on the lines represent the maximum permissible values of $\Trh$ which correspond to the maximum allowable values of $\lO$ from~\cref{Table:GH_stability}. 
The colored stripes on the right of the figure indicate that those values of $\Trh$ are not allowed from the stability analysis.

\begin{table}[H]
\centering
\caption{\it $m_\phi, \G_\phi, \Trh$, and $\hubble_I$ for the benchmark values from~\cref{Table:benchmark_values_GH}.}
\label{Table:reheating_values_GH}
\begin{tabular}{| c ||  c | c|  c | c|} 
 \hline
{\it Benchmark} & $m_\phi/\mpl$ & $\G_\phi \mpl$ & $\Trh/\lambda_{12}$ & ${\cal H}_I/\mpl$ \\
&   (\cref{Eq:mass_of_inflaton_GH}) & (\cref{Eq:decay-width-of-inflaton}) & (\cref{Eq:definition of reheating temperature}) & (\cref{Eq:HI-GH})  \\
[0.5ex]
 \hline\hline
 GH-BM1 &   $8.5812\times 10^{-6}$ & $ 4636.73\lambda_{12}^2$ & $ 30.0576$ & $2.7228\times 10^{-6}$\\ 
 \hline
   GH-BM2 &   $6.1351 \times 10^{-6}$ & $ 6485.42\lambda_{12}^2$ & $35.5482 $ & $1.6682\times 10^{-7}$\\ 
 \hline
   GH-BM3 &   $ 7.4321 \times 10^{-6}$ & $5353.64 \lambda_{12}^2$ & $ 32.2978$ & $3.5200\times 10^{-6}$\\ 
 \hline
\end{tabular}
\end{table}
%
%

%
%
%
%
%
Next, we consider the production of \dm~during reheating. 
If $n_\c$ is the number density of $\c$, then the conservation equation of the comoving number density, $N_\c$ which is defined as $n_\c \, \cs^3$, is
\eq{\label{Eq:Boltzaman equation for comoving number density}%
\frac{\td N_\c}{\td t} = \cs^3 \, \gamma  \,,
}
%
in which $t$ is the physical time, and $\g\equiv \g(t)$ which is the rate of production of $\c$, has a quartic mass-dimension.

When the temperature of the universe is $\Tmax> T> \Trh$, then~\cite{Bernal:2021qrl}
\be\label{Eq:Hubble parameter during reheating}
{\cal H} = \frac{\pi}{3} \sqrt{\frac{\gs}{10}} \frac{T^4}{M_P\, \Trh^2}\,.
\ee 
During that period, the energy density of the universe is dominated by $\r_\phi$. Thus, from the first Friedman equation, we get
\be\label{Eq:rho_phi}
\r_{\phi} = \frac{ \pi^2  \gs }{30 } \frac{T^8}{ \Trh^4} \,.
\ee
If we combine~\cref{Eq:arXiv:1201.6164v1} with~\cref{Eq:rho_phi}, and then using~\cref{Eq:Hubble parameter during reheating} into~\cref{Eq:Boltzaman equation for comoving number density}, we obtain
\be\label{Eq:dNchidT}
\frac{\td N_\c}{\td T} = - \frac{8 \mpl }{\pi} \lt(\frac{10}{  \gs}\rt)^{1/2} \frac{\Trh^{10}}{T^{13}} 
\, \cs^3 (\Trh) \, \g  \,.
\ee 

To filter out the effect of the expansion of the universe on the time evolution of $n_\c$~\cite{Lisanti:2016jxe}, 
we define \dm~yield, $\Yc$, as $\Yc = {n_\c(T)}/{s(T)}$, where $s(T)=\lt({2 \pi^2}/{45}\rt) \gss \,T^3 $ and $\gss$ refers to the effective number of degrees of freedom contributing to the entropy of the relativistic fluid. We assume that entropy density per comoving volume remains conserved once the reheating era is over. 
The energy density of $\c$ continues to decrease with time, contributing to the cold dark matter (\cdm) density of the present universe. Present day \cdm~yield can be calculated as~\cite{Garrett:2010hd}
\be\label{Eq:present day CDM yield}
 Y_{{\rm CDM},0}  = \frac{\Omega_{\rm CDM}\, \r_{\rm crit} }{s_0 \, m_\c}=\frac{4.3. \times 10^{-10}}{\mc} \,,
\ee 
where $m_\c$ is expressed in $\GeV$ and we have used scaling factor for Hubble expansion rate $h_{\cmb}\approx0.674$, cold dark matter density of the universe $\Omega_{\rm CDM}=0.12\, h_{\cmb}^{-2}$, present day entropy  density $s_0=2891.2 \unit{cm}^{-3}$, and critical density of the Universe $\r_{\rm crit}=1.053\times 10^{-5}\, h_{\cmb}^{2}\, \GeV  \unit{cm}^{-3}$ (as we choose $k_B=c=1$ unit) from Ref.~\cite{ParticleDataGroup:2022pth}. Now, we estimate the yield of $\c$ produced through decay or via scattering, and then compare with $ Y_{{\rm CDM},0}$ to determine the extent to which the produced $\c$ contributes to the total \cdm~density of the present universe.

\subsubsection{Inflaton decay}
If $\c$ is produced exclusively via the decay of inflaton ($\phi\to \cc$), then
%
\be
\g = 2 \, \Br \, \G \,  \frac{\r_{\phi}}{m_{\phi}} \,.
\ee 
When we substitute this in~\cref{Eq:dNchidT}, 
along with the assumption that the \dm~yield remains unaltered after the end of reheating to present day, 
we obtain the expression for the \dm~Yield from the decay of inflaton as,
%
\eq{
\Yco&=\frac{N_\c (\Trh)}{s(\Trh)\, \cs^3(\Trh)}\\
&\simeq \frac{3}{\pi} \frac{\gs}{\gss} \sqrt{\frac{10}{\gs}} \frac{\mpl \, \Gamma}{m_{\phi}\, \Trh} \text{Br}%
= \frac{3}{\pi} \frac{\gs}{\gss} \sqrt{\frac{10}{\gs}} \frac{\mpl }{ \Trh}  \frac{(\yc)^2}{8 \pi} \\
 &= 1.163\times 10^{-2} \mpl \frac{\yc^2 }{\Trh}\,. \label{Eq:ychi0-decay-total}
}
%
Here, we assume $\gss=\gs$. Equating~\cref{Eq:ychi0-decay-total} with~\cref{Eq:present day CDM yield}, we get 
the condition to generate the complete \cdm~energy density in terms of the DM mass to be
\be\label{Eq:eq to plot 2}
\Trh  \seq  
6.49 \times 10^{25}\, \yc^2 \, \mc \,.
\ee 

\begin{figure}[H]
    \centering    
%
\includegraphics[width=0.48\linewidth]{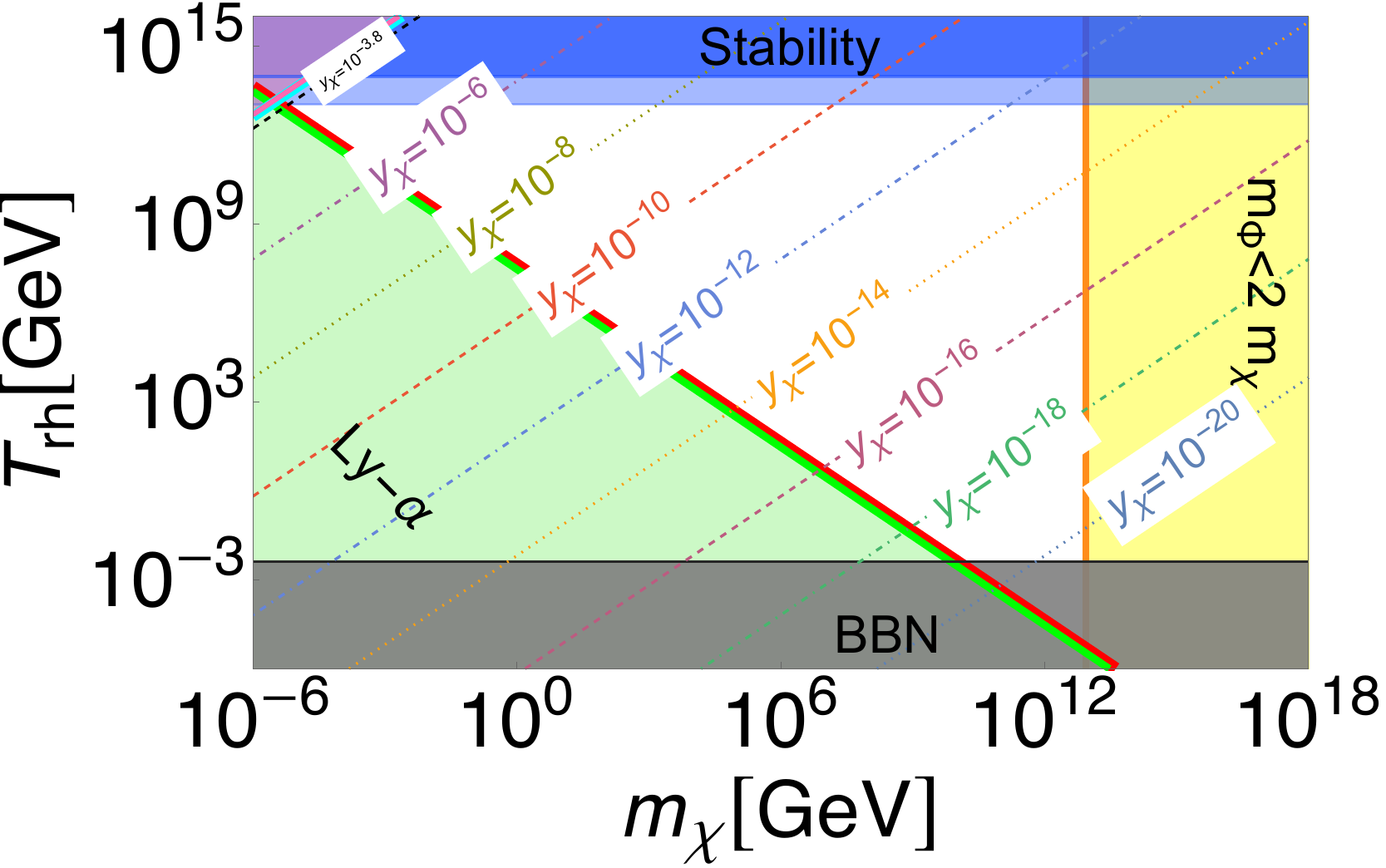}
    \caption{\it \raggedright \label{Fig:CDM_Yield_Decay-GH}
    The plot showcases the variation in the permissible region on $(\Trh, m_\chi)$ plane as well as allowed range of $\yc$, specifically for benchmark values `GH-BM1' and `GH-BM2'. The unshaded region on this plane represents the allowed region, and the dashed or dashed-dotted lines that intersect it correspond to~\cref{Eq:eq to plot 2} for different values of $y_\chi$ satisfying present-day CDM density. The regions shaded in light neon blue (horizontal stripe at the top of the figure) and yellow (vertical stripe on the right-hand side) represent the bounds from the permissible upper bound on $\lO$ (from stability analysis) and the maximum possible value of $\mc$, while the purple, and green shaded area show the bounds on $\yc$ from stability analysis, and Ly-$\alpha$  bound on the mass of \dm~\cite{Bernal:2021qrl} (see also~\cref{Eq:Lyman-alpha-bound}) for `GH-BM2'. The same bounds for `GH-BM1' are presented by deep neon blue-shaded region, and orange solid line, cyan and red colored solid lines. The gray shaded stripe at the bottom is for bound from \bbn~(i.e. $\Trh$ should be $\gsim 4\unit{MeV}$). This figure shows that $\yc\sim 10^{-8}$ is allowed for `GH-BM1', but not for `GH-BM2'.
    }
\end{figure}

Lines for different values of $\yc$ from~\cref{Eq:eq to plot 2} are shown as discontinuous inclined lines on $(\Trh, \mc)$ plane in log-log scale in~\cref{Fig:CDM_Yield_Decay-GH}. 
These lines actually correspond to $\chi$ being produced from the decay of inflaton and also producing the total \cdm~density of the present-day universe. \cref{Fig:CDM_Yield_Decay-GH} actually shows how the permissible region which is shown as white region on $(\Trh, \mc)$ plane, varies with benchmarks. 
Bounds on this plane are from: the largest permissible value of $\lO$ and $\yc$ from stability analysis of~\cref{Table:GH_stability} (neon blue-colored stripe at the top of the plot and purple-colored wedge-shaped region at the left-top of the plot for `GH-BM1', while deep neon blue-colored stripe at the top of the plot and cyan-colored wedge-shaped region at the left-top of the plot for `GH-BM2'), BBN temperature: $\Trh\gsim 4 \unit{MeV}$ (gray colored horizontal stripe at the bottom of the plot), and Ly-$\alpha$ bound from~\cref{Eq:Lyman-alpha-bound}: $\Trh\gsim (2 m_\phi)/\mc$ (region shaded with green color for `GH-BM1' and red color (line) for `GH-BM2') and the maximum possible value of $\mc=m_\phi/2$ (vertical stripe at the right, shaded with yellow color for `GH-BM1' and orange color for `GH-BM2').
The allowed range of $\yc$ shown by inclined, discontinuous lines passing through the white unshaded region. While the allowed range of $\yc$ are nearly the same for both chosen benchmark values, the range of $\yc$ for `\cw-BM2' is narrower. For example, $\yc\sim 10^{-3.8}$ is not allowed for `\cw-BM2' but permissible for `\cw-BM1'.  
The allowed range of $\mc$ for $3.1738\times 10^{-6} \GeV \lsim \mc \lsim 1.0297\times 10^{13}\GeV$ for `GH-BM1' and  $2.7836 \times 10^{-6}\GeV \lsim \mc \lsim 7.3621\times 10^{12}\GeV$ for `GH-BM2'. If both $\yc$ and $\mc$ fall within these ranges, then it can satisfy two conditions simultaneously: (i) produced solely through the decay of inflation, and (ii) produce the total \cdm~density of the universe. 
All the aforementioned  bounds vary for three benchmark values mentioned in~\cref{Table:benchmark_values_GH}. One of the two reason for this variation is different values of $m_\phi$, which is already mentioned in~\cref{Table:reheating_values_GH}, with 
$m_\phi$ for `GH-BM1'$>$ for `GH-BM2'. Additionally, the upper limit of the possible value of $\lO/\mpl$ from stability analysis for `GH-BM1'$>$`GH-BM2'. 
These factor imposes lower limit on $\mc$ as $2783.64 \GeV$ for `GH-BM2'.
However, on this $(\Trh, \mc)$ plane more stringent bounds can be drawn from gravitino production. 
 Nevertheless, we ignore this bound in our analysis as we do not take supersymmetry into account.

%
%

To derive the Ly-$\alpha$ bound used in~\cref{Fig:CDM_Yield_Decay-GH} it is assumed that $\chi$ being feebly interacting \bsm~particle, its momentum only decreases due to red-shift and thus~\cite{Bernal:2021qrl} 
\eq{
p_{0} = \frac{\cs_{\rm rh}}{\cs_0} p_{\rm rh} \simeq 3 \times 10^{-14} \, \frac{m_\phi}{\Trh} \GeV\,,
}
where $p_{\rm rh}$ is the momentum of $\c$ at the time of reheating (i.e. at temperature $\Trh$) and $p_{\rm rh}$ is assumed as the initial momentum of the \dm~particles. Similarly, $p_0$, $\cs_0$, and $\cs_{\rm rh}$ are the momentum, cosmological scale factor today and at the time of reheating, respectively. The maximum possible value of $p_{\rm rh}$ is $p_{\rm rh}=m_\phi/2$. If we approximate $p_0= \mc\, v_\c$, where $v_\c$ is the present day velocity of $\c$ particles, then, using upper bound on $v_x/c\lsim 1.8\times 10^{-8}$ and $m_c\gsim 3.5 \unit{keV}$ (for further details about bound on the mass of warm dark matter from Ly-$\alpha$ see references within~\cite{Ghoshal:2022jeo}) such that $\c$ as a warm dark matter particle does not negatively impact on structure formation, we can obtain that 
\be\label{Eq:Lyman-alpha-bound}
\mc \gsim 2 \frac{m_\phi}{\Trh}\,,
\ee
where $\mc$ is expressed in $\unit{keV}$. 

\subsubsection{\dm~from scattering}
\label{Sec:DM from scattering GH}
In this subsection, we look at three important 2-to-2 scattering mechanisms for \dm~production: from scattering of non-relativistic inflaton with graviton as the mediator, scattering of \sm~particles with graviton as mediator, and scattering of \sm~particles with inflaton as mediator. If $Y_{IS,0}$, $Y_{\sm g,0}$, and $Y_{\sm i,0}$ are respectively the \dm~yield produced only via those three scattering channels, then~\cite{Bernal:2021qrl}
\begin{align} \label{Eq:yield-DM-scattering-inflaton-graviton}
	    Y_{IS,0} &\simeq \frac{\gs^2}{81920\gss} \sqrt{\frac{10}{\gs}} \left(\frac{\Trh}{\mpl}\right)^3 \left[\left(\frac{\Tmax}{\Trh}\right)^4 - 1 \right] \frac{m_{\chi}^2}{m_{\phi}^2} \left(1-\frac{m_{\chi}^2}{m_{\phi}^2}\right)^{3/2}\,. \\
 \label{Eq:Ysmg0}
 	Y_{\sm g,0} &= 
	\begin{cases}
	\frac{45 \, \alpha_{\dm}}{2\pi^3 \gss} \sqrt{\frac{10}{\gs}} \left(\frac{\Trh}{M_P}\right)^3, \qquad \text{ for } m_\chi \ll \Trh\,,\\
	\frac{45 \,\alpha_{\dm}}{2\pi^3 \gss} \sqrt{\frac{10}{\gs}} \frac{\Trh^7}{M_P^3\, m_\chi^4}\,, \qquad \text{ for } \Tmax \gg m_\chi \gg \Trh\,.
	\end{cases}\\
	%
Y_{\sm i,0} &\simeq \frac{135\,  y^2_{\chi}\, \lambda_{12}^2}{4 \pi^8\, \gss} \sqrt{\frac{10}{g_\star}}\, \frac{M_P\, \Trh}{m_{\phi}^4}\, , \qquad \text{ for } \Trh \ll m_{\phi}, \Trh > T \,.
\label{Eq:smo}
\end{align}

In~\cref{Eq:Ysmg0}, $\alpha_\dm =1.1\times 10^{-3}$~\cite{Bernal:2018qlk} and it is related to the coupling of gravitational interaction. In~\cref{Eq:yield-DM-scattering-inflaton-graviton}, $Y_{IS,0}$ is increasing function of $\Trh$. 
Now, from the stability analysis of~\cref{Table:GH_stability}, maximum permissible value for $\Trh$ are 
\eq{\label{Eq:GH-Trh-MaxPossible}
{\Trh}_{\text{, allowable}}=
\begin{cases}
    9.2027\times 10^{13}\GeV \, \quad \text{for `GH-BM1'}\,,\\
    1.059\times 10^{13}\GeV \, \quad \text{for `GH-BM2'}\,,\\
    9.0297\times 10^{13}\GeV \, \quad \text{for `GH-BM3'}\,.
\end{cases}
}
For $\Trh\sim {\Trh}_{\text{, allowable}}$, $\Tmax/\Trh\sim 3.2581$ for `GH-BM1', $\sim 4.7783$ for `GH-BM2', and $\sim 3.5072$ for `GH-BM3'. Now,
if we choose 
$\Trh\sim {\Trh}_{\text{, allowable}}$ from~\cref{Eq:GH-Trh-MaxPossible}, 
then, in order to achieve $Y_{IS,0}\sim  Y_{{\rm CDM},0}$, it is required that $\mc$ should be ${\mc}^{IS,0}$ where
\eq{\label{Eq:GH-large mc is needed-InflatonScattering}
{\mc}^{IS,0} \sim \begin{cases}
    4.1722\times 10^{10}\GeV \, \quad \text{for `GH-BM1'}\,,\\
    1.7358\times 10^{11}\GeV \, \quad \text{for `GH-BM2'}\,,\\
    3.4993\times 10^{10}\GeV \, \quad \text{for `GH-BM3'}\,.
\end{cases}
}
Thus, to achieve comparable values of $Y_{IS,0}$ and $ Y_{{\rm CDM},0}$, 
we need $\mc> {\mc}^{IS,0}$ for $\Trh<{\Trh}_{\text{, allowable}}$. 
Hence, if $\c$ is produced exclusively only through the 2-to-2 scattering of inflaton via graviton mediation and contributes completely to the total \cdm~relic density, the required condition is $\mc\gsim  {\mc}^{IS,0}$ for $\Trh\lsim {\Trh}_{\text{, allowable}}$.


For \dm~production solely through scattering of \sm~particles with graviton as mediator and for $(m_\chi \gg \Trh)$ and
with $\Trh\sim {\Trh}_{\text{, allowable}}$
\eq{
Y_{\sm g,0}\sim
\begin{cases}
    1.2903\times 10^{-19} \, \quad \text{for `GH-BM1'}\,,\\
    1.9663\times 10^{-22}\, \quad \text{for `GH-BM2'}\,,\\
    1.2189\times 10^{-19}\, \quad \text{for `GH-BM3'}\,.
\end{cases}
}
Consequently, in order to satisfy $Y_{IS,0}\sim  Y_{{\rm CDM},0}$, and for $\Trh\lsim {\Trh}_{\text{, allowable}}$, it is required that 
\eq{\label{Eq:GH-large mc is needed}
\mc \gsim
\begin{cases}
    3.3326\times 10^{9}\GeV\, \quad  \text{for `GH-BM1'}\,,\\
    2.1868\times 10^{12}\GeV\, \quad  \text{for `GH-BM2'}\,,\\
    3.5279\times 10^{9} \GeV\, \quad \text{for `GH-BM3'}\,.
\end{cases}
}
Therefore, from~\cref{Eq:GH-large mc is needed} we can deduce that \dm~particles produced solely though this specific scattering channel (\two) can contribute to the total \cdm~density, provided that $\mc$ is very high.
However, for $(\Tmax \gg m_\chi \gg \Trh)$ and $Y_{\sm g,0}=Y_{{\rm CDM},0}$, we get
\eq{\label{Eq:TERM}
\frac{\Trh^{\sm g,0}}{\GeV}= 7.1389 \times 10^{12} \, {f}_{\Trh}^{-3/4}\,,
}
where we have assumed $\Trh=\mc \, {f}_{\Trh}$, ${f}_{\Trh}$ being a fractional number much less than $1$. As the value of ${f}_{\Trh}$ decreases, $\Trh^{\sm g,0}$ increases. For ${f}_{\Trh}\sim 0.6$, $\Trh^{\sm g,0} \sim {\cal O}(10^{13})$. However, for  $\Trh\sim {\cal O}(10^{13})$, $\Tmax/\Trh\sim 3$ (see~\cref{Eq:GH-Trh-MaxPossible}). Therefore, it seems less probable, that satisfying the condition $\Tmax\gg \mc\gg \Trh$, $\c$ produced through this scattering process (\three) contributes $100\%$ of the total \cdm~density.

%

Now we consider production of $\c$ via 2-to-2 scattering of SM particles
with inflaton as mediator (along with the constraint $\Trh\ll m_\phi$). The condition that $Y_{\sm i,0}$ (from~\cref{Eq:smo}) contributes a fraction $f_\c$ ($f_\c$ is a dimensionless fractional number) of total \cdm~relic density, then 
we obtain (in $\GeV$)
\eq{\label{Eq:game_changer}
{\mc}^{\sm i,0} = 4.2164 \times 10^{-5}
\,\frac{
 m_\phi^4}{\Trh,\,\mpl \, \yc^2\, \lO^2}\,f_\c\,.
}
Therefore, as values of $\yc$ and $\lO$ decreases the value of ${\mc}^{\sm i,0}$ increases. Now, using
$\Trh= f_\phi \, m_\phi$ where $f_\phi$ is fractional ($\ll 1$) dimensionless number, in~\cref{Eq:definition of reheating temperature}, we get (in $\GeV$)
\eq{\label{Eq:game_changer1}
{\mc}^{\sm i,0} = 3.2689 \times 10^{-7} \,\frac{f_\c}{\yc^2\, f_\phi^3}\,.
}
For 
the maximum permissible values of $\yc$ (from~\cref{Table:GH_stability})
${\mc}^{\sm i,0}$ becomes   
\eq{\label{Eq:GH-con}
{\mc}^{\sm i,0} =
\begin{cases}
    5.2090\, \frac{f_\c}{f_\phi^3}  \GeV \, \quad \text{(for `GH-BM1')}\,,\\
    4.4239\, \frac{f_\c}{f_\phi^3}  \GeV \, \quad \text{(for `GH-BM2')} \,,\\
    2.3828\, \frac{f_\c}{f_\phi^3}  \GeV \, \quad \text{(for `GH-BM2')}\,.
\end{cases}
}
If $f_\c\sim 1$ and $f_\phi\sim 10^{-2}$, then for ${\mc}^{\sm i,0}\sim {\cal O}(10^8)\GeV$, $\c$
 produced through this scattering channel (\four) can contribute total \cdm~relic density.



Therefore, if $\c$ is produced solely via either of the three scattering processes we have considered (i.e. \one,\two, and \four), for GH inflation, then produced $\c$ can contribute up to $100\%$ of the total \cdm~relic density of the present universe.

\section{Coleman-Weinberg inflation}
\label{Sec:CW}

\begin{figure}[H]
    \centering    
    \includegraphics[height=6cm,width=0.48\linewidth]{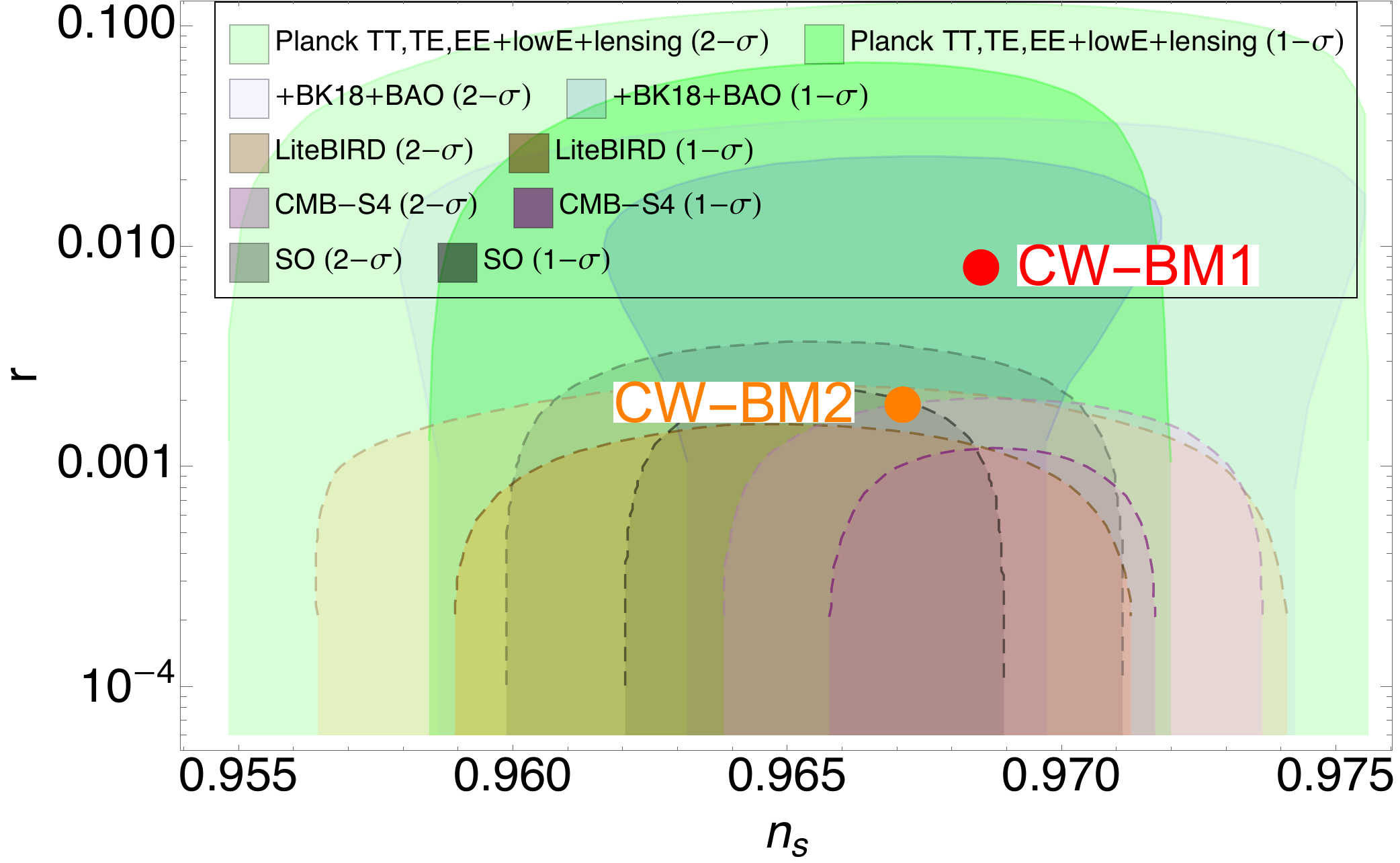} 
    \caption{\it \raggedright \label{Fig:NonminimalCW-ns-r-bound} 
    The predicted values of $n_s, r$ for the two benchmark values from~\cref{Table:Benchmark-CW} for \cw~slow roll inflationary scenario are displayed as colored circular dots.
    In addition, this figure also displays $2-\sigma$ and $1-\sigma$ $(n_s, r)$ best-fit contours from the \Planck~2018+\BICEP, as well as additional upcoming \cmb~observations mentioned earlier in~\cref{Fig:GH-ns-r-bound}. 
    }
\end{figure}

If we consider a \sm~gauge singlet \bsm~field as the inflaton with quartic potential, the presence of self-interaction or Yukawa interaction in the Lagrangian density can give rise to quantum corrections. When such quantum loop corrections originating from interaction terms, UV completions~\cite{Racioppi:2018zoy}, etc., are included, the effective potential for inflaton is given by~\cite{Barenboim:2013wra,Kallosh:2019jnl}
\begin{align}\label{eq:CW_potential}
V_{\cw}(\phi)= \frac{A \cq^4}{4}+A \phi ^4 \left[\log \left(\frac{\phi }{\cq}\right)-\frac{1}{4}\right]\,,
\end{align}
where $\cq$ represents the renormalization scale, and the inclusion of the term $\frac{A \cq^4}{4}$ is to ensure that the value of the potential is $\gsim 0$, even at its minimum. This Coleman-Weinberg (\cw) inflationary scenario involves inflation beginning near $\phi=0$ and then the inflaton travels towards $\phi_{\rm min}$, the minimum of the potential of~\cref{eq:CW_potential}.
The benchmark values of \cw~inflationary scenario are mentioned in
~\cref{Table:Benchmark-CW} and the $n_s-r$ predictions for the two benchmark values from~\cref{Table:Benchmark-CW} are shown in \cref{Fig:NonminimalCW-ns-r-bound}. The predicted value $r$ for benchmark `\cw-BM1'~and `\cw-BM2'~fall within $1-\sigma$ and $2-\sigma$ best-fit contour of \Planck2018+\BICEP3 (2022)+\KeckArray2018 data and SO analysis, respectively.

\begin{table}[H]
\centering
\caption{\it Benchmark values for \cw~inflationary model (for $\ncmb \approx 60$).}
\label{Table:Benchmark-CW}
\begin{tabular}{|c||c | c | c | c | c | c |   } 
 \hline
{\it Benchmark}&$\cq/\mpl$ &  $\phi_{*}/\mpl$ & $\phi_{\rm end}/\mpl$ &  $n_s$ & $r\times 10^{3}$ & $A \times 10^{14}$ \\ %
 \hline 
 \hline 
\cw-BM1  & $16.4800$ &  $ 4.8450$ & $ 8.4229$ & $0.9685$ & $7.9849$ & $  1.4075$ \\
 \hline
\cw-BM2 & $ 12.0500$&  $2.3000  $& $4.1520 $& $0.9671  $& $1.9091$ & $1.1370$  \\
 \hline
\end{tabular}
\end{table}
%
%

%
%

\subsection{Stability analysis for \cw~inflation}
Here, we follow steps similar to that followed in~\cref{Sec:GH-Stability analysis}, to determine the maximum permissible value of $\lambda_{12}$ and $\yc$, considering $V_{\rm tree}(\phi)\equiv \vcw(\phi)$. Subsequently, we get
\begin{align}
 V_{\rm tree}^{\prime} (\phi) &\equiv \vcw'(\phi) =  4 A \, \phi ^3 \left(\log \left(\frac{\phi }{v}\right)-\frac{1}{4}\right)+A \, \phi ^3\,, \\
V_{\rm tree}^{\prime\prime} (\phi) &\equiv \vcw''(\phi) = 
12 A\, \phi ^2 \left(\log \left(\frac{\phi }{v}\right)-\frac{1}{4}\right)+7 A\, \phi ^2
   \,.
\end{align}
The maximum allowable values for $\lO$ and $\yc$ are presented in~\cref{Table:CW_stability}. The results from the table lead us to the conclusion that the upper limits for these couplings are as follows: 
$\yc < 5.5257 \times 10^{-4}$, $< 5.8920 \times 10^{-4}$, and $\lO/\mpl < 2.8864 \times 10^{-6}$, $< 1.4345 \times 10^{-6}$ for `\cw-BM1', and `\cw-BM2', respectively. 
\begin{table}[H]
\begin{center}
\caption{Allowed upper limit of $\yc$ and $\lO$ for the benchmark values from~\cref{Table:Benchmark-CW}.}
\label{Table:CW_stability}
\begin{tabular}{|c|| c| c| c|c|} 
 \hline
{\it Benchmark}  &\multicolumn{2}{|c|}{stability for $\yc$} & \multicolumn{2}{|c|}{stability for $\lO$}\\ [0.5ex] 
  \cline{2-5} 
   & about $\mu=\phi_*$ & about $\mu=\phi_{\rm end}$ & about $\mu=\phi_*$ & about $\mu=\phi_{\rm end}$ \\ [0.5ex] 
 \hline\hline
\cw-BM1 & $\yc<6.4530\times 10^{-4}$ & $\yc<
 5.5257\times 10^{-4}$ & $\lO/\mpl<2.8864\times 10^{-6}$ & $\lO/\mpl<3.6784\times 10^{-6}$ \\ 
 \hline
\cw-BM2 & $\yc<6.3004\times 10^{-4}$ & $\yc<5.8920\times 10^{-4}$ & $\lO/\mpl<1.4345\times 10^{-6}$ & $\lO/\mpl<2.0618\times 10^{-6}$ \\
 \hline
\end{tabular}
\end{center}
\end{table}


\subsection{Reheating and production of \dm~from inflaton decay}
\begin{figure}[H]
    \centering    \includegraphics[width=0.48\linewidth]{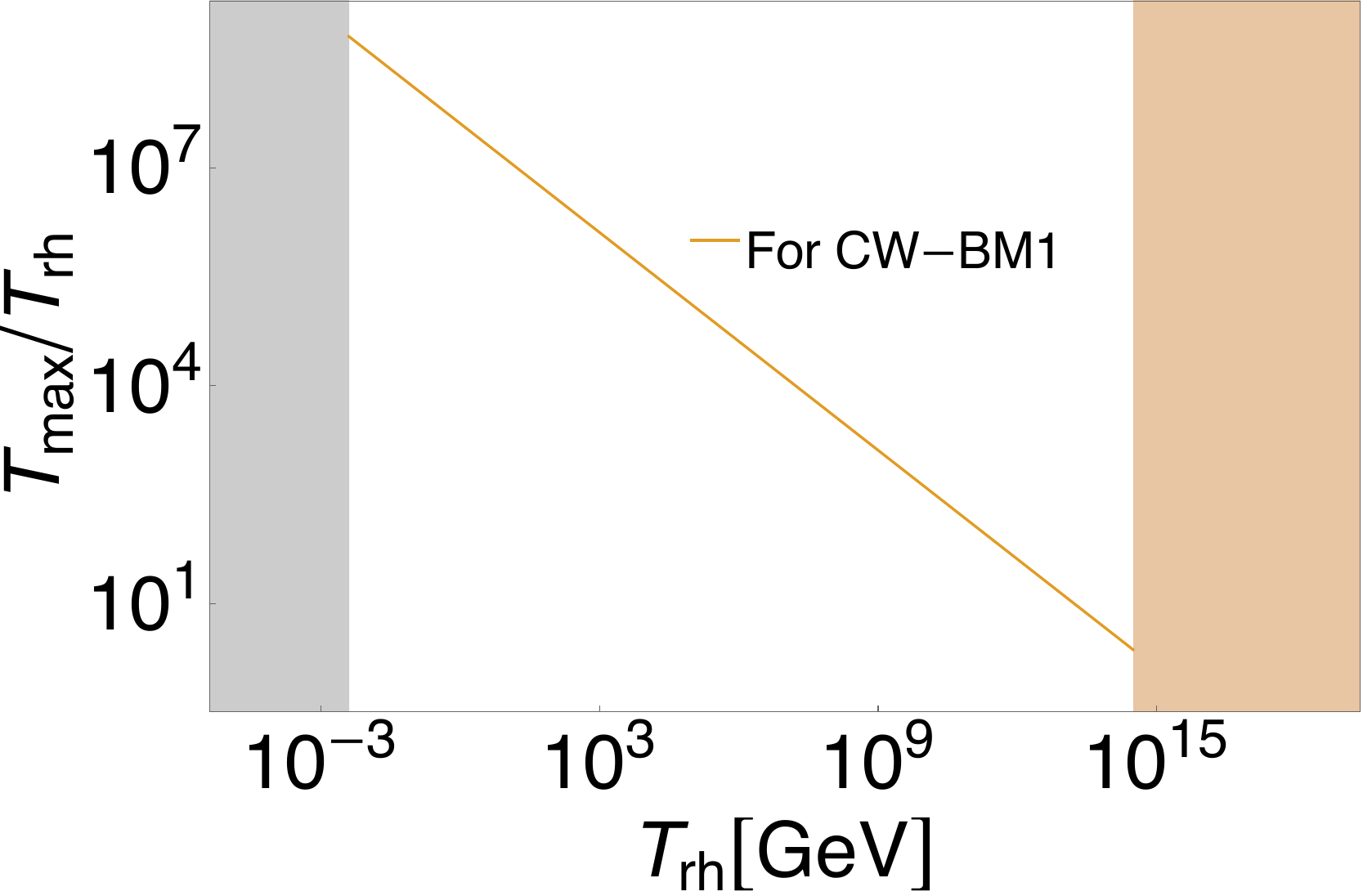}
    \caption{\it \raggedright \label{Fig:Tmax/Trh-CW}
    This figure showcases the variation of $T_{max}/\Trh$ against $\Trh$ for benchmark value `\cw-BM1'~from~\cref{Table:Benchmark-CW}. 
    The vertical stripe on the left of the plot, highlighted in gray, displays the lower bound on $\Trh$ i.e. $\Trh\gsim 4 \unit{MeV}$. The colorful stripe on the right of the plot shows that there is no permissible value of $\Trh$ based on stability analysis of the corresponding benchmark value from~\cref{Table:CW_stability}.  
    }
\end{figure}

In this section, we assume akin to what we have made for GH inflationary scenario in~\cref{Sec:GH-Reheating} that following the end of slow roll inflationary phase, inflaton quickly descends to the minimum  of the potential $\vcw(\phi)$ and subsequently  oscillates quasi-periodically about a quadratic minimum. The minimum of  $\vcw(\phi)$ is located at $\phi=\cq$, and $m_\phi, \Gamma_\phi, \Trh$, and $\hubble_I$ for \cw~inflationary scenario are listed in~\cref{Table:reheating_CW}. 
$\hubble_I$ for \cw~inflationary model is defined as
\eq{
\label{Eq:HI-CW}
{\cal H}_I= \sqrt{\frac{\vcw(\phi_{\rm end})}{3\mpl^2}}\,.
}
From~\cref{Table:reheating_CW} we see that the difference between the values of ${\cal H}_I/\mpl$ is not very large for both benchmarks `\cw-BM1' and `\cw-BM2'. The same is also true for the parameters $\Trh/\lO$ and $m_\phi/\mpl$ for the two chosen benchmark values. 
Therefore, we present $\Tmax/\Trh$ (defined in~\cref{Eq:TmaxTrh-ratio}) against $\Trh$ as solid line only for `\cw-BM1' in~\cref{Fig:Tmax/Trh-CW}. The gray-colored vertical stripe on the left of the figure shows a limit on $\Trh$ due to the requirement that $\Trh$ must be $\geq 4 \unit{MeV}$.  On the right-hand side of the figure, there is a colored stripe that signifies that those $\Trh$ values are not permitted based on the stability analysis from~\cref{Table:CW_stability}. The value of $\Tmax/\Trh$ is highest for $\Trh\sim 4 \MeV$ and the ratio decreases as the value of $
\Trh$ increases. This pattern closely resembles the observed behavior illustrated in~\cref{Fig:Tmax/Trh-GH} for the GH inflationary scenario. The maximum value of $\Tmax/\Trh$ is $\sim 6.4796\times 10^8$ for `\cw-BM1' (and the maximum value of $\Tmax/\Trh \sim 5.5401\times 10^8$ for `\cw-BM2'~at $\Trh\sim 4 \unit{MeV}$). Furthermore, lower bound on $\Trh$ leads to establishing the minimum permissible value of $\lO$ which are $\lO/\mpl\gsim 3.7431\times 10^{-23}$ for `\cw-BM1', and $\lO/\mpl\gsim 3.0344\times 10^{-23}$ for `\cw-BM2'.

\begin{table}[H]
\centering
\caption{\it $m_\phi, \G_\phi, \Trh$, and $\hubble_I$ for the benchmark values from~\cref{Table:Benchmark-CW}.}
\label{Table:reheating_CW}
\begin{tabular}{| c ||  c | c|  c | c|} 
 \hline
{\it Benchmark} & $m_\phi/\mpl$ & $\G_\phi \mpl$ & $\Trh/\lambda_{12}$ & ${\cal H}_I/\mpl$ \\
&   (\cref{Eq:mass_of_inflaton_GH}) & (\cref{Eq:decay-width-of-inflaton}) & (\cref{Eq:definition of reheating temperature}) & (\cref{Eq:HI-CW})  \\
[0.5ex]
 \hline\hline
 \cw-BM1 &   $3.9103\times 10^{-6}$ & $ 10175.3\lambda_{12}^2$ & $ 44.5268$ & $8.0473\times 10^{-6}$\\ 
 \hline
   \cw-BM2 &   $2.5698 \times 10^{-6}$ & $ 15483.5\lambda_{12}^2$ & $54.9265 $ & $4.3006\times 10^{-6}$\\ 
 \hline
\end{tabular}
\end{table}
%
%

%
%
%
%
%
%

\begin{figure}[H]
    \centering    
   \includegraphics[width=0.48\linewidth]{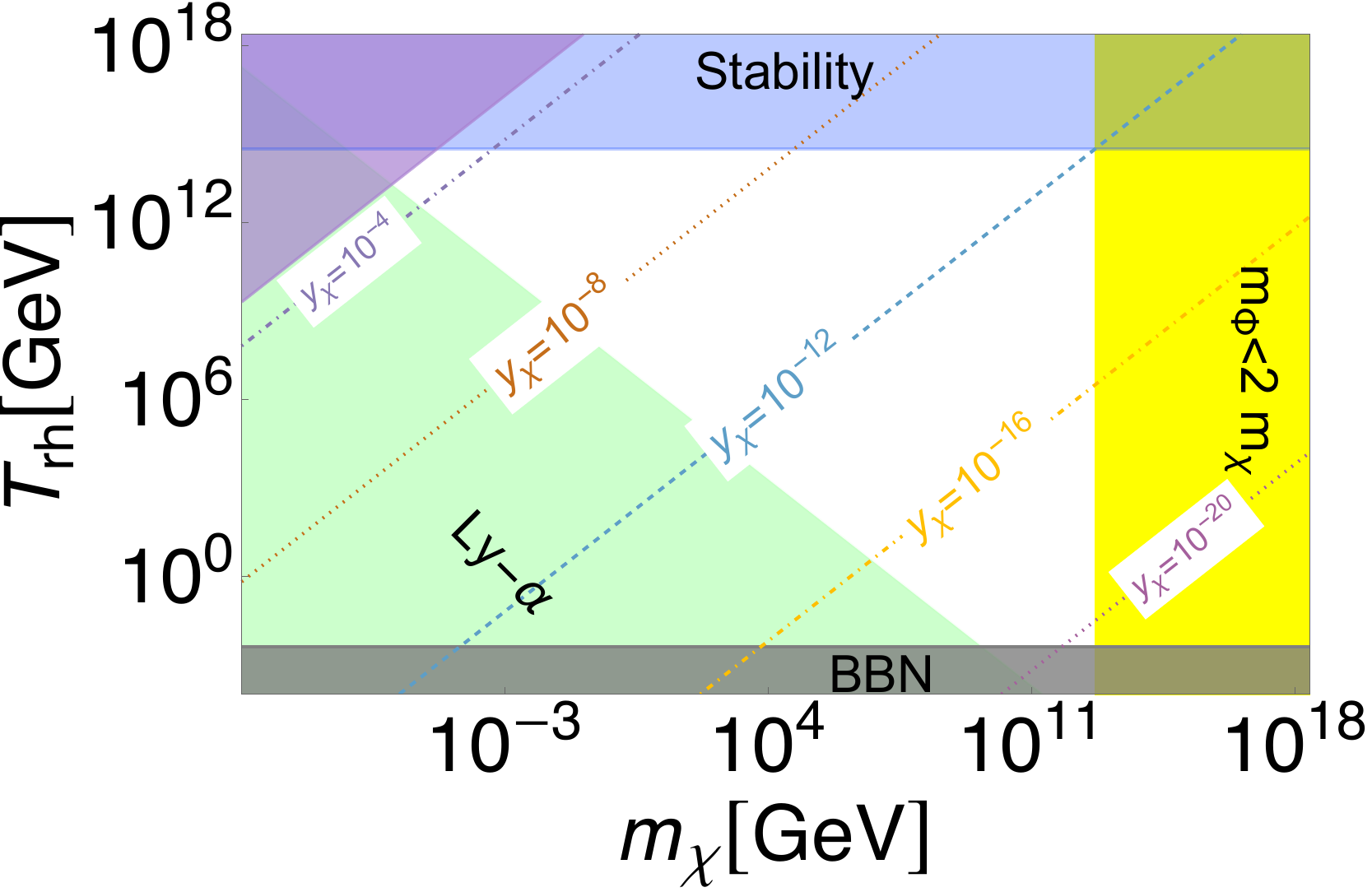}
    \caption{\it \raggedright \label{Fig:CDM_Yield_Decay-CW}
    The region without any color on $(\Trh, m_\chi)$ plane, depicted on log-log scale, indicates the allowed region for the benchmark value `\cw-BM1'~from~\cref{Table:Benchmark-CW}. The dashed or dashed-dotted lines that intersect the white region correspond to~\cref{Eq:eq to plot 2} for various values of $y_\chi$ that satisfy present-day \cdm~relic density. The colored regions on this plane represent various bounds – stability analysis: the maximum allowed value of $\Trh$ from the upper limit of $\lO$ from~\cref{Table:CW_stability} (neo blue colored horizontal stripe at the top) and the maximum allowed value of $\yc$ from~\cref{Table:CW_stability} (purple-colored wedge-shaped region at the top-left), from Ly-$\alpha$ bound on the mass of \dm~\cref{Eq:Lyman-alpha-bound}: green-colored region, and $\mc\lsim m_\phi/2$: yellow colored vertical stripe on the right. The gray colored stripe at the bottom depicts that $\Trh$ should be $\gsim 4\unit{MeV}$.
    }
\end{figure}

Similarly,~\cref{Fig:CDM_Yield_Decay-CW} shows $(\Trh, \mc)$ plane for \cw~inflation, particularly for benchmark `\cw-BM1'. The discontinuous lines correspond to~\cref{Eq:eq to plot 2}. The bounds on this plane are similar to~\cref{Fig:CDM_Yield_Decay-GH}: the largest permissible value of $\lO$ and $\yc$ from stability analysis of~\cref{Table:CW_stability} (neon blue-colored horizontal stripe at the top of the plot and purple-colored wedge-shaped region at the left-top corner of the plot), BBN temperature: $\Trh\gsim 4 \unit{MeV}$ (gray colored horizontal stripe at the bottom of the plot), and Ly-$\alpha$ bound from~\cref{Eq:Lyman-alpha-bound}: $\Trh\gsim (2 m_\phi)/\mc$ (region shaded with green color) and maximum possible value of $\mc=m_\phi/2$ (vertical stripe at the right, shaded with yellow color). The unshaded region is allowed on $(\Trh, \mc)$ plane and thus if $10^{-4}\gsim\yc\gsim  10^{-20}$ ($9.7318\times 10^{-7}\GeV \lsim \mc \lsim 4.6924\times 10^{12}\GeV$), then $\c$ produced only via the decay of inflaton for the benchmark `\cw-BM1'~in \cw~inflationary scenario can contribute up to $100\%$ of the entire \cdm~relic density of the present universe. Since, $m_\phi$ and ${\cal H}_I$ are almost similar for `\cw-BM1'~and `\cw-BM2'~(see~\cref{Table:reheating_CW}), Ly$-\alpha$ and $\mc \lsim m_\phi/2$ are almost similar for both benchmarks. However, significant difference on the allowed region on $(\Trh,\mc)$ for the two benchmark values comes the maximum permissible value of $\Trh$ from stability analysis of~\cref{Table:CW_stability}. As a result, the permitted range of $\mc$ for `\cw-BM2'~is almost similar $\sim 7.3988\times 10^{-7} \GeV \lsim \mc \lsim 3.0837\times 10^{12}\GeV$.

%
%
%

\subsubsection{DM from scattering}
Similar to GH inflationary scenario, there is an upper limit on $\lO$ from the stability analysis in this case (\cw~inflation). Therefore, upper limit on $\Trh$ from stability analysis
\eq{\label{Eq:CW_max_allowed_Trh}
{\Trh}_{\text{, allowable}}= 
\begin{cases}
    3.0845\times 10^{14} \, \quad \text{for `\cw-BM1'}\,, \\
    1.8911 \times 10^{14} \, \quad \text{for `\cw-BM2'}\,.
\end{cases}
}
And for such maximum possible values of $\Trh$,   $\Tmax/{\Trh}_{, \text{allowable}}>1$ is maintained.
Now, for $\Trh\sim {\Trh}_{, \text{allowable}} $, and the condition $Y_{IS,0}\sim Y_{{\rm CDM},0}$ (from~\cref{Eq:yield-DM-scattering-inflaton-graviton,Eq:eq to plot 2}), the value of $\mc$ is
\eq{\label{Eq:NMCW-large mc is needed-InflatonScattering}
{\mc}^{IS,0}\sim
\begin{cases}
    1.1602 \times 10^{10}\GeV \, \quad \text{for `\cw-BM1'}\,,\\
    1.2677\times 10^{10}\GeV \, \quad \text{for `\cw-BM2'}\,.
\end{cases}
}
Therefore, for \cw, the required condition  to make $Y_{IS,0}\sim Y_{{\rm CDM},0}$ is $\mc\gsim {\mc}^{IS,0}$ for $\Trh\lsim {\Trh}_{, \text{allowable}}$.

In the scenario where \dm~particles are produced through the scattering process involving SM particles and the mediator being a graviton (with $\mc\ll\Trh$) (\cref{Eq:Ysmg0}), assuming $\Trh\sim {\Trh}_{, \text{allowable}}$, we obtain
\eq{
Y_{\sm g,0}=
\begin{cases}
    4.8586 \times 10^{-18} \, \quad \text{(for `\cw-BM1')}\,,\\
    1.1196 \times 10^{-18}  \, \quad \text{(for `\cw-BM2')}\,.
\end{cases}
}
And thus to make $Y_{\sm g,0}\sim Y_{{\rm CDM},0}$ (from~\cref{Eq:present day CDM yield}), we get
\eq{\label{Eq:CW-large mc is needed}
\mc \sim 
\begin{cases}
    8.8503 \times 10^{7} \GeV \, \quad \text{(for `\cw-BM1')}\,,\\
    3.8407 \times 10^{8} \GeV \, \quad \text{(for `\cw-BM2')}\,.
\end{cases}
}
Thus, to satisfy the \cdm~yield, 
$\mc$ should be greater than or equal to the values mentioned in~\cref{Eq:CW-large mc is needed}.

However, it appears that the conclusion for \cw~is similar to GH inflation for \three~following~\cref{Eq:TERM}. This is because $\Tmax/\Trh\sim 40$ for $\Trh \sim 10^{12}\GeV$ and $\Tmax/{\Trh}_{, \text{allowable}}\sim 2$ for both `\cw-BM1'~and `\cw-BM2'~(see~\cref{Eq:CW_max_allowed_Trh}).

For the production of $\c$ via  2-to-2 scattering of SM particles
with inflaton as mediator (along with the constraint $\Trh\ll m_\phi$), the condition that $Y_{\sm i,0}$ contributes $f_\c$ fraction of total \cdm~relic density (\cref{Eq:game_changer1}) along with the maximum permissible values of $\yc$ (from~\cref{Table:CW_stability})
\eq{\label{Eq:CW-con}
{\mc}^{\sm i,0} =
\begin{cases}
    1.0706\, \frac{f_\c}{f_\phi^3}  \GeV \, \quad \text{(for `\cw-BM1')}\,,\\
    0.9416
    \, \frac{f_\c}{f_\phi^3}  \GeV \, \quad \text{(for `\cw-BM2')}\,.
\end{cases}
}
Therefore, just like in GH inflationary scenario, if $f_\c /f_\phi^3 \sim 1$, then $\mc\sim {\cal O}(1)\GeV$, while for $f_\c /f_\phi^3 \sim 10^8$, $\mc\sim {\cal O}(10^7 -10^8)\GeV$ is required.


\newpage

\section{Discussion and Conclusion }
\label{Sec:Conclusion}
In this study, we considered two distinct slow roll single-field inflationary scenarios: GH and \cw~with inflaton minimally coupled to gravity and suggested production of a non-thermal vector-like fermionic particle $\c$ during reheating era.
For each inflationary model, we considered a set of benchmark values and for those set of benchmark values, we explored the permissible parameter space encompassing $\yc$, $\mc$, assuming that $\c$ is a potential candidate for the \cdm~of the present universe. 
Salient features of our findings are as follows:
\begin{itemize}
 
    \item For our chosen benchmarks, we found that $r$ can be as small as $2.69\times 10^{-6}$ for small-field GH inflationary scenario (see~\cref{Table:benchmark_values_GH}), and this predicted value fits inside $1-\sigma$ contour on $(n_s,r)$ plane of \Planck2018+\BICEP3+\KeckArray2018 as well as upcoming SO experiment. On the other hand, the lowest value of $r$ we found for \cw~inflation is $\sim 1.9091\times 10^{-3}$, which can be corroborated by upcoming SO (see~\cref{Table:Benchmark-CW}).

   \item Using stability analysis and bound from \bbn~temperature, we computed the permissible upper limit for the coupling $\yc$ and permissible range for $\lO$ (defined in~\cref{Eq:reheating lagrangian}) for GH and \cw~inflation and found $\yc<{\cal O}\qty(10^{-4})$ and ${\cal O}(10^{-5})\lsim \lO \lsim {\cal O}(10^{12})$ (in $\GeV$)(see~\cref{Table:GH_stability,Table:CW_stability}), and these limits  varies with various benchmarks as well as inflationary models. The highest permissible value for $\yc$ is found for `CW-BM2' and largest allowed range of $\lO$ for the `CW-BM1' among all chosen benchmarks in both inflationary scenarios.

   \item From~\cref{Table:reheating_values_GH,Table:reheating_CW}, we infer that the value of $m_\phi$ is subject to alteration depending on benchmarks (and thus with inflationary parameters, including $r$) and inflationary scenarios, being highest ($\sim 2.0595\times 10^{13}\GeV$) for `GH-BM1'~and lowest for `\cw-BM2'~($\sim 6.1675\times 10^{12}\GeV$), consequently affecting $\Trh$, Ly-$\alpha$ bound and maximum possible value of $\mc$. 

   \item \cref{Fig:Tmax/Trh-GH,Fig:Tmax/Trh-CW} illustrates the variation of $\Tmax/\Trh$ against $\Trh$ with the highest value of $\Tmax/\Trh \sim {\cal O}\qty(10^8)$ at $\Trh\sim 4\unit{MeV}$. However, $\Tmax/\Trh$ changes based on chosen benchmarks and thus depends on inflationary parameters, e.g. $n_s$, $r$, etc. 

   \item Ensuring that $\c$ is produced only through inflaton decay and contributes $100\%$ of the total relic density of \cdm~of the present universe,~\cref{Fig:CDM_Yield_Decay-GH,Fig:CDM_Yield_Decay-CW} demonstrates that the allowed range of $\yc$  is $10^{-4}\gsim \yc\gsim 10^{-20}$ for both GH and \cw~inflationary scenarios. However, the allowed space on the $(\Trh, \mc)$ plane, which determines the allowed range of $\yc$, varies with benchmark values (for example, see~\cref{Fig:CDM_Yield_Decay-GH}) and thus depends on inflationary parameters.

   \item Assuming that $\c$ is produced via 2-to-2 scattering channels involving \sm~particles 
   with $\mc\ll \Trh$ or non-relativistic inflaton (both processes are mediated by graviton) or through 
   \four, 
    it was found that the individual contributions from these scattering channels for each inflationary scenario can account for $100\%$ of the total
   \cdm~yield of the present universe if $\mc\gsim {\cal O}(10^{10}) \GeV$ (see~\cref{Eq:GH-large mc is needed-InflatonScattering,Eq:NMCW-large mc is needed-InflatonScattering}) and $\mc \gsim {\cal O}(10^{9}) \GeV$ (see~\cref{Eq:GH-large mc is needed,Eq:CW-large mc is needed}) and $\mc \gsim {\cal O}(10^{8}) \GeV$ (for $\Trh\sim 10^{-2} m_\phi$, see~\cref{Eq:GH-con,Eq:CW-con}) , respectively. However, depending on the 
   inflationary parameters along with the specific model of inflation being considered, the precise range of $\mc$ undergoes variation. Nonetheless, the scattering channel involving  \sm~particles mediated by graviton, where $\Tmax \gg \mc \gg \Trh$, appears to be less efficient in generating the overall relic density of \cdm.

\end{itemize}

 In this article, we considered generalized version of Hilltop inflation and Coleman-Weinberg inflationary scenario with non-minimal coupling to curvature scalar. Our study was focused on the reheating era which is preceded by slow roll inflation happening near the maximum of the potential. Along with the inflaton, we incorporated a fermionic \sm~singlet field into our analysis, which is expected to contribute to the \cdm~density of the present universe. The verification of such a model could possibly be achieved through the detection BB mode by future \cmb~experiments, e.g. \cmbsfour, SO, and so on~\cite{CMB-S4:2020lpa,Hazumi:2019lys,Adak:2021lbu,SPT:2019nip,POLARBEAR:2015ixw,ACT:2020gnv,Harrington:2016jrz,LSPE:2020uos, Mennella:2019cwk, SimonsObservatory:2018koc,SPIDER:2021ncy}.  Presence of interaction with inflaton and \dm~particles may leave imprint on non-Gaussianities on \cmb~power spectrum and this can provide an alternative path to test our theory. However, exploring these aspects extends beyond the scope of this work.

\medskip

\section*{Acknowledgement}
The authors appreciate the insightful exchanges with 	Qaisar Shafi. Work of S.P. is funded by RSF Grant 19-42-02004. The research by M.K. was carried out in Southern Federal University with financial support of the Ministry of Science and Higher Education of the Russian Federation (State contract GZ0110/23-10-IF). Z.L. has been supported by the Polish National Science Center grant 2017/27/B/ ST2/02531.


\bibliographystyle{apsrev4-1}
\bibliography{CW-new-try.bib}

\begin{thebibliography}{104}%
\makeatletter
\providecommand \@ifxundefined [1]{%
 \@ifx{#1\undefined}
}%
\providecommand \@ifnum [1]{%
 \ifnum #1\expandafter \@firstoftwo
 \else \expandafter \@secondoftwo
 \fi
}%
\providecommand \@ifx [1]{%
 \ifx #1\expandafter \@firstoftwo
 \else \expandafter \@secondoftwo
 \fi
}%
\providecommand \natexlab [1]{#1}%
\providecommand \enquote  [1]{``#1''}%
\providecommand \bibnamefont  [1]{#1}%
\providecommand \bibfnamefont [1]{#1}%
\providecommand \citenamefont [1]{#1}%
\providecommand \href@noop [0]{\@secondoftwo}%
\providecommand \href [0]{\begingroup \@sanitize@url \@href}%
\providecommand \@href[1]{\@@startlink{#1}\@@href}%
\providecommand \@@href[1]{\endgroup#1\@@endlink}%
\providecommand \@sanitize@url [0]{\catcode `\\12\catcode `\$12\catcode `\&12\catcode `\#12\catcode `\^12\catcode `\_12\catcode `\%12\relax}%
\providecommand \@@startlink[1]{}%
\providecommand \@@endlink[0]{}%
\providecommand \url  [0]{\begingroup\@sanitize@url \@url }%
\providecommand \@url [1]{\endgroup\@href {#1}{\urlprefix }}%
\providecommand \urlprefix  [0]{URL }%
\providecommand \Eprint [0]{\href }%
\providecommand \doibase [0]{http://dx.doi.org/}%
\providecommand \selectlanguage [0]{\@gobble}%
\providecommand \bibinfo  [0]{\@secondoftwo}%
\providecommand \bibfield  [0]{\@secondoftwo}%
\providecommand \translation [1]{[#1]}%
\providecommand \BibitemOpen [0]{}%
\providecommand \bibitemStop [0]{}%
\providecommand \bibitemNoStop [0]{.\EOS\space}%
\providecommand \EOS [0]{\spacefactor3000\relax}%
\providecommand \BibitemShut  [1]{\csname bibitem#1\endcsname}%
\let\auto@bib@innerbib\@empty
\bibitem [{\citenamefont {Carr}\ and\ \citenamefont {Kuhnel}(2020)}]{Carr:2020xqk}%
  \BibitemOpen
  \bibfield  {author} {\bibinfo {author} {\bibfnamefont {B.}~\bibnamefont {Carr}}\ and\ \bibinfo {author} {\bibfnamefont {F.}~\bibnamefont {Kuhnel}},\ }\href {\doibase 10.1146/annurev-nucl-050520-125911} {\bibfield  {journal} {\bibinfo  {journal} {Ann. Rev. Nucl. Part. Sci.}\ }\textbf {\bibinfo {volume} {70}},\ \bibinfo {pages} {355} (\bibinfo {year} {2020})},\ \Eprint {http://arxiv.org/abs/2006.02838} {arXiv:2006.02838 [astro-ph.CO]} \BibitemShut {NoStop}%
\bibitem [{\citenamefont {Dolgov}\ and\ \citenamefont {Porey}(2019)}]{Dolgov:2019vlq}%
  \BibitemOpen
  \bibfield  {author} {\bibinfo {author} {\bibfnamefont {A.~D.}\ \bibnamefont {Dolgov}}\ and\ \bibinfo {author} {\bibfnamefont {S.}~\bibnamefont {Porey}},\ }\href@noop {} {\bibfield  {journal} {\bibinfo  {journal} {Bulg. Astron. J.}\ }\textbf {\bibinfo {volume} {34}},\ \bibinfo {pages} {2021} (\bibinfo {year} {2019})},\ \Eprint {http://arxiv.org/abs/1905.10972} {arXiv:1905.10972 [astro-ph.CO]} \BibitemShut {NoStop}%
\bibitem [{\citenamefont {McDonald}(2002)}]{McDonald:2001vt}%
  \BibitemOpen
  \bibfield  {author} {\bibinfo {author} {\bibfnamefont {J.}~\bibnamefont {McDonald}},\ }\href {\doibase 10.1103/PhysRevLett.88.091304} {\bibfield  {journal} {\bibinfo  {journal} {Phys. Rev. Lett.}\ }\textbf {\bibinfo {volume} {88}},\ \bibinfo {pages} {091304} (\bibinfo {year} {2002})},\ \Eprint {http://arxiv.org/abs/hep-ph/0106249} {arXiv:hep-ph/0106249} \BibitemShut {NoStop}%
\bibitem [{\citenamefont {Choi}\ and\ \citenamefont {Roszkowski}(2005)}]{Choi:2005vq}%
  \BibitemOpen
  \bibfield  {author} {\bibinfo {author} {\bibfnamefont {K.-Y.}\ \bibnamefont {Choi}}\ and\ \bibinfo {author} {\bibfnamefont {L.}~\bibnamefont {Roszkowski}},\ }\href {\doibase 10.1063/1.2149672} {\bibfield  {journal} {\bibinfo  {journal} {AIP Conf. Proc.}\ }\textbf {\bibinfo {volume} {805}},\ \bibinfo {pages} {30} (\bibinfo {year} {2005})},\ \Eprint {http://arxiv.org/abs/hep-ph/0511003} {arXiv:hep-ph/0511003} \BibitemShut {NoStop}%
\bibitem [{\citenamefont {Kusenko}(2006)}]{Kusenko:2006rh}%
  \BibitemOpen
  \bibfield  {author} {\bibinfo {author} {\bibfnamefont {A.}~\bibnamefont {Kusenko}},\ }\href {\doibase 10.1103/PhysRevLett.97.241301} {\bibfield  {journal} {\bibinfo  {journal} {Phys. Rev. Lett.}\ }\textbf {\bibinfo {volume} {97}},\ \bibinfo {pages} {241301} (\bibinfo {year} {2006})},\ \Eprint {http://arxiv.org/abs/hep-ph/0609081} {arXiv:hep-ph/0609081} \BibitemShut {NoStop}%
\bibitem [{\citenamefont {Petraki}\ and\ \citenamefont {Kusenko}(2008)}]{Petraki:2007gq}%
  \BibitemOpen
  \bibfield  {author} {\bibinfo {author} {\bibfnamefont {K.}~\bibnamefont {Petraki}}\ and\ \bibinfo {author} {\bibfnamefont {A.}~\bibnamefont {Kusenko}},\ }\href {\doibase 10.1103/PhysRevD.77.065014} {\bibfield  {journal} {\bibinfo  {journal} {Phys. Rev. D}\ }\textbf {\bibinfo {volume} {77}},\ \bibinfo {pages} {065014} (\bibinfo {year} {2008})},\ \Eprint {http://arxiv.org/abs/0711.4646} {arXiv:0711.4646 [hep-ph]} \BibitemShut {NoStop}%
\bibitem [{\citenamefont {Hall}\ \emph {et~al.}(2010)\citenamefont {Hall}, \citenamefont {Jedamzik}, \citenamefont {March-Russell},\ and\ \citenamefont {West}}]{Hall:2009bx}%
  \BibitemOpen
  \bibfield  {author} {\bibinfo {author} {\bibfnamefont {L.~J.}\ \bibnamefont {Hall}}, \bibinfo {author} {\bibfnamefont {K.}~\bibnamefont {Jedamzik}}, \bibinfo {author} {\bibfnamefont {J.}~\bibnamefont {March-Russell}}, \ and\ \bibinfo {author} {\bibfnamefont {S.~M.}\ \bibnamefont {West}},\ }\href {\doibase 10.1007/JHEP03(2010)080} {\bibfield  {journal} {\bibinfo  {journal} {JHEP}\ }\textbf {\bibinfo {volume} {03}},\ \bibinfo {pages} {080} (\bibinfo {year} {2010})},\ \Eprint {http://arxiv.org/abs/0911.1120} {arXiv:0911.1120 [hep-ph]} \BibitemShut {NoStop}%
\bibitem [{\citenamefont {Bernal}\ \emph {et~al.}(2017)\citenamefont {Bernal}, \citenamefont {Heikinheimo}, \citenamefont {Tenkanen}, \citenamefont {Tuominen},\ and\ \citenamefont {Vaskonen}}]{Bernal:2017kxu}%
  \BibitemOpen
  \bibfield  {author} {\bibinfo {author} {\bibfnamefont {N.}~\bibnamefont {Bernal}}, \bibinfo {author} {\bibfnamefont {M.}~\bibnamefont {Heikinheimo}}, \bibinfo {author} {\bibfnamefont {T.}~\bibnamefont {Tenkanen}}, \bibinfo {author} {\bibfnamefont {K.}~\bibnamefont {Tuominen}}, \ and\ \bibinfo {author} {\bibfnamefont {V.}~\bibnamefont {Vaskonen}},\ }\href {\doibase 10.1142/S0217751X1730023X} {\bibfield  {journal} {\bibinfo  {journal} {Int. J. Mod. Phys. A}\ }\textbf {\bibinfo {volume} {32}},\ \bibinfo {pages} {1730023} (\bibinfo {year} {2017})},\ \Eprint {http://arxiv.org/abs/1706.07442} {arXiv:1706.07442 [hep-ph]} \BibitemShut {NoStop}%
\bibitem [{\citenamefont {Haque}\ and\ \citenamefont {Maity}(2022)}]{Haque:2021mab}%
  \BibitemOpen
  \bibfield  {author} {\bibinfo {author} {\bibfnamefont {M.~R.}\ \bibnamefont {Haque}}\ and\ \bibinfo {author} {\bibfnamefont {D.}~\bibnamefont {Maity}},\ }\href {\doibase 10.1103/PhysRevD.106.023506} {\bibfield  {journal} {\bibinfo  {journal} {Phys. Rev. D}\ }\textbf {\bibinfo {volume} {106}},\ \bibinfo {pages} {023506} (\bibinfo {year} {2022})},\ \Eprint {http://arxiv.org/abs/2112.14668} {arXiv:2112.14668 [hep-ph]} \BibitemShut {NoStop}%
\bibitem [{\citenamefont {Haque}\ and\ \citenamefont {Maity}(2023)}]{Haque:2022kez}%
  \BibitemOpen
  \bibfield  {author} {\bibinfo {author} {\bibfnamefont {M.~R.}\ \bibnamefont {Haque}}\ and\ \bibinfo {author} {\bibfnamefont {D.}~\bibnamefont {Maity}},\ }\href {\doibase 10.1103/PhysRevD.107.043531} {\bibfield  {journal} {\bibinfo  {journal} {Phys. Rev. D}\ }\textbf {\bibinfo {volume} {107}},\ \bibinfo {pages} {043531} (\bibinfo {year} {2023})},\ \Eprint {http://arxiv.org/abs/2201.02348} {arXiv:2201.02348 [hep-ph]} \BibitemShut {NoStop}%
\bibitem [{\citenamefont {Haque}\ \emph {et~al.}(2023)\citenamefont {Haque}, \citenamefont {Maity},\ and\ \citenamefont {Mondal}}]{Haque:2023yra}%
  \BibitemOpen
  \bibfield  {author} {\bibinfo {author} {\bibfnamefont {M.~R.}\ \bibnamefont {Haque}}, \bibinfo {author} {\bibfnamefont {D.}~\bibnamefont {Maity}}, \ and\ \bibinfo {author} {\bibfnamefont {R.}~\bibnamefont {Mondal}},\ }\href {\doibase 10.1007/JHEP09(2023)012} {\bibfield  {journal} {\bibinfo  {journal} {JHEP}\ }\textbf {\bibinfo {volume} {09}},\ \bibinfo {pages} {012} (\bibinfo {year} {2023})},\ \Eprint {http://arxiv.org/abs/2301.01641} {arXiv:2301.01641 [hep-ph]} \BibitemShut {NoStop}%
\bibitem [{\citenamefont {Baer}\ \emph {et~al.}(2015)\citenamefont {Baer}, \citenamefont {Choi}, \citenamefont {Kim},\ and\ \citenamefont {Roszkowski}}]{Baer:2014eja}%
  \BibitemOpen
  \bibfield  {author} {\bibinfo {author} {\bibfnamefont {H.}~\bibnamefont {Baer}}, \bibinfo {author} {\bibfnamefont {K.-Y.}\ \bibnamefont {Choi}}, \bibinfo {author} {\bibfnamefont {J.~E.}\ \bibnamefont {Kim}}, \ and\ \bibinfo {author} {\bibfnamefont {L.}~\bibnamefont {Roszkowski}},\ }\href {\doibase 10.1016/j.physrep.2014.10.002} {\bibfield  {journal} {\bibinfo  {journal} {Phys. Rept.}\ }\textbf {\bibinfo {volume} {555}},\ \bibinfo {pages} {1} (\bibinfo {year} {2015})},\ \Eprint {http://arxiv.org/abs/1407.0017} {arXiv:1407.0017 [hep-ph]} \BibitemShut {NoStop}%
\bibitem [{\citenamefont {Garny}\ \emph {et~al.}(2016)\citenamefont {Garny}, \citenamefont {Sandora},\ and\ \citenamefont {Sloth}}]{Garny:2015sjg}%
  \BibitemOpen
  \bibfield  {author} {\bibinfo {author} {\bibfnamefont {M.}~\bibnamefont {Garny}}, \bibinfo {author} {\bibfnamefont {M.}~\bibnamefont {Sandora}}, \ and\ \bibinfo {author} {\bibfnamefont {M.~S.}\ \bibnamefont {Sloth}},\ }\href {\doibase 10.1103/PhysRevLett.116.101302} {\bibfield  {journal} {\bibinfo  {journal} {Phys. Rev. Lett.}\ }\textbf {\bibinfo {volume} {116}},\ \bibinfo {pages} {101302} (\bibinfo {year} {2016})},\ \Eprint {http://arxiv.org/abs/1511.03278} {arXiv:1511.03278 [hep-ph]} \BibitemShut {NoStop}%
\bibitem [{\citenamefont {Tang}\ and\ \citenamefont {Wu}(2016)}]{Tang:2016vch}%
  \BibitemOpen
  \bibfield  {author} {\bibinfo {author} {\bibfnamefont {Y.}~\bibnamefont {Tang}}\ and\ \bibinfo {author} {\bibfnamefont {Y.-L.}\ \bibnamefont {Wu}},\ }\href {\doibase 10.1016/j.physletb.2016.05.045} {\bibfield  {journal} {\bibinfo  {journal} {Phys. Lett. B}\ }\textbf {\bibinfo {volume} {758}},\ \bibinfo {pages} {402} (\bibinfo {year} {2016})},\ \Eprint {http://arxiv.org/abs/1604.04701} {arXiv:1604.04701 [hep-ph]} \BibitemShut {NoStop}%
\bibitem [{\citenamefont {Tang}\ and\ \citenamefont {Wu}(2017)}]{Tang:2017hvq}%
  \BibitemOpen
  \bibfield  {author} {\bibinfo {author} {\bibfnamefont {Y.}~\bibnamefont {Tang}}\ and\ \bibinfo {author} {\bibfnamefont {Y.-L.}\ \bibnamefont {Wu}},\ }\href {\doibase 10.1016/j.physletb.2017.10.034} {\bibfield  {journal} {\bibinfo  {journal} {Phys. Lett. B}\ }\textbf {\bibinfo {volume} {774}},\ \bibinfo {pages} {676} (\bibinfo {year} {2017})},\ \Eprint {http://arxiv.org/abs/1708.05138} {arXiv:1708.05138 [hep-ph]} \BibitemShut {NoStop}%
\bibitem [{\citenamefont {Garny}\ \emph {et~al.}(2018)\citenamefont {Garny}, \citenamefont {Palessandro}, \citenamefont {Sandora},\ and\ \citenamefont {Sloth}}]{Garny:2017kha}%
  \BibitemOpen
  \bibfield  {author} {\bibinfo {author} {\bibfnamefont {M.}~\bibnamefont {Garny}}, \bibinfo {author} {\bibfnamefont {A.}~\bibnamefont {Palessandro}}, \bibinfo {author} {\bibfnamefont {M.}~\bibnamefont {Sandora}}, \ and\ \bibinfo {author} {\bibfnamefont {M.~S.}\ \bibnamefont {Sloth}},\ }\href {\doibase 10.1088/1475-7516/2018/02/027} {\bibfield  {journal} {\bibinfo  {journal} {JCAP}\ }\textbf {\bibinfo {volume} {02}},\ \bibinfo {pages} {027} (\bibinfo {year} {2018})},\ \Eprint {http://arxiv.org/abs/1709.09688} {arXiv:1709.09688 [hep-ph]} \BibitemShut {NoStop}%
\bibitem [{\citenamefont {Bernal}\ \emph {et~al.}(2018{\natexlab{a}})\citenamefont {Bernal}, \citenamefont {Dutra}, \citenamefont {Mambrini}, \citenamefont {Olive}, \citenamefont {Peloso},\ and\ \citenamefont {Pierre}}]{Bernal:2018qlk}%
  \BibitemOpen
  \bibfield  {author} {\bibinfo {author} {\bibfnamefont {N.}~\bibnamefont {Bernal}}, \bibinfo {author} {\bibfnamefont {M.}~\bibnamefont {Dutra}}, \bibinfo {author} {\bibfnamefont {Y.}~\bibnamefont {Mambrini}}, \bibinfo {author} {\bibfnamefont {K.}~\bibnamefont {Olive}}, \bibinfo {author} {\bibfnamefont {M.}~\bibnamefont {Peloso}}, \ and\ \bibinfo {author} {\bibfnamefont {M.}~\bibnamefont {Pierre}},\ }\href {\doibase 10.1103/PhysRevD.97.115020} {\bibfield  {journal} {\bibinfo  {journal} {Phys. Rev. D}\ }\textbf {\bibinfo {volume} {97}},\ \bibinfo {pages} {115020} (\bibinfo {year} {2018}{\natexlab{a}})},\ \Eprint {http://arxiv.org/abs/1803.01866} {arXiv:1803.01866 [hep-ph]} \BibitemShut {NoStop}%
\bibitem [{\citenamefont {Paul}\ \emph {et~al.}(2019)\citenamefont {Paul}, \citenamefont {Ghoshal}, \citenamefont {Chatterjee},\ and\ \citenamefont {Pal}}]{Paul:2018njm}%
  \BibitemOpen
  \bibfield  {author} {\bibinfo {author} {\bibfnamefont {A.}~\bibnamefont {Paul}}, \bibinfo {author} {\bibfnamefont {A.}~\bibnamefont {Ghoshal}}, \bibinfo {author} {\bibfnamefont {A.}~\bibnamefont {Chatterjee}}, \ and\ \bibinfo {author} {\bibfnamefont {S.}~\bibnamefont {Pal}},\ }\href {\doibase 10.1140/epjc/s10052-019-7348-5} {\bibfield  {journal} {\bibinfo  {journal} {Eur. Phys. J. C}\ }\textbf {\bibinfo {volume} {79}},\ \bibinfo {pages} {818} (\bibinfo {year} {2019})},\ \Eprint {http://arxiv.org/abs/1808.09706} {arXiv:1808.09706 [astro-ph.CO]} \BibitemShut {NoStop}%
\bibitem [{\citenamefont {Ghoshal}\ \emph {et~al.}(2023{\natexlab{a}})\citenamefont {Ghoshal}, \citenamefont {Lalak},\ and\ \citenamefont {Porey}}]{Ghoshal:2023phi}%
  \BibitemOpen
  \bibfield  {author} {\bibinfo {author} {\bibfnamefont {A.}~\bibnamefont {Ghoshal}}, \bibinfo {author} {\bibfnamefont {Z.}~\bibnamefont {Lalak}}, \ and\ \bibinfo {author} {\bibfnamefont {S.}~\bibnamefont {Porey}},\ }\href {\doibase 10.1103/PhysRevD.108.063030} {\bibfield  {journal} {\bibinfo  {journal} {Phys. Rev. D}\ }\textbf {\bibinfo {volume} {108}},\ \bibinfo {pages} {063030} (\bibinfo {year} {2023}{\natexlab{a}})},\ \Eprint {http://arxiv.org/abs/2302.03268} {arXiv:2302.03268 [hep-ph]} \BibitemShut {NoStop}%
\bibitem [{\citenamefont {Starobinsky}(1980)}]{Starobinsky:1980te}%
  \BibitemOpen
  \bibfield  {author} {\bibinfo {author} {\bibfnamefont {A.~A.}\ \bibnamefont {Starobinsky}},\ }\href {\doibase 10.1016/0370-2693(80)90670-X} {\bibfield  {journal} {\bibinfo  {journal} {Phys. Lett. B}\ }\textbf {\bibinfo {volume} {91}},\ \bibinfo {pages} {99} (\bibinfo {year} {1980})}\BibitemShut {NoStop}%
\bibitem [{\citenamefont {Guth}(1981)}]{Guth:1980zm}%
  \BibitemOpen
  \bibfield  {author} {\bibinfo {author} {\bibfnamefont {A.~H.}\ \bibnamefont {Guth}},\ }\href {\doibase 10.1103/PhysRevD.23.347} {\bibfield  {journal} {\bibinfo  {journal} {Phys. Rev. D}\ }\textbf {\bibinfo {volume} {23}},\ \bibinfo {pages} {347} (\bibinfo {year} {1981})}\BibitemShut {NoStop}%
\bibitem [{\citenamefont {Linde}(1982)}]{Linde:1981mu}%
  \BibitemOpen
  \bibfield  {author} {\bibinfo {author} {\bibfnamefont {A.~D.}\ \bibnamefont {Linde}},\ }\href {\doibase 10.1016/0370-2693(82)91219-9} {\bibfield  {journal} {\bibinfo  {journal} {Phys. Lett. B}\ }\textbf {\bibinfo {volume} {108}},\ \bibinfo {pages} {389} (\bibinfo {year} {1982})}\BibitemShut {NoStop}%
\bibitem [{\citenamefont {Albrecht}\ and\ \citenamefont {Steinhardt}(1982)}]{Albrecht:1982wi}%
  \BibitemOpen
  \bibfield  {author} {\bibinfo {author} {\bibfnamefont {A.}~\bibnamefont {Albrecht}}\ and\ \bibinfo {author} {\bibfnamefont {P.~J.}\ \bibnamefont {Steinhardt}},\ }\href {\doibase 10.1103/PhysRevLett.48.1220} {\bibfield  {journal} {\bibinfo  {journal} {Phys. Rev. Lett.}\ }\textbf {\bibinfo {volume} {48}},\ \bibinfo {pages} {1220} (\bibinfo {year} {1982})}\BibitemShut {NoStop}%
\bibitem [{\citenamefont {Lerner}\ and\ \citenamefont {McDonald}(2009)}]{Lerner:2009xg}%
  \BibitemOpen
  \bibfield  {author} {\bibinfo {author} {\bibfnamefont {R.~N.}\ \bibnamefont {Lerner}}\ and\ \bibinfo {author} {\bibfnamefont {J.}~\bibnamefont {McDonald}},\ }\href {\doibase 10.1103/PhysRevD.80.123507} {\bibfield  {journal} {\bibinfo  {journal} {Phys. Rev. D}\ }\textbf {\bibinfo {volume} {80}},\ \bibinfo {pages} {123507} (\bibinfo {year} {2009})},\ \Eprint {http://arxiv.org/abs/0909.0520} {arXiv:0909.0520 [hep-ph]} \BibitemShut {NoStop}%
\bibitem [{\citenamefont {Kahlhoefer}\ and\ \citenamefont {McDonald}(2015)}]{Kahlhoefer:2015jma}%
  \BibitemOpen
  \bibfield  {author} {\bibinfo {author} {\bibfnamefont {F.}~\bibnamefont {Kahlhoefer}}\ and\ \bibinfo {author} {\bibfnamefont {J.}~\bibnamefont {McDonald}},\ }\href {\doibase 10.1088/1475-7516/2015/11/015} {\bibfield  {journal} {\bibinfo  {journal} {JCAP}\ }\textbf {\bibinfo {volume} {11}},\ \bibinfo {pages} {015} (\bibinfo {year} {2015})},\ \Eprint {http://arxiv.org/abs/1507.03600} {arXiv:1507.03600 [astro-ph.CO]} \BibitemShut {NoStop}%
\bibitem [{\citenamefont {Clark}\ \emph {et~al.}(2009)\citenamefont {Clark}, \citenamefont {Liu}, \citenamefont {Love},\ and\ \citenamefont {ter Veldhuis}}]{Clark:2009dc}%
  \BibitemOpen
  \bibfield  {author} {\bibinfo {author} {\bibfnamefont {T.~E.}\ \bibnamefont {Clark}}, \bibinfo {author} {\bibfnamefont {B.}~\bibnamefont {Liu}}, \bibinfo {author} {\bibfnamefont {S.~T.}\ \bibnamefont {Love}}, \ and\ \bibinfo {author} {\bibfnamefont {T.}~\bibnamefont {ter Veldhuis}},\ }\href {\doibase 10.1103/PhysRevD.80.075019} {\bibfield  {journal} {\bibinfo  {journal} {Phys. Rev. D}\ }\textbf {\bibinfo {volume} {80}},\ \bibinfo {pages} {075019} (\bibinfo {year} {2009})},\ \Eprint {http://arxiv.org/abs/0906.5595} {arXiv:0906.5595 [hep-ph]} \BibitemShut {NoStop}%
\bibitem [{\citenamefont {Khoze}(2013)}]{Khoze:2013uia}%
  \BibitemOpen
  \bibfield  {author} {\bibinfo {author} {\bibfnamefont {V.~V.}\ \bibnamefont {Khoze}},\ }\href {\doibase 10.1007/JHEP11(2013)215} {\bibfield  {journal} {\bibinfo  {journal} {JHEP}\ }\textbf {\bibinfo {volume} {11}},\ \bibinfo {pages} {215} (\bibinfo {year} {2013})},\ \Eprint {http://arxiv.org/abs/1308.6338} {arXiv:1308.6338 [hep-ph]} \BibitemShut {NoStop}%
\bibitem [{\citenamefont {Almeida}\ \emph {et~al.}(2019)\citenamefont {Almeida}, \citenamefont {Bernal}, \citenamefont {Rubio},\ and\ \citenamefont {Tenkanen}}]{Almeida:2018oid}%
  \BibitemOpen
  \bibfield  {author} {\bibinfo {author} {\bibfnamefont {J.~P.~B.}\ \bibnamefont {Almeida}}, \bibinfo {author} {\bibfnamefont {N.}~\bibnamefont {Bernal}}, \bibinfo {author} {\bibfnamefont {J.}~\bibnamefont {Rubio}}, \ and\ \bibinfo {author} {\bibfnamefont {T.}~\bibnamefont {Tenkanen}},\ }\href {\doibase 10.1088/1475-7516/2019/03/012} {\bibfield  {journal} {\bibinfo  {journal} {JCAP}\ }\textbf {\bibinfo {volume} {03}},\ \bibinfo {pages} {012} (\bibinfo {year} {2019})},\ \Eprint {http://arxiv.org/abs/1811.09640} {arXiv:1811.09640 [hep-ph]} \BibitemShut {NoStop}%
\bibitem [{\citenamefont {Bernal}\ \emph {et~al.}(2018{\natexlab{b}})\citenamefont {Bernal}, \citenamefont {Chatterjee},\ and\ \citenamefont {Paul}}]{Bernal:2018hjm}%
  \BibitemOpen
  \bibfield  {author} {\bibinfo {author} {\bibfnamefont {N.}~\bibnamefont {Bernal}}, \bibinfo {author} {\bibfnamefont {A.}~\bibnamefont {Chatterjee}}, \ and\ \bibinfo {author} {\bibfnamefont {A.}~\bibnamefont {Paul}},\ }\href {\doibase 10.1088/1475-7516/2018/12/020} {\bibfield  {journal} {\bibinfo  {journal} {JCAP}\ }\textbf {\bibinfo {volume} {12}},\ \bibinfo {pages} {020} (\bibinfo {year} {2018}{\natexlab{b}})},\ \Eprint {http://arxiv.org/abs/1809.02338} {arXiv:1809.02338 [hep-ph]} \BibitemShut {NoStop}%
\bibitem [{\citenamefont {Aravind}\ \emph {et~al.}(2016)\citenamefont {Aravind}, \citenamefont {Xiao},\ and\ \citenamefont {Yu}}]{Aravind:2015xst}%
  \BibitemOpen
  \bibfield  {author} {\bibinfo {author} {\bibfnamefont {A.}~\bibnamefont {Aravind}}, \bibinfo {author} {\bibfnamefont {M.}~\bibnamefont {Xiao}}, \ and\ \bibinfo {author} {\bibfnamefont {J.-H.}\ \bibnamefont {Yu}},\ }\href {\doibase 10.1103/PhysRevD.93.123513} {\bibfield  {journal} {\bibinfo  {journal} {Phys. Rev. D}\ }\textbf {\bibinfo {volume} {93}},\ \bibinfo {pages} {123513} (\bibinfo {year} {2016})},\ \bibinfo {note} {[Erratum: Phys.Rev.D 96, 069901 (2017)]},\ \Eprint {http://arxiv.org/abs/1512.09126} {arXiv:1512.09126 [hep-ph]} \BibitemShut {NoStop}%
\bibitem [{\citenamefont {Ballesteros}\ \emph {et~al.}(2017)\citenamefont {Ballesteros}, \citenamefont {Redondo}, \citenamefont {Ringwald},\ and\ \citenamefont {Tamarit}}]{Ballesteros:2016xej}%
  \BibitemOpen
  \bibfield  {author} {\bibinfo {author} {\bibfnamefont {G.}~\bibnamefont {Ballesteros}}, \bibinfo {author} {\bibfnamefont {J.}~\bibnamefont {Redondo}}, \bibinfo {author} {\bibfnamefont {A.}~\bibnamefont {Ringwald}}, \ and\ \bibinfo {author} {\bibfnamefont {C.}~\bibnamefont {Tamarit}},\ }\href {\doibase 10.1088/1475-7516/2017/08/001} {\bibfield  {journal} {\bibinfo  {journal} {JCAP}\ }\textbf {\bibinfo {volume} {08}},\ \bibinfo {pages} {001} (\bibinfo {year} {2017})},\ \Eprint {http://arxiv.org/abs/1610.01639} {arXiv:1610.01639 [hep-ph]} \BibitemShut {NoStop}%
\bibitem [{\citenamefont {Borah}\ \emph {et~al.}(2019)\citenamefont {Borah}, \citenamefont {Dev},\ and\ \citenamefont {Kumar}}]{Borah:2018rca}%
  \BibitemOpen
  \bibfield  {author} {\bibinfo {author} {\bibfnamefont {D.}~\bibnamefont {Borah}}, \bibinfo {author} {\bibfnamefont {P.~S.~B.}\ \bibnamefont {Dev}}, \ and\ \bibinfo {author} {\bibfnamefont {A.}~\bibnamefont {Kumar}},\ }\href {\doibase 10.1103/PhysRevD.99.055012} {\bibfield  {journal} {\bibinfo  {journal} {Phys. Rev. D}\ }\textbf {\bibinfo {volume} {99}},\ \bibinfo {pages} {055012} (\bibinfo {year} {2019})},\ \Eprint {http://arxiv.org/abs/1810.03645} {arXiv:1810.03645 [hep-ph]} \BibitemShut {NoStop}%
\bibitem [{\citenamefont {Hamada}\ \emph {et~al.}(2014)\citenamefont {Hamada}, \citenamefont {Kawai},\ and\ \citenamefont {Oda}}]{Hamada:2014xka}%
  \BibitemOpen
  \bibfield  {author} {\bibinfo {author} {\bibfnamefont {Y.}~\bibnamefont {Hamada}}, \bibinfo {author} {\bibfnamefont {H.}~\bibnamefont {Kawai}}, \ and\ \bibinfo {author} {\bibfnamefont {K.-y.}\ \bibnamefont {Oda}},\ }\href {\doibase 10.1007/JHEP07(2014)026} {\bibfield  {journal} {\bibinfo  {journal} {JHEP}\ }\textbf {\bibinfo {volume} {07}},\ \bibinfo {pages} {026} (\bibinfo {year} {2014})},\ \Eprint {http://arxiv.org/abs/1404.6141} {arXiv:1404.6141 [hep-ph]} \BibitemShut {NoStop}%
\bibitem [{\citenamefont {Choubey}\ and\ \citenamefont {Kumar}(2017)}]{Choubey:2017hsq}%
  \BibitemOpen
  \bibfield  {author} {\bibinfo {author} {\bibfnamefont {S.}~\bibnamefont {Choubey}}\ and\ \bibinfo {author} {\bibfnamefont {A.}~\bibnamefont {Kumar}},\ }\href {\doibase 10.1007/JHEP11(2017)080} {\bibfield  {journal} {\bibinfo  {journal} {JHEP}\ }\textbf {\bibinfo {volume} {11}},\ \bibinfo {pages} {080} (\bibinfo {year} {2017})},\ \Eprint {http://arxiv.org/abs/1707.06587} {arXiv:1707.06587 [hep-ph]} \BibitemShut {NoStop}%
\bibitem [{\citenamefont {Cline}\ \emph {et~al.}(2020)\citenamefont {Cline}, \citenamefont {Puel},\ and\ \citenamefont {Toma}}]{Cline:2020mdt}%
  \BibitemOpen
  \bibfield  {author} {\bibinfo {author} {\bibfnamefont {J.~M.}\ \bibnamefont {Cline}}, \bibinfo {author} {\bibfnamefont {M.}~\bibnamefont {Puel}}, \ and\ \bibinfo {author} {\bibfnamefont {T.}~\bibnamefont {Toma}},\ }\href {\doibase 10.1007/JHEP05(2020)039} {\bibfield  {journal} {\bibinfo  {journal} {JHEP}\ }\textbf {\bibinfo {volume} {05}},\ \bibinfo {pages} {039} (\bibinfo {year} {2020})},\ \Eprint {http://arxiv.org/abs/2001.11505} {arXiv:2001.11505 [hep-ph]} \BibitemShut {NoStop}%
\bibitem [{\citenamefont {Tenkanen}(2016)}]{Tenkanen:2016twd}%
  \BibitemOpen
  \bibfield  {author} {\bibinfo {author} {\bibfnamefont {T.}~\bibnamefont {Tenkanen}},\ }\href {\doibase 10.1007/JHEP09(2016)049} {\bibfield  {journal} {\bibinfo  {journal} {JHEP}\ }\textbf {\bibinfo {volume} {09}},\ \bibinfo {pages} {049} (\bibinfo {year} {2016})},\ \Eprint {http://arxiv.org/abs/1607.01379} {arXiv:1607.01379 [hep-ph]} \BibitemShut {NoStop}%
\bibitem [{\citenamefont {Abe}\ \emph {et~al.}(2021)\citenamefont {Abe}, \citenamefont {Toma},\ and\ \citenamefont {Yoshioka}}]{Abe:2020ldj}%
  \BibitemOpen
  \bibfield  {author} {\bibinfo {author} {\bibfnamefont {Y.}~\bibnamefont {Abe}}, \bibinfo {author} {\bibfnamefont {T.}~\bibnamefont {Toma}}, \ and\ \bibinfo {author} {\bibfnamefont {K.}~\bibnamefont {Yoshioka}},\ }\href {\doibase 10.1007/JHEP03(2021)130} {\bibfield  {journal} {\bibinfo  {journal} {JHEP}\ }\textbf {\bibinfo {volume} {03}},\ \bibinfo {pages} {130} (\bibinfo {year} {2021})},\ \Eprint {http://arxiv.org/abs/2012.10286} {arXiv:2012.10286 [hep-ph]} \BibitemShut {NoStop}%
\bibitem [{\citenamefont {Okada}\ \emph {et~al.}(2020)\citenamefont {Okada}, \citenamefont {Raut},\ and\ \citenamefont {Shafi}}]{Okada:2020cvq}%
  \BibitemOpen
  \bibfield  {author} {\bibinfo {author} {\bibfnamefont {N.}~\bibnamefont {Okada}}, \bibinfo {author} {\bibfnamefont {D.}~\bibnamefont {Raut}}, \ and\ \bibinfo {author} {\bibfnamefont {Q.}~\bibnamefont {Shafi}},\ }\href {\doibase 10.1140/epjc/s10052-020-8343-6} {\bibfield  {journal} {\bibinfo  {journal} {Eur. Phys. J. C}\ }\textbf {\bibinfo {volume} {80}},\ \bibinfo {pages} {1056} (\bibinfo {year} {2020})},\ \Eprint {http://arxiv.org/abs/2002.07110} {arXiv:2002.07110 [hep-ph]} \BibitemShut {NoStop}%
\bibitem [{\citenamefont {Das}\ \emph {et~al.}(2022)\citenamefont {Das}, \citenamefont {Gola}, \citenamefont {Mandal},\ and\ \citenamefont {Sinha}}]{Das:2022oyx}%
  \BibitemOpen
  \bibfield  {author} {\bibinfo {author} {\bibfnamefont {A.}~\bibnamefont {Das}}, \bibinfo {author} {\bibfnamefont {S.}~\bibnamefont {Gola}}, \bibinfo {author} {\bibfnamefont {S.}~\bibnamefont {Mandal}}, \ and\ \bibinfo {author} {\bibfnamefont {N.}~\bibnamefont {Sinha}},\ }\href {\doibase 10.1016/j.physletb.2022.137117} {\bibfield  {journal} {\bibinfo  {journal} {Phys. Lett. B}\ }\textbf {\bibinfo {volume} {829}},\ \bibinfo {pages} {137117} (\bibinfo {year} {2022})},\ \Eprint {http://arxiv.org/abs/2202.01443} {arXiv:2202.01443 [hep-ph]} \BibitemShut {NoStop}%
\bibitem [{\citenamefont {Das}\ \emph {et~al.}(2016)\citenamefont {Das}, \citenamefont {Oda}, \citenamefont {Okada},\ and\ \citenamefont {Takahashi}}]{Das:2016zue}%
  \BibitemOpen
  \bibfield  {author} {\bibinfo {author} {\bibfnamefont {A.}~\bibnamefont {Das}}, \bibinfo {author} {\bibfnamefont {S.}~\bibnamefont {Oda}}, \bibinfo {author} {\bibfnamefont {N.}~\bibnamefont {Okada}}, \ and\ \bibinfo {author} {\bibfnamefont {D.-s.}\ \bibnamefont {Takahashi}},\ }\href {\doibase 10.1103/PhysRevD.93.115038} {\bibfield  {journal} {\bibinfo  {journal} {Phys. Rev. D}\ }\textbf {\bibinfo {volume} {93}},\ \bibinfo {pages} {115038} (\bibinfo {year} {2016})},\ \Eprint {http://arxiv.org/abs/1605.01157} {arXiv:1605.01157 [hep-ph]} \BibitemShut {NoStop}%
\bibitem [{\citenamefont {Das}\ \emph {et~al.}(2017)\citenamefont {Das}, \citenamefont {Okada},\ and\ \citenamefont {Papapietro}}]{Das:2015nwk}%
  \BibitemOpen
  \bibfield  {author} {\bibinfo {author} {\bibfnamefont {A.}~\bibnamefont {Das}}, \bibinfo {author} {\bibfnamefont {N.}~\bibnamefont {Okada}}, \ and\ \bibinfo {author} {\bibfnamefont {N.}~\bibnamefont {Papapietro}},\ }\href {\doibase 10.1140/epjc/s10052-017-4683-2} {\bibfield  {journal} {\bibinfo  {journal} {Eur. Phys. J. C}\ }\textbf {\bibinfo {volume} {77}},\ \bibinfo {pages} {122} (\bibinfo {year} {2017})},\ \Eprint {http://arxiv.org/abs/1509.01466} {arXiv:1509.01466 [hep-ph]} \BibitemShut {NoStop}%
\bibitem [{\citenamefont {Daido}\ \emph {et~al.}(2017)\citenamefont {Daido}, \citenamefont {Takahashi},\ and\ \citenamefont {Yin}}]{Daido:2017wwb}%
  \BibitemOpen
  \bibfield  {author} {\bibinfo {author} {\bibfnamefont {R.}~\bibnamefont {Daido}}, \bibinfo {author} {\bibfnamefont {F.}~\bibnamefont {Takahashi}}, \ and\ \bibinfo {author} {\bibfnamefont {W.}~\bibnamefont {Yin}},\ }\href {\doibase 10.1088/1475-7516/2017/05/044} {\bibfield  {journal} {\bibinfo  {journal} {JCAP}\ }\textbf {\bibinfo {volume} {05}},\ \bibinfo {pages} {044} (\bibinfo {year} {2017})},\ \Eprint {http://arxiv.org/abs/1702.03284} {arXiv:1702.03284 [hep-ph]} \BibitemShut {NoStop}%
\bibitem [{\citenamefont {Shaposhnikov}\ and\ \citenamefont {Tkachev}(2006)}]{Shaposhnikov:2006xi}%
  \BibitemOpen
  \bibfield  {author} {\bibinfo {author} {\bibfnamefont {M.}~\bibnamefont {Shaposhnikov}}\ and\ \bibinfo {author} {\bibfnamefont {I.}~\bibnamefont {Tkachev}},\ }\href {\doibase 10.1016/j.physletb.2006.06.063} {\bibfield  {journal} {\bibinfo  {journal} {Phys. Lett. B}\ }\textbf {\bibinfo {volume} {639}},\ \bibinfo {pages} {414} (\bibinfo {year} {2006})},\ \Eprint {http://arxiv.org/abs/hep-ph/0604236} {arXiv:hep-ph/0604236} \BibitemShut {NoStop}%
\bibitem [{\citenamefont {Davoudiasl}\ \emph {et~al.}(2005)\citenamefont {Davoudiasl}, \citenamefont {Kitano}, \citenamefont {Li},\ and\ \citenamefont {Murayama}}]{Davoudiasl:2004be}%
  \BibitemOpen
  \bibfield  {author} {\bibinfo {author} {\bibfnamefont {H.}~\bibnamefont {Davoudiasl}}, \bibinfo {author} {\bibfnamefont {R.}~\bibnamefont {Kitano}}, \bibinfo {author} {\bibfnamefont {T.}~\bibnamefont {Li}}, \ and\ \bibinfo {author} {\bibfnamefont {H.}~\bibnamefont {Murayama}},\ }\href {\doibase 10.1016/j.physletb.2005.01.026} {\bibfield  {journal} {\bibinfo  {journal} {Phys. Lett. B}\ }\textbf {\bibinfo {volume} {609}},\ \bibinfo {pages} {117} (\bibinfo {year} {2005})},\ \Eprint {http://arxiv.org/abs/hep-ph/0405097} {arXiv:hep-ph/0405097} \BibitemShut {NoStop}%
\bibitem [{\citenamefont {Hooper}\ \emph {et~al.}(2019)\citenamefont {Hooper}, \citenamefont {Krnjaic}, \citenamefont {Long},\ and\ \citenamefont {Mcdermott}}]{Hooper:2018buz}%
  \BibitemOpen
  \bibfield  {author} {\bibinfo {author} {\bibfnamefont {D.}~\bibnamefont {Hooper}}, \bibinfo {author} {\bibfnamefont {G.}~\bibnamefont {Krnjaic}}, \bibinfo {author} {\bibfnamefont {A.~J.}\ \bibnamefont {Long}}, \ and\ \bibinfo {author} {\bibfnamefont {S.~D.}\ \bibnamefont {Mcdermott}},\ }\href {\doibase 10.1103/PhysRevLett.122.091802} {\bibfield  {journal} {\bibinfo  {journal} {Phys. Rev. Lett.}\ }\textbf {\bibinfo {volume} {122}},\ \bibinfo {pages} {091802} (\bibinfo {year} {2019})},\ \Eprint {http://arxiv.org/abs/1807.03308} {arXiv:1807.03308 [hep-ph]} \BibitemShut {NoStop}%
\bibitem [{\citenamefont {Ema}\ \emph {et~al.}(2017)\citenamefont {Ema}, \citenamefont {Hamaguchi}, \citenamefont {Moroi},\ and\ \citenamefont {Nakayama}}]{Ema:2016ops}%
  \BibitemOpen
  \bibfield  {author} {\bibinfo {author} {\bibfnamefont {Y.}~\bibnamefont {Ema}}, \bibinfo {author} {\bibfnamefont {K.}~\bibnamefont {Hamaguchi}}, \bibinfo {author} {\bibfnamefont {T.}~\bibnamefont {Moroi}}, \ and\ \bibinfo {author} {\bibfnamefont {K.}~\bibnamefont {Nakayama}},\ }\href {\doibase 10.1007/JHEP01(2017)096} {\bibfield  {journal} {\bibinfo  {journal} {JHEP}\ }\textbf {\bibinfo {volume} {01}},\ \bibinfo {pages} {096} (\bibinfo {year} {2017})},\ \Eprint {http://arxiv.org/abs/1612.05492} {arXiv:1612.05492 [hep-ph]} \BibitemShut {NoStop}%
\bibitem [{\citenamefont {Barman}\ \emph {et~al.}(2023{\natexlab{a}})\citenamefont {Barman}, \citenamefont {Ghoshal}, \citenamefont {Grzadkowski},\ and\ \citenamefont {Socha}}]{Barman:2023ktz}%
  \BibitemOpen
  \bibfield  {author} {\bibinfo {author} {\bibfnamefont {B.}~\bibnamefont {Barman}}, \bibinfo {author} {\bibfnamefont {A.}~\bibnamefont {Ghoshal}}, \bibinfo {author} {\bibfnamefont {B.}~\bibnamefont {Grzadkowski}}, \ and\ \bibinfo {author} {\bibfnamefont {A.}~\bibnamefont {Socha}},\ }\href {\doibase 10.1007/JHEP07(2023)231} {\bibfield  {journal} {\bibinfo  {journal} {JHEP}\ }\textbf {\bibinfo {volume} {07}},\ \bibinfo {pages} {231} (\bibinfo {year} {2023}{\natexlab{a}})},\ \Eprint {http://arxiv.org/abs/2305.00027} {arXiv:2305.00027 [hep-ph]} \BibitemShut {NoStop}%
\bibitem [{\citenamefont {Barman}\ \emph {et~al.}(2023{\natexlab{b}})\citenamefont {Barman}, \citenamefont {Bhupal~Dev},\ and\ \citenamefont {Ghoshal}}]{Barman:2022scg}%
  \BibitemOpen
  \bibfield  {author} {\bibinfo {author} {\bibfnamefont {B.}~\bibnamefont {Barman}}, \bibinfo {author} {\bibfnamefont {P.~S.}\ \bibnamefont {Bhupal~Dev}}, \ and\ \bibinfo {author} {\bibfnamefont {A.}~\bibnamefont {Ghoshal}},\ }\href {\doibase 10.1103/PhysRevD.108.035037} {\bibfield  {journal} {\bibinfo  {journal} {Phys. Rev. D}\ }\textbf {\bibinfo {volume} {108}},\ \bibinfo {pages} {035037} (\bibinfo {year} {2023}{\natexlab{b}})},\ \Eprint {http://arxiv.org/abs/2210.07739} {arXiv:2210.07739 [hep-ph]} \BibitemShut {NoStop}%
\bibitem [{\citenamefont {Barman}\ and\ \citenamefont {Ghoshal}(2022{\natexlab{a}})}]{Barman:2022njh}%
  \BibitemOpen
  \bibfield  {author} {\bibinfo {author} {\bibfnamefont {B.}~\bibnamefont {Barman}}\ and\ \bibinfo {author} {\bibfnamefont {A.}~\bibnamefont {Ghoshal}},\ }\href {\doibase 10.1088/1475-7516/2022/10/082} {\bibfield  {journal} {\bibinfo  {journal} {JCAP}\ }\textbf {\bibinfo {volume} {10}},\ \bibinfo {pages} {082} (\bibinfo {year} {2022}{\natexlab{a}})},\ \Eprint {http://arxiv.org/abs/2203.13269} {arXiv:2203.13269 [hep-ph]} \BibitemShut {NoStop}%
\bibitem [{\citenamefont {Barman}\ \emph {et~al.}(2022{\natexlab{a}})\citenamefont {Barman}, \citenamefont {Ghosh}, \citenamefont {Ghoshal},\ and\ \citenamefont {Mukherjee}}]{Barman:2021yaz}%
  \BibitemOpen
  \bibfield  {author} {\bibinfo {author} {\bibfnamefont {B.}~\bibnamefont {Barman}}, \bibinfo {author} {\bibfnamefont {P.}~\bibnamefont {Ghosh}}, \bibinfo {author} {\bibfnamefont {A.}~\bibnamefont {Ghoshal}}, \ and\ \bibinfo {author} {\bibfnamefont {L.}~\bibnamefont {Mukherjee}},\ }\href {\doibase 10.1088/1475-7516/2022/08/049} {\bibfield  {journal} {\bibinfo  {journal} {JCAP}\ }\textbf {\bibinfo {volume} {08}},\ \bibinfo {pages} {049} (\bibinfo {year} {2022}{\natexlab{a}})},\ \Eprint {http://arxiv.org/abs/2112.12798} {arXiv:2112.12798 [hep-ph]} \BibitemShut {NoStop}%
\bibitem [{\citenamefont {Barman}\ and\ \citenamefont {Ghoshal}(2022{\natexlab{b}})}]{Barman:2021lot}%
  \BibitemOpen
  \bibfield  {author} {\bibinfo {author} {\bibfnamefont {B.}~\bibnamefont {Barman}}\ and\ \bibinfo {author} {\bibfnamefont {A.}~\bibnamefont {Ghoshal}},\ }\href {\doibase 10.1088/1475-7516/2022/03/003} {\bibfield  {journal} {\bibinfo  {journal} {JCAP}\ }\textbf {\bibinfo {volume} {03}},\ \bibinfo {pages} {003} (\bibinfo {year} {2022}{\natexlab{b}})},\ \Eprint {http://arxiv.org/abs/2109.03259} {arXiv:2109.03259 [hep-ph]} \BibitemShut {NoStop}%
\bibitem [{\citenamefont {Ghosh}\ \emph {et~al.}(2023)\citenamefont {Ghosh}, \citenamefont {Ghoshal},\ and\ \citenamefont {Jeesun}}]{Ghosh:2023tyz}%
  \BibitemOpen
  \bibfield  {author} {\bibinfo {author} {\bibfnamefont {D.~K.}\ \bibnamefont {Ghosh}}, \bibinfo {author} {\bibfnamefont {A.}~\bibnamefont {Ghoshal}}, \ and\ \bibinfo {author} {\bibfnamefont {S.}~\bibnamefont {Jeesun}},\ }\href@noop {} {\  (\bibinfo {year} {2023})},\ \Eprint {http://arxiv.org/abs/2305.09188} {arXiv:2305.09188 [hep-ph]} \BibitemShut {NoStop}%
\bibitem [{\citenamefont {Chakraborty}\ \emph {et~al.}(2023)\citenamefont {Chakraborty}, \citenamefont {Haque}, \citenamefont {Maity},\ and\ \citenamefont {Mondal}}]{Chakraborty:2023ocr}%
  \BibitemOpen
  \bibfield  {author} {\bibinfo {author} {\bibfnamefont {A.}~\bibnamefont {Chakraborty}}, \bibinfo {author} {\bibfnamefont {M.~R.}\ \bibnamefont {Haque}}, \bibinfo {author} {\bibfnamefont {D.}~\bibnamefont {Maity}}, \ and\ \bibinfo {author} {\bibfnamefont {R.}~\bibnamefont {Mondal}},\ }\href {\doibase 10.1103/PhysRevD.108.023515} {\bibfield  {journal} {\bibinfo  {journal} {Phys. Rev. D}\ }\textbf {\bibinfo {volume} {108}},\ \bibinfo {pages} {023515} (\bibinfo {year} {2023})},\ \Eprint {http://arxiv.org/abs/2304.13637} {arXiv:2304.13637 [astro-ph.CO]} \BibitemShut {NoStop}%
\bibitem [{\citenamefont {Ghoshal}\ \emph {et~al.}(2022{\natexlab{a}})\citenamefont {Ghoshal}, \citenamefont {Heurtier},\ and\ \citenamefont {Paul}}]{Ghoshal:2022ruy}%
  \BibitemOpen
  \bibfield  {author} {\bibinfo {author} {\bibfnamefont {A.}~\bibnamefont {Ghoshal}}, \bibinfo {author} {\bibfnamefont {L.}~\bibnamefont {Heurtier}}, \ and\ \bibinfo {author} {\bibfnamefont {A.}~\bibnamefont {Paul}},\ }\href {\doibase 10.1007/JHEP12(2022)105} {\bibfield  {journal} {\bibinfo  {journal} {JHEP}\ }\textbf {\bibinfo {volume} {12}},\ \bibinfo {pages} {105} (\bibinfo {year} {2022}{\natexlab{a}})},\ \Eprint {http://arxiv.org/abs/2208.01670} {arXiv:2208.01670 [hep-ph]} \BibitemShut {NoStop}%
\bibitem [{\citenamefont {Berbig}\ and\ \citenamefont {Ghoshal}(2023)}]{Berbig:2023yyy}%
  \BibitemOpen
  \bibfield  {author} {\bibinfo {author} {\bibfnamefont {M.}~\bibnamefont {Berbig}}\ and\ \bibinfo {author} {\bibfnamefont {A.}~\bibnamefont {Ghoshal}},\ }\href {\doibase 10.1007/JHEP05(2023)172} {\bibfield  {journal} {\bibinfo  {journal} {JHEP}\ }\textbf {\bibinfo {volume} {05}},\ \bibinfo {pages} {172} (\bibinfo {year} {2023})},\ \Eprint {http://arxiv.org/abs/2301.05672} {arXiv:2301.05672 [hep-ph]} \BibitemShut {NoStop}%
\bibitem [{\citenamefont {Ghoshal}\ and\ \citenamefont {Saha}(2022)}]{Ghoshal:2022jdt}%
  \BibitemOpen
  \bibfield  {author} {\bibinfo {author} {\bibfnamefont {A.}~\bibnamefont {Ghoshal}}\ and\ \bibinfo {author} {\bibfnamefont {P.}~\bibnamefont {Saha}},\ }\href@noop {} {\  (\bibinfo {year} {2022})},\ \Eprint {http://arxiv.org/abs/2203.14424} {arXiv:2203.14424 [hep-ph]} \BibitemShut {NoStop}%
\bibitem [{\citenamefont {Ghoshal}\ \emph {et~al.}(2023{\natexlab{b}})\citenamefont {Ghoshal}, \citenamefont {Khlopov}, \citenamefont {Lalak},\ and\ \citenamefont {Porey}}]{Ghoshal:2023jvf}%
  \BibitemOpen
  \bibfield  {author} {\bibinfo {author} {\bibfnamefont {A.}~\bibnamefont {Ghoshal}}, \bibinfo {author} {\bibfnamefont {M.~Y.}\ \bibnamefont {Khlopov}}, \bibinfo {author} {\bibfnamefont {Z.}~\bibnamefont {Lalak}}, \ and\ \bibinfo {author} {\bibfnamefont {S.}~\bibnamefont {Porey}},\ }\href@noop {} {\  (\bibinfo {year} {2023}{\natexlab{b}})},\ \Eprint {http://arxiv.org/abs/2306.08675} {arXiv:2306.08675 [hep-ph]} \BibitemShut {NoStop}%
\bibitem [{\citenamefont {Bernal}\ and\ \citenamefont {Xu}(2021)}]{Bernal:2021qrl}%
  \BibitemOpen
  \bibfield  {author} {\bibinfo {author} {\bibfnamefont {N.}~\bibnamefont {Bernal}}\ and\ \bibinfo {author} {\bibfnamefont {Y.}~\bibnamefont {Xu}},\ }\href {\doibase 10.1140/epjc/s10052-021-09694-5} {\bibfield  {journal} {\bibinfo  {journal} {Eur. Phys. J. C}\ }\textbf {\bibinfo {volume} {81}},\ \bibinfo {pages} {877} (\bibinfo {year} {2021})},\ \Eprint {http://arxiv.org/abs/2106.03950} {arXiv:2106.03950 [hep-ph]} \BibitemShut {NoStop}%
\bibitem [{\citenamefont {Ghoshal}\ \emph {et~al.}(2022{\natexlab{b}})\citenamefont {Ghoshal}, \citenamefont {Lambiase}, \citenamefont {Pal}, \citenamefont {Paul},\ and\ \citenamefont {Porey}}]{Ghoshal:2022jeo}%
  \BibitemOpen
  \bibfield  {author} {\bibinfo {author} {\bibfnamefont {A.}~\bibnamefont {Ghoshal}}, \bibinfo {author} {\bibfnamefont {G.}~\bibnamefont {Lambiase}}, \bibinfo {author} {\bibfnamefont {S.}~\bibnamefont {Pal}}, \bibinfo {author} {\bibfnamefont {A.}~\bibnamefont {Paul}}, \ and\ \bibinfo {author} {\bibfnamefont {S.}~\bibnamefont {Porey}},\ }\href {\doibase 10.1007/JHEP09(2022)231} {\bibfield  {journal} {\bibinfo  {journal} {JHEP}\ }\textbf {\bibinfo {volume} {09}},\ \bibinfo {pages} {231} (\bibinfo {year} {2022}{\natexlab{b}})},\ \Eprint {http://arxiv.org/abs/2206.10648} {arXiv:2206.10648 [hep-ph]} \BibitemShut {NoStop}%
\bibitem [{\citenamefont {Ghoshal}\ \emph {et~al.}(2023{\natexlab{c}})\citenamefont {Ghoshal}, \citenamefont {Lambiase}, \citenamefont {Pal}, \citenamefont {Paul},\ and\ \citenamefont {Porey}}]{Ghoshal:2023noe}%
  \BibitemOpen
  \bibfield  {author} {\bibinfo {author} {\bibfnamefont {A.}~\bibnamefont {Ghoshal}}, \bibinfo {author} {\bibfnamefont {G.}~\bibnamefont {Lambiase}}, \bibinfo {author} {\bibfnamefont {S.}~\bibnamefont {Pal}}, \bibinfo {author} {\bibfnamefont {A.}~\bibnamefont {Paul}}, \ and\ \bibinfo {author} {\bibfnamefont {S.}~\bibnamefont {Porey}},\ }\href {\doibase 10.3390/sym15020543} {\bibfield  {journal} {\bibinfo  {journal} {Symmetry}\ }\textbf {\bibinfo {volume} {15}},\ \bibinfo {pages} {543} (\bibinfo {year} {2023}{\natexlab{c}})}\BibitemShut {NoStop}%
\bibitem [{\citenamefont {Ghoshal}\ \emph {et~al.}(2022{\natexlab{c}})\citenamefont {Ghoshal}, \citenamefont {Lambiase}, \citenamefont {Pal}, \citenamefont {Paul},\ and\ \citenamefont {Porey}}]{Ghoshal:2022aoh}%
  \BibitemOpen
  \bibfield  {author} {\bibinfo {author} {\bibfnamefont {A.}~\bibnamefont {Ghoshal}}, \bibinfo {author} {\bibfnamefont {G.}~\bibnamefont {Lambiase}}, \bibinfo {author} {\bibfnamefont {S.}~\bibnamefont {Pal}}, \bibinfo {author} {\bibfnamefont {A.}~\bibnamefont {Paul}}, \ and\ \bibinfo {author} {\bibfnamefont {S.}~\bibnamefont {Porey}},\ }in\ \href@noop {} {\emph {\bibinfo {booktitle} {{25th Workshop on What Comes Beyond the Standard Models?}}}}\ (\bibinfo {year} {2022})\ \Eprint {http://arxiv.org/abs/2211.15061} {arXiv:2211.15061 [astro-ph.CO]} \BibitemShut {NoStop}%
\bibitem [{\citenamefont {Ghoshal}\ \emph {et~al.}(2024)\citenamefont {Ghoshal}, \citenamefont {Lalak}, \citenamefont {Pal},\ and\ \citenamefont {Porey}}]{Ghoshal:2024ycp}%
  \BibitemOpen
  \bibfield  {author} {\bibinfo {author} {\bibfnamefont {A.}~\bibnamefont {Ghoshal}}, \bibinfo {author} {\bibfnamefont {Z.}~\bibnamefont {Lalak}}, \bibinfo {author} {\bibfnamefont {S.}~\bibnamefont {Pal}}, \ and\ \bibinfo {author} {\bibfnamefont {S.}~\bibnamefont {Porey}},\ }\href@noop {} {\  (\bibinfo {year} {2024})},\ \Eprint {http://arxiv.org/abs/2401.17262} {arXiv:2401.17262 [astro-ph.CO]} \BibitemShut {NoStop}%
\bibitem [{\citenamefont {Akrami}\ \emph {et~al.}(2020)\citenamefont {Akrami} \emph {et~al.}}]{Planck:2018jri}%
  \BibitemOpen
  \bibfield  {author} {\bibinfo {author} {\bibfnamefont {Y.}~\bibnamefont {Akrami}} \emph {et~al.} (\bibinfo {collaboration} {Planck}),\ }\href {\doibase 10.1051/0004-6361/201833887} {\bibfield  {journal} {\bibinfo  {journal} {Astron. Astrophys.}\ }\textbf {\bibinfo {volume} {641}},\ \bibinfo {pages} {A10} (\bibinfo {year} {2020})},\ \Eprint {http://arxiv.org/abs/1807.06211} {arXiv:1807.06211 [astro-ph.CO]} \BibitemShut {NoStop}%
\bibitem [{\citenamefont {Kallosh}\ and\ \citenamefont {Linde}(2019{\natexlab{a}})}]{Kallosh:2019jnl}%
  \BibitemOpen
  \bibfield  {author} {\bibinfo {author} {\bibfnamefont {R.}~\bibnamefont {Kallosh}}\ and\ \bibinfo {author} {\bibfnamefont {A.}~\bibnamefont {Linde}},\ }\href {\doibase 10.1088/1475-7516/2019/09/030} {\bibfield  {journal} {\bibinfo  {journal} {JCAP}\ }\textbf {\bibinfo {volume} {09}},\ \bibinfo {pages} {030} (\bibinfo {year} {2019}{\natexlab{a}})},\ \Eprint {http://arxiv.org/abs/1906.02156} {arXiv:1906.02156 [hep-th]} \BibitemShut {NoStop}%
\bibitem [{\citenamefont {Kallosh}\ and\ \citenamefont {Linde}(2019{\natexlab{b}})}]{Kallosh:2019hzo}%
  \BibitemOpen
  \bibfield  {author} {\bibinfo {author} {\bibfnamefont {R.}~\bibnamefont {Kallosh}}\ and\ \bibinfo {author} {\bibfnamefont {A.}~\bibnamefont {Linde}},\ }\href {\doibase 10.1103/PhysRevD.100.123523} {\bibfield  {journal} {\bibinfo  {journal} {Phys. Rev. D}\ }\textbf {\bibinfo {volume} {100}},\ \bibinfo {pages} {123523} (\bibinfo {year} {2019}{\natexlab{b}})},\ \Eprint {http://arxiv.org/abs/1909.04687} {arXiv:1909.04687 [hep-th]} \BibitemShut {NoStop}%
\bibitem [{\citenamefont {Kallosh}\ and\ \citenamefont {Linde}(2021)}]{Kallosh:2021mnu}%
  \BibitemOpen
  \bibfield  {author} {\bibinfo {author} {\bibfnamefont {R.}~\bibnamefont {Kallosh}}\ and\ \bibinfo {author} {\bibfnamefont {A.}~\bibnamefont {Linde}},\ }\href {\doibase 10.1088/1475-7516/2021/12/008} {\bibfield  {journal} {\bibinfo  {journal} {JCAP}\ }\textbf {\bibinfo {volume} {12}},\ \bibinfo {pages} {008} (\bibinfo {year} {2021})},\ \Eprint {http://arxiv.org/abs/2110.10902} {arXiv:2110.10902 [astro-ph.CO]} \BibitemShut {NoStop}%
\bibitem [{\citenamefont {Lillepalu}\ and\ \citenamefont {Racioppi}(2023)}]{Lillepalu:2022knx}%
  \BibitemOpen
  \bibfield  {author} {\bibinfo {author} {\bibfnamefont {H.~G.}\ \bibnamefont {Lillepalu}}\ and\ \bibinfo {author} {\bibfnamefont {A.}~\bibnamefont {Racioppi}},\ }\href {\doibase 10.1140/epjp/s13360-023-04512-1} {\bibfield  {journal} {\bibinfo  {journal} {Eur. Phys. J. Plus}\ }\textbf {\bibinfo {volume} {138}},\ \bibinfo {pages} {894} (\bibinfo {year} {2023})},\ \Eprint {http://arxiv.org/abs/2211.02426} {arXiv:2211.02426 [astro-ph.CO]} \BibitemShut {NoStop}%
\bibitem [{\citenamefont {Hoffmann}\ and\ \citenamefont {Sloan}(2021)}]{Hoffmann:2021vty}%
  \BibitemOpen
  \bibfield  {author} {\bibinfo {author} {\bibfnamefont {J.}~\bibnamefont {Hoffmann}}\ and\ \bibinfo {author} {\bibfnamefont {D.}~\bibnamefont {Sloan}},\ }\href {\doibase 10.1103/PhysRevD.104.123542} {\bibfield  {journal} {\bibinfo  {journal} {Phys. Rev. D}\ }\textbf {\bibinfo {volume} {104}},\ \bibinfo {pages} {123542} (\bibinfo {year} {2021})},\ \Eprint {http://arxiv.org/abs/2108.13316} {arXiv:2108.13316 [gr-qc]} \BibitemShut {NoStop}%
\bibitem [{\citenamefont {Boubekeur}\ and\ \citenamefont {Lyth}(2005)}]{Boubekeur:2005zm}%
  \BibitemOpen
  \bibfield  {author} {\bibinfo {author} {\bibfnamefont {L.}~\bibnamefont {Boubekeur}}\ and\ \bibinfo {author} {\bibfnamefont {D.~H.}\ \bibnamefont {Lyth}},\ }\href {\doibase 10.1088/1475-7516/2005/07/010} {\bibfield  {journal} {\bibinfo  {journal} {JCAP}\ }\textbf {\bibinfo {volume} {07}},\ \bibinfo {pages} {010} (\bibinfo {year} {2005})},\ \Eprint {http://arxiv.org/abs/hep-ph/0502047} {arXiv:hep-ph/0502047} \BibitemShut {NoStop}%
\bibitem [{\citenamefont {German}(2021)}]{German:2020rpn}%
  \BibitemOpen
  \bibfield  {author} {\bibinfo {author} {\bibfnamefont {G.}~\bibnamefont {German}},\ }\href {\doibase 10.1088/1475-7516/2021/02/034} {\bibfield  {journal} {\bibinfo  {journal} {JCAP}\ }\textbf {\bibinfo {volume} {02}},\ \bibinfo {pages} {034} (\bibinfo {year} {2021})},\ \Eprint {http://arxiv.org/abs/2011.12804} {arXiv:2011.12804 [astro-ph.CO]} \BibitemShut {NoStop}%
\bibitem [{\citenamefont {Dimopoulos}(2020)}]{Dimopoulos:2020kol}%
  \BibitemOpen
  \bibfield  {author} {\bibinfo {author} {\bibfnamefont {K.}~\bibnamefont {Dimopoulos}},\ }\href {\doibase 10.1016/j.physletb.2020.135688} {\bibfield  {journal} {\bibinfo  {journal} {Phys. Lett. B}\ }\textbf {\bibinfo {volume} {809}},\ \bibinfo {pages} {135688} (\bibinfo {year} {2020})},\ \Eprint {http://arxiv.org/abs/2006.06029} {arXiv:2006.06029 [hep-ph]} \BibitemShut {NoStop}%
\bibitem [{\citenamefont {Baumann}(2022)}]{Baumann:2022mni}%
  \BibitemOpen
  \bibfield  {author} {\bibinfo {author} {\bibfnamefont {D.}~\bibnamefont {Baumann}},\ }\href {\doibase 10.1017/9781108937092} {\emph {\bibinfo {title} {{Cosmology}}}}\ (\bibinfo  {publisher} {Cambridge University Press},\ \bibinfo {year} {2022})\BibitemShut {NoStop}%
\bibitem [{\citenamefont {Aghanim}\ \emph {et~al.}(2020)\citenamefont {Aghanim} \emph {et~al.}}]{Aghanim:2018eyx}%
  \BibitemOpen
  \bibfield  {author} {\bibinfo {author} {\bibfnamefont {N.}~\bibnamefont {Aghanim}} \emph {et~al.} (\bibinfo {collaboration} {Planck}),\ }\href {\doibase 10.1051/0004-6361/201833910} {\bibfield  {journal} {\bibinfo  {journal} {Astron. Astrophys.}\ }\textbf {\bibinfo {volume} {641}},\ \bibinfo {pages} {A6} (\bibinfo {year} {2020})},\ \bibinfo {note} {[Erratum: Astron.Astrophys. 652, C4 (2021)]},\ \Eprint {http://arxiv.org/abs/1807.06209} {arXiv:1807.06209 [astro-ph.CO]} \BibitemShut {NoStop}%
\bibitem [{\citenamefont {Workman}\ \emph {et~al.}(2022)\citenamefont {Workman} \emph {et~al.}}]{ParticleDataGroup:2022pth}%
  \BibitemOpen
  \bibfield  {author} {\bibinfo {author} {\bibfnamefont {R.~L.}\ \bibnamefont {Workman}} \emph {et~al.} (\bibinfo {collaboration} {Particle Data Group}),\ }\href {\doibase 10.1093/ptep/ptac097} {\bibfield  {journal} {\bibinfo  {journal} {PTEP}\ }\textbf {\bibinfo {volume} {2022}},\ \bibinfo {pages} {083C01} (\bibinfo {year} {2022})}\BibitemShut {NoStop}%
\bibitem [{\citenamefont {Ade}\ \emph {et~al.}(2022{\natexlab{a}})\citenamefont {Ade} \emph {et~al.}}]{BICEPKeck:2022mhb}%
  \BibitemOpen
  \bibfield  {author} {\bibinfo {author} {\bibfnamefont {P.~A.~R.}\ \bibnamefont {Ade}} \emph {et~al.} (\bibinfo {collaboration} {BICEP/Keck}),\ }in\ \href@noop {} {\emph {\bibinfo {booktitle} {{56th Rencontres de Moriond on Cosmology}}}}\ (\bibinfo {year} {2022})\ \Eprint {http://arxiv.org/abs/2203.16556} {arXiv:2203.16556 [astro-ph.CO]} \BibitemShut {NoStop}%
\bibitem [{\citenamefont {Ade}\ \emph {et~al.}(2021)\citenamefont {Ade} \emph {et~al.}}]{BICEP:2021xfz}%
  \BibitemOpen
  \bibfield  {author} {\bibinfo {author} {\bibfnamefont {P.~A.~R.}\ \bibnamefont {Ade}} \emph {et~al.} (\bibinfo {collaboration} {BICEP, Keck}),\ }\href {\doibase 10.1103/PhysRevLett.127.151301} {\bibfield  {journal} {\bibinfo  {journal} {Phys. Rev. Lett.}\ }\textbf {\bibinfo {volume} {127}},\ \bibinfo {pages} {151301} (\bibinfo {year} {2021})},\ \Eprint {http://arxiv.org/abs/2110.00483} {arXiv:2110.00483 [astro-ph.CO]} \BibitemShut {NoStop}%
\bibitem [{\citenamefont {Campeti}\ and\ \citenamefont {Komatsu}(2022)}]{Campeti:2022vom}%
  \BibitemOpen
  \bibfield  {author} {\bibinfo {author} {\bibfnamefont {P.}~\bibnamefont {Campeti}}\ and\ \bibinfo {author} {\bibfnamefont {E.}~\bibnamefont {Komatsu}},\ }\href {\doibase 10.3847/1538-4357/ac9ea3} {\bibfield  {journal} {\bibinfo  {journal} {Astrophys. J.}\ }\textbf {\bibinfo {volume} {941}},\ \bibinfo {pages} {110} (\bibinfo {year} {2022})},\ \Eprint {http://arxiv.org/abs/2205.05617} {arXiv:2205.05617 [astro-ph.CO]} \BibitemShut {NoStop}%
\bibitem [{\citenamefont {Hazumi}\ \emph {et~al.}(2020)\citenamefont {Hazumi} \emph {et~al.}}]{LiteBIRD:2020khw}%
  \BibitemOpen
  \bibfield  {author} {\bibinfo {author} {\bibfnamefont {M.}~\bibnamefont {Hazumi}} \emph {et~al.} (\bibinfo {collaboration} {LiteBIRD}),\ }\href {\doibase 10.1117/12.2563050} {\bibfield  {journal} {\bibinfo  {journal} {Proc. SPIE Int. Soc. Opt. Eng.}\ }\textbf {\bibinfo {volume} {11443}},\ \bibinfo {pages} {114432F} (\bibinfo {year} {2020})},\ \Eprint {http://arxiv.org/abs/2101.12449} {arXiv:2101.12449 [astro-ph.IM]} \BibitemShut {NoStop}%
\bibitem [{\citenamefont {Abazajian}\ \emph {et~al.}(2016)\citenamefont {Abazajian} \emph {et~al.}}]{CMB-S4:2016ple}%
  \BibitemOpen
  \bibfield  {author} {\bibinfo {author} {\bibfnamefont {K.~N.}\ \bibnamefont {Abazajian}} \emph {et~al.} (\bibinfo {collaboration} {CMB-S4}),\ }\href@noop {} {\  (\bibinfo {year} {2016})},\ \Eprint {http://arxiv.org/abs/1610.02743} {arXiv:1610.02743 [astro-ph.CO]} \BibitemShut {NoStop}%
\bibitem [{\citenamefont {Ade}\ \emph {et~al.}(2019)\citenamefont {Ade} \emph {et~al.}}]{SimonsObservatory:2018koc}%
  \BibitemOpen
  \bibfield  {author} {\bibinfo {author} {\bibfnamefont {P.}~\bibnamefont {Ade}} \emph {et~al.} (\bibinfo {collaboration} {Simons Observatory}),\ }\href {\doibase 10.1088/1475-7516/2019/02/056} {\bibfield  {journal} {\bibinfo  {journal} {JCAP}\ }\textbf {\bibinfo {volume} {02}},\ \bibinfo {pages} {056} (\bibinfo {year} {2019})},\ \Eprint {http://arxiv.org/abs/1808.07445} {arXiv:1808.07445 [astro-ph.CO]} \BibitemShut {NoStop}%
\bibitem [{\citenamefont {Drees}\ and\ \citenamefont {Xu}(2021)}]{Drees:2021wgd}%
  \BibitemOpen
  \bibfield  {author} {\bibinfo {author} {\bibfnamefont {M.}~\bibnamefont {Drees}}\ and\ \bibinfo {author} {\bibfnamefont {Y.}~\bibnamefont {Xu}},\ }\href {\doibase 10.1088/1475-7516/2021/09/012} {\bibfield  {journal} {\bibinfo  {journal} {JCAP}\ }\textbf {\bibinfo {volume} {09}},\ \bibinfo {pages} {012} (\bibinfo {year} {2021})},\ \Eprint {http://arxiv.org/abs/2104.03977} {arXiv:2104.03977 [hep-ph]} \BibitemShut {NoStop}%
\bibitem [{\citenamefont {Coleman}\ and\ \citenamefont {Weinberg}(1973)}]{Coleman:1973jx}%
  \BibitemOpen
  \bibfield  {author} {\bibinfo {author} {\bibfnamefont {S.~R.}\ \bibnamefont {Coleman}}\ and\ \bibinfo {author} {\bibfnamefont {E.~J.}\ \bibnamefont {Weinberg}},\ }\href {\doibase 10.1103/PhysRevD.7.1888} {\bibfield  {journal} {\bibinfo  {journal} {Phys. Rev. D}\ }\textbf {\bibinfo {volume} {7}},\ \bibinfo {pages} {1888} (\bibinfo {year} {1973})}\BibitemShut {NoStop}%
\bibitem [{\citenamefont {Enqvist}(2012)}]{Enqvist:2012qc}%
  \BibitemOpen
  \bibfield  {author} {\bibinfo {author} {\bibfnamefont {K.}~\bibnamefont {Enqvist}},\ }in\ \href@noop {} {\emph {\bibinfo {booktitle} {{2010 European School of High Energy Physics}}}}\ (\bibinfo {year} {2012})\ \Eprint {http://arxiv.org/abs/1201.6164} {arXiv:1201.6164 [gr-qc]} \BibitemShut {NoStop}%
\bibitem [{\citenamefont {Garcia}\ \emph {et~al.}(2020)\citenamefont {Garcia}, \citenamefont {Kaneta}, \citenamefont {Mambrini},\ and\ \citenamefont {Olive}}]{Garcia:2020eof}%
  \BibitemOpen
  \bibfield  {author} {\bibinfo {author} {\bibfnamefont {M.~A.~G.}\ \bibnamefont {Garcia}}, \bibinfo {author} {\bibfnamefont {K.}~\bibnamefont {Kaneta}}, \bibinfo {author} {\bibfnamefont {Y.}~\bibnamefont {Mambrini}}, \ and\ \bibinfo {author} {\bibfnamefont {K.~A.}\ \bibnamefont {Olive}},\ }\href {\doibase 10.1103/PhysRevD.101.123507} {\bibfield  {journal} {\bibinfo  {journal} {Phys. Rev. D}\ }\textbf {\bibinfo {volume} {101}},\ \bibinfo {pages} {123507} (\bibinfo {year} {2020})},\ \Eprint {http://arxiv.org/abs/2004.08404} {arXiv:2004.08404 [hep-ph]} \BibitemShut {NoStop}%
\bibitem [{\citenamefont {Giudice}\ \emph {et~al.}(2001)\citenamefont {Giudice}, \citenamefont {Kolb},\ and\ \citenamefont {Riotto}}]{Giudice:2000ex}%
  \BibitemOpen
  \bibfield  {author} {\bibinfo {author} {\bibfnamefont {G.~F.}\ \bibnamefont {Giudice}}, \bibinfo {author} {\bibfnamefont {E.~W.}\ \bibnamefont {Kolb}}, \ and\ \bibinfo {author} {\bibfnamefont {A.}~\bibnamefont {Riotto}},\ }\href {\doibase 10.1103/PhysRevD.64.023508} {\bibfield  {journal} {\bibinfo  {journal} {Phys. Rev. D}\ }\textbf {\bibinfo {volume} {64}},\ \bibinfo {pages} {023508} (\bibinfo {year} {2001})},\ \Eprint {http://arxiv.org/abs/hep-ph/0005123} {arXiv:hep-ph/0005123} \BibitemShut {NoStop}%
\bibitem [{\citenamefont {Chung}\ \emph {et~al.}(1999)\citenamefont {Chung}, \citenamefont {Kolb},\ and\ \citenamefont {Riotto}}]{Chung:1998rq}%
  \BibitemOpen
  \bibfield  {author} {\bibinfo {author} {\bibfnamefont {D.~J.~H.}\ \bibnamefont {Chung}}, \bibinfo {author} {\bibfnamefont {E.~W.}\ \bibnamefont {Kolb}}, \ and\ \bibinfo {author} {\bibfnamefont {A.}~\bibnamefont {Riotto}},\ }\href {\doibase 10.1103/PhysRevD.60.063504} {\bibfield  {journal} {\bibinfo  {journal} {Phys. Rev. D}\ }\textbf {\bibinfo {volume} {60}},\ \bibinfo {pages} {063504} (\bibinfo {year} {1999})},\ \Eprint {http://arxiv.org/abs/hep-ph/9809453} {arXiv:hep-ph/9809453} \BibitemShut {NoStop}%
\bibitem [{\citenamefont {Mambrini}\ and\ \citenamefont {Olive}(2021)}]{Mambrini:2021zpp}%
  \BibitemOpen
  \bibfield  {author} {\bibinfo {author} {\bibfnamefont {Y.}~\bibnamefont {Mambrini}}\ and\ \bibinfo {author} {\bibfnamefont {K.~A.}\ \bibnamefont {Olive}},\ }\href {\doibase 10.1103/PhysRevD.103.115009} {\bibfield  {journal} {\bibinfo  {journal} {Phys. Rev. D}\ }\textbf {\bibinfo {volume} {103}},\ \bibinfo {pages} {115009} (\bibinfo {year} {2021})},\ \Eprint {http://arxiv.org/abs/2102.06214} {arXiv:2102.06214 [hep-ph]} \BibitemShut {NoStop}%
\bibitem [{\citenamefont {Barman}\ \emph {et~al.}(2022{\natexlab{b}})\citenamefont {Barman}, \citenamefont {Bernal}, \citenamefont {Xu},\ and\ \citenamefont {Zapata}}]{Barman:2022tzk}%
  \BibitemOpen
  \bibfield  {author} {\bibinfo {author} {\bibfnamefont {B.}~\bibnamefont {Barman}}, \bibinfo {author} {\bibfnamefont {N.}~\bibnamefont {Bernal}}, \bibinfo {author} {\bibfnamefont {Y.}~\bibnamefont {Xu}}, \ and\ \bibinfo {author} {\bibfnamefont {O.}~\bibnamefont {Zapata}},\ }\href {\doibase 10.1088/1475-7516/2022/07/019} {\bibfield  {journal} {\bibinfo  {journal} {JCAP}\ }\textbf {\bibinfo {volume} {07}},\ \bibinfo {pages} {019} (\bibinfo {year} {2022}{\natexlab{b}})},\ \Eprint {http://arxiv.org/abs/2202.12906} {arXiv:2202.12906 [hep-ph]} \BibitemShut {NoStop}%
\bibitem [{\citenamefont {Lozanov}(2019)}]{Lozanov:2019jxc}%
  \BibitemOpen
  \bibfield  {author} {\bibinfo {author} {\bibfnamefont {K.~D.}\ \bibnamefont {Lozanov}},\ }\href@noop {} {\  (\bibinfo {year} {2019})},\ \Eprint {http://arxiv.org/abs/1907.04402} {arXiv:1907.04402 [astro-ph.CO]} \BibitemShut {NoStop}%
\bibitem [{\citenamefont {Kolb}\ \emph {et~al.}(2003)\citenamefont {Kolb}, \citenamefont {Notari},\ and\ \citenamefont {Riotto}}]{Kolb:2003ke}%
  \BibitemOpen
  \bibfield  {author} {\bibinfo {author} {\bibfnamefont {E.~W.}\ \bibnamefont {Kolb}}, \bibinfo {author} {\bibfnamefont {A.}~\bibnamefont {Notari}}, \ and\ \bibinfo {author} {\bibfnamefont {A.}~\bibnamefont {Riotto}},\ }\href {\doibase 10.1103/PhysRevD.68.123505} {\bibfield  {journal} {\bibinfo  {journal} {Phys. Rev. D}\ }\textbf {\bibinfo {volume} {68}},\ \bibinfo {pages} {123505} (\bibinfo {year} {2003})},\ \Eprint {http://arxiv.org/abs/hep-ph/0307241} {arXiv:hep-ph/0307241} \BibitemShut {NoStop}%
\bibitem [{\citenamefont {Lisanti}(2017)}]{Lisanti:2016jxe}%
  \BibitemOpen
  \bibfield  {author} {\bibinfo {author} {\bibfnamefont {M.}~\bibnamefont {Lisanti}},\ }in\ \href {\doibase 10.1142/9789813149441_0007} {\emph {\bibinfo {booktitle} {{Theoretical Advanced Study Institute in Elementary Particle Physics}: {New Frontiers in Fields and Strings}}}}\ (\bibinfo {year} {2017})\ pp.\ \bibinfo {pages} {399--446},\ \Eprint {http://arxiv.org/abs/1603.03797} {arXiv:1603.03797 [hep-ph]} \BibitemShut {NoStop}%
\bibitem [{\citenamefont {Garrett}\ and\ \citenamefont {Duda}(2011)}]{Garrett:2010hd}%
  \BibitemOpen
  \bibfield  {author} {\bibinfo {author} {\bibfnamefont {K.}~\bibnamefont {Garrett}}\ and\ \bibinfo {author} {\bibfnamefont {G.}~\bibnamefont {Duda}},\ }\href {\doibase 10.1155/2011/968283} {\bibfield  {journal} {\bibinfo  {journal} {Adv. Astron.}\ }\textbf {\bibinfo {volume} {2011}},\ \bibinfo {pages} {968283} (\bibinfo {year} {2011})},\ \Eprint {http://arxiv.org/abs/1006.2483} {arXiv:1006.2483 [hep-ph]} \BibitemShut {NoStop}%
\bibitem [{\citenamefont {Racioppi}(2018)}]{Racioppi:2018zoy}%
  \BibitemOpen
  \bibfield  {author} {\bibinfo {author} {\bibfnamefont {A.}~\bibnamefont {Racioppi}},\ }\href {\doibase 10.1103/PhysRevD.97.123514} {\bibfield  {journal} {\bibinfo  {journal} {Phys. Rev. D}\ }\textbf {\bibinfo {volume} {97}},\ \bibinfo {pages} {123514} (\bibinfo {year} {2018})},\ \Eprint {http://arxiv.org/abs/1801.08810} {arXiv:1801.08810 [astro-ph.CO]} \BibitemShut {NoStop}%
\bibitem [{\citenamefont {Barenboim}\ \emph {et~al.}(2014)\citenamefont {Barenboim}, \citenamefont {Chun},\ and\ \citenamefont {Lee}}]{Barenboim:2013wra}%
  \BibitemOpen
  \bibfield  {author} {\bibinfo {author} {\bibfnamefont {G.}~\bibnamefont {Barenboim}}, \bibinfo {author} {\bibfnamefont {E.~J.}\ \bibnamefont {Chun}}, \ and\ \bibinfo {author} {\bibfnamefont {H.~M.}\ \bibnamefont {Lee}},\ }\href {\doibase 10.1016/j.physletb.2014.01.039} {\bibfield  {journal} {\bibinfo  {journal} {Phys. Lett. B}\ }\textbf {\bibinfo {volume} {730}},\ \bibinfo {pages} {81} (\bibinfo {year} {2014})},\ \Eprint {http://arxiv.org/abs/1309.1695} {arXiv:1309.1695 [hep-ph]} \BibitemShut {NoStop}%
\bibitem [{\citenamefont {Abazajian}\ \emph {et~al.}(2022)\citenamefont {Abazajian} \emph {et~al.}}]{CMB-S4:2020lpa}%
  \BibitemOpen
  \bibfield  {author} {\bibinfo {author} {\bibfnamefont {K.}~\bibnamefont {Abazajian}} \emph {et~al.} (\bibinfo {collaboration} {CMB-S4}),\ }\href {\doibase 10.3847/1538-4357/ac1596} {\bibfield  {journal} {\bibinfo  {journal} {Astrophys. J.}\ }\textbf {\bibinfo {volume} {926}},\ \bibinfo {pages} {54} (\bibinfo {year} {2022})},\ \Eprint {http://arxiv.org/abs/2008.12619} {arXiv:2008.12619 [astro-ph.CO]} \BibitemShut {NoStop}%
\bibitem [{\citenamefont {Hazumi}\ \emph {et~al.}(2019)\citenamefont {Hazumi} \emph {et~al.}}]{Hazumi:2019lys}%
  \BibitemOpen
  \bibfield  {author} {\bibinfo {author} {\bibfnamefont {M.}~\bibnamefont {Hazumi}} \emph {et~al.},\ }\href {\doibase 10.1007/s10909-019-02150-5} {\bibfield  {journal} {\bibinfo  {journal} {J. Low Temp. Phys.}\ }\textbf {\bibinfo {volume} {194}},\ \bibinfo {pages} {443} (\bibinfo {year} {2019})}\BibitemShut {NoStop}%
\bibitem [{\citenamefont {Adak}\ \emph {et~al.}(2022)\citenamefont {Adak}, \citenamefont {Sen}, \citenamefont {Basak}, \citenamefont {Delabrouille}, \citenamefont {Ghosh}, \citenamefont {Rotti}, \citenamefont {Mart\'\i{}nez-Solaeche},\ and\ \citenamefont {Souradeep}}]{Adak:2021lbu}%
  \BibitemOpen
  \bibfield  {author} {\bibinfo {author} {\bibfnamefont {D.}~\bibnamefont {Adak}}, \bibinfo {author} {\bibfnamefont {A.}~\bibnamefont {Sen}}, \bibinfo {author} {\bibfnamefont {S.}~\bibnamefont {Basak}}, \bibinfo {author} {\bibfnamefont {J.}~\bibnamefont {Delabrouille}}, \bibinfo {author} {\bibfnamefont {T.}~\bibnamefont {Ghosh}}, \bibinfo {author} {\bibfnamefont {A.}~\bibnamefont {Rotti}}, \bibinfo {author} {\bibfnamefont {G.}~\bibnamefont {Mart\'\i{}nez-Solaeche}}, \ and\ \bibinfo {author} {\bibfnamefont {T.}~\bibnamefont {Souradeep}},\ }\href {\doibase 10.1093/mnras/stac1474} {\bibfield  {journal} {\bibinfo  {journal} {Mon. Not. Roy. Astron. Soc.}\ }\textbf {\bibinfo {volume} {514}},\ \bibinfo {pages} {3002} (\bibinfo {year} {2022})},\ \Eprint {http://arxiv.org/abs/2110.12362} {arXiv:2110.12362 [astro-ph.CO]} \BibitemShut {NoStop}%
\bibitem [{\citenamefont {Sayre}\ \emph {et~al.}(2020)\citenamefont {Sayre} \emph {et~al.}}]{SPT:2019nip}%
  \BibitemOpen
  \bibfield  {author} {\bibinfo {author} {\bibfnamefont {J.~T.}\ \bibnamefont {Sayre}} \emph {et~al.} (\bibinfo {collaboration} {SPT}),\ }\href {\doibase 10.1103/PhysRevD.101.122003} {\bibfield  {journal} {\bibinfo  {journal} {Phys. Rev. D}\ }\textbf {\bibinfo {volume} {101}},\ \bibinfo {pages} {122003} (\bibinfo {year} {2020})},\ \Eprint {http://arxiv.org/abs/1910.05748} {arXiv:1910.05748 [astro-ph.CO]} \BibitemShut {NoStop}%
\bibitem [{\citenamefont {Suzuki}\ \emph {et~al.}(2016)\citenamefont {Suzuki} \emph {et~al.}}]{POLARBEAR:2015ixw}%
  \BibitemOpen
  \bibfield  {author} {\bibinfo {author} {\bibfnamefont {A.}~\bibnamefont {Suzuki}} \emph {et~al.} (\bibinfo {collaboration} {POLARBEAR}),\ }\href {\doibase 10.1007/s10909-015-1425-4} {\bibfield  {journal} {\bibinfo  {journal} {J. Low Temp. Phys.}\ }\textbf {\bibinfo {volume} {184}},\ \bibinfo {pages} {805} (\bibinfo {year} {2016})},\ \Eprint {http://arxiv.org/abs/1512.07299} {arXiv:1512.07299 [astro-ph.IM]} \BibitemShut {NoStop}%
\bibitem [{\citenamefont {Aiola}\ \emph {et~al.}(2020)\citenamefont {Aiola} \emph {et~al.}}]{ACT:2020gnv}%
  \BibitemOpen
  \bibfield  {author} {\bibinfo {author} {\bibfnamefont {S.}~\bibnamefont {Aiola}} \emph {et~al.} (\bibinfo {collaboration} {ACT}),\ }\href {\doibase 10.1088/1475-7516/2020/12/047} {\bibfield  {journal} {\bibinfo  {journal} {JCAP}\ }\textbf {\bibinfo {volume} {12}},\ \bibinfo {pages} {047} (\bibinfo {year} {2020})},\ \Eprint {http://arxiv.org/abs/2007.07288} {arXiv:2007.07288 [astro-ph.CO]} \BibitemShut {NoStop}%
\bibitem [{\citenamefont {Harrington}\ \emph {et~al.}(2016)\citenamefont {Harrington} \emph {et~al.}}]{Harrington:2016jrz}%
  \BibitemOpen
  \bibfield  {author} {\bibinfo {author} {\bibfnamefont {K.}~\bibnamefont {Harrington}} \emph {et~al.},\ }\href {\doibase 10.1117/12.2233125} {\bibfield  {journal} {\bibinfo  {journal} {Proc. SPIE Int. Soc. Opt. Eng.}\ }\textbf {\bibinfo {volume} {9914}},\ \bibinfo {pages} {99141K} (\bibinfo {year} {2016})},\ \Eprint {http://arxiv.org/abs/1608.08234} {arXiv:1608.08234 [astro-ph.IM]} \BibitemShut {NoStop}%
\bibitem [{\citenamefont {Addamo}\ \emph {et~al.}(2021)\citenamefont {Addamo} \emph {et~al.}}]{LSPE:2020uos}%
  \BibitemOpen
  \bibfield  {author} {\bibinfo {author} {\bibfnamefont {G.}~\bibnamefont {Addamo}} \emph {et~al.} (\bibinfo {collaboration} {LSPE}),\ }\href {\doibase 10.1088/1475-7516/2021/08/008} {\bibfield  {journal} {\bibinfo  {journal} {JCAP}\ }\textbf {\bibinfo {volume} {08}},\ \bibinfo {pages} {008} (\bibinfo {year} {2021})},\ \Eprint {http://arxiv.org/abs/2008.11049} {arXiv:2008.11049 [astro-ph.IM]} \BibitemShut {NoStop}%
\bibitem [{\citenamefont {Mennella}\ \emph {et~al.}(2019)\citenamefont {Mennella} \emph {et~al.}}]{Mennella:2019cwk}%
  \BibitemOpen
  \bibfield  {author} {\bibinfo {author} {\bibfnamefont {A.}~\bibnamefont {Mennella}} \emph {et~al.},\ }\href {\doibase 10.3390/universe5020042} {\bibfield  {journal} {\bibinfo  {journal} {Universe}\ }\textbf {\bibinfo {volume} {5}},\ \bibinfo {pages} {42} (\bibinfo {year} {2019})}\BibitemShut {NoStop}%
\bibitem [{\citenamefont {Ade}\ \emph {et~al.}(2022{\natexlab{b}})\citenamefont {Ade} \emph {et~al.}}]{SPIDER:2021ncy}%
  \BibitemOpen
  \bibfield  {author} {\bibinfo {author} {\bibfnamefont {P.~A.~R.}\ \bibnamefont {Ade}} \emph {et~al.} (\bibinfo {collaboration} {SPIDER}),\ }\href {\doibase 10.3847/1538-4357/ac20df} {\bibfield  {journal} {\bibinfo  {journal} {Astrophys. J.}\ }\textbf {\bibinfo {volume} {927}},\ \bibinfo {pages} {174} (\bibinfo {year} {2022}{\natexlab{b}})},\ \Eprint {http://arxiv.org/abs/2103.13334} {arXiv:2103.13334 [astro-ph.CO]} \BibitemShut {NoStop}%
\end{thebibliography}%
\end{document}